\documentclass[10pt,twocolumn,letterpaper]{article}

\usepackage[pagenumbers]{iccv} 

\definecolor{Green}{rgb}{0.0, 0.5, 0.0}

\newcommand{\red}[1]{{\color{red}#1}}

\newcommand{\vect}[1]{\ensuremath{\mathbf{#1}}}
\newcommand{\myparagraph}[1]{\vspace{0.5em}\noindent\textbf{#1}}

\definecolor{tabfirst}{rgb}{0.96, 0.77, 0.77} 
\definecolor{tabsecond}{rgb}{0.98 , 0.93, 0.77} 
\definecolor{tabthird}{rgb}{1, 1, 0.7} 

\newcommand{\tann}[1]{\textbf{\color{blue}tan: #1}}
\newcommand{\tree}[1]{\textbf{\color{violet}tree: #1}}

\usepackage{mathtools}
\DeclarePairedDelimiter{\parens}{\lparen}{\rparen}

\DeclarePairedDelimiter\abs{\lvert}{\rvert}

\usepackage{colortbl}
\usepackage{tabu}
\usepackage{multirow}
\usepackage{dsfont}
\usepackage{xcolor}

\usepackage[super]{nth}
\usepackage{lipsum}
\usepackage{algpseudocode}
\usepackage[ruled,vlined,linesnumbered]{algorithm2e}
\usepackage{amsmath}

\usepackage{caption}
\usepackage{makecell}
\usepackage{float}
\usepackage{placeins}

\usepackage{footmisc}

\definecolor{iccvblue}{rgb}{0.21,0.49,0.74}
\usepackage[pagebackref,breaklinks,colorlinks,allcolors=iccvblue]{hyperref}

\def\paperID{7811} 
\def\confName{ICCV}
\def\confYear{2025}

\title{LUSD: Localized Update Score Distillation for Text-Guided Image Editing}

\newcommand{\vistec}{%
  $^1$
}

\newcommand{\scb}{%
  $^2$
}

\newcommand{\siriraj}{%
  $^3$
}

\newcommand{\pixiv}{%
  $^4$
}

\newcommand{\treeaff}{%
  $^{1,3}$
}

\newcommand{\tanaff}{%
  $^{1,2}$
}

\renewcommand{\thefootnote}{\fnsymbol{footnote}}
\def\authorBlock{
    Worameth Chinchuthakun\footnotemark[1]\hspace{5pt}\footnotemark[2]\hspace{5pt}\tanaff \qquad
    Tossaporn Saengja\footnotemark[1]\hspace{5pt}\footnotemark[2]\hspace{5pt}\treeaff \qquad\quad
    Nontawat Tritrong\hspace{1pt}\vistec \\
    Pitchaporn Rewatbowornwong\hspace{1pt}\vistec \quad\quad
    Pramook Khungurn\hspace{1pt}\pixiv \qquad\quad
    Supasorn Suwajanakorn\hspace{1pt}\vistec \\
    \vspace{-0.65em}\\
    \vistec \hspace{-0.4em} VISTEC \qquad
    \scb \hspace{-0.4em} Siam Commercial Bank \qquad
    \siriraj \hspace{-0.4em} Faculty of Medicine Siriraj Hospital \qquad
    \pixiv \hspace{-0.4em} Pixiv \\ 
    
}
\author{\authorBlock}

\begin{document}
\maketitle
\begin{figure}[!t]
    \centering
    \vspace{-0.9em}
    \includegraphics[width=0.95\columnwidth]{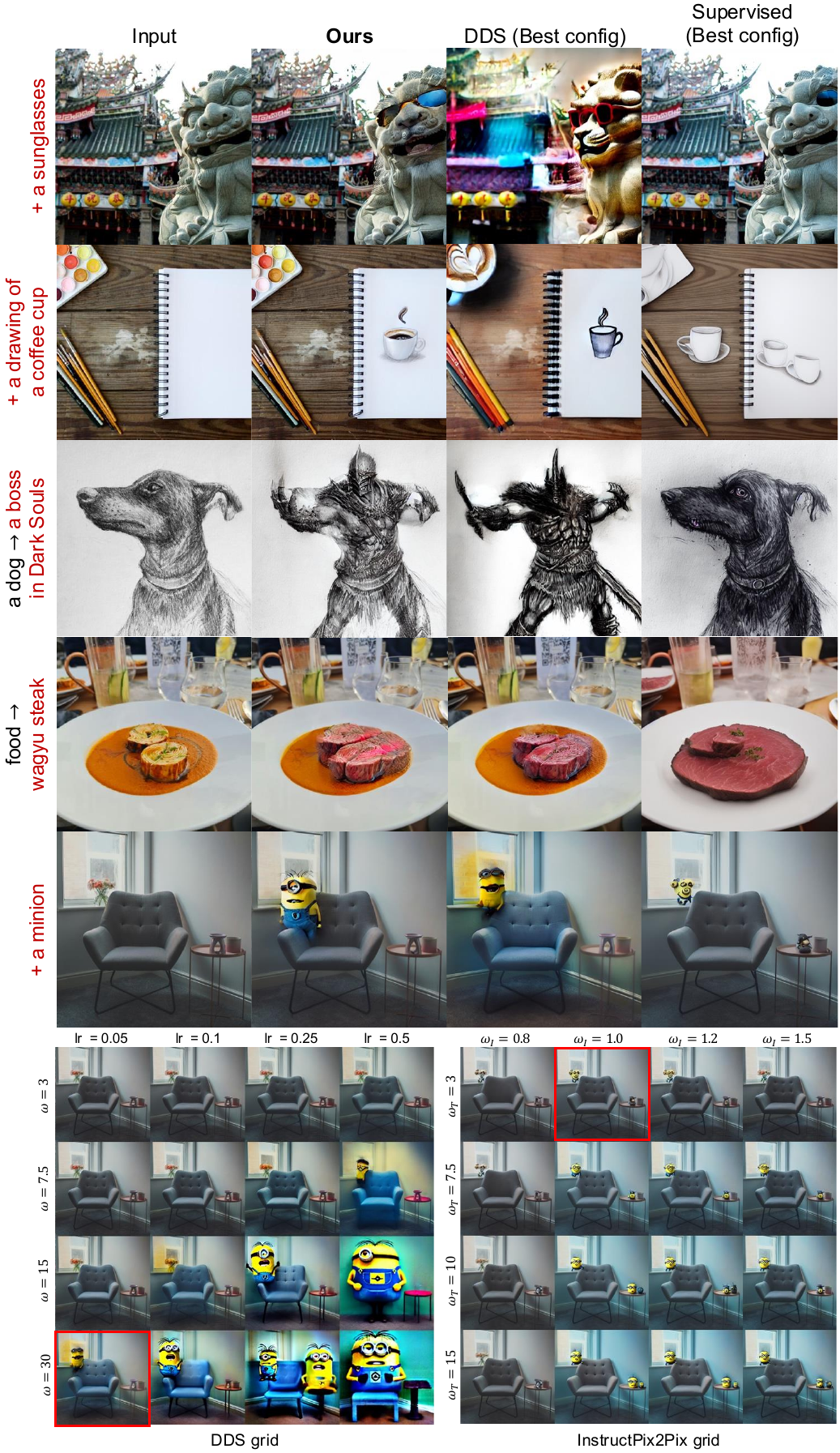}
        \vspace{-0.5em}
    \caption{We propose a novel score distillation technique for object insertion and image editing tasks. Compared to existing score distillation methods, (e.g., DDS~\cite{hertz2023delta}) and supervised methods (e.g., InstructPix2Pix~\cite{brooks2022instructpix2pix}), our LUSD achieves a higher success rate with superior background preservation.
    We compare the best results (prioritizing object appearance) from state-of-the-art methods, optimized through hyperparameter tuning (highlighted by red boxes), with our results all generated using a \emph{single} configuration. Hypertuning grids for other images are in Appendix \ref{supp:tuning_grid}.}
    \label{fig:teaser}
    \vspace{-12pt}
\end{figure}

\vspace{-1em}
\begin{abstract}
\footnotetext[1]{Authors contributed equally to this work.}
\footnotetext[2]{Work done during research assistantships at VISTEC.}
While diffusion models show promising results in image editing given a target prompt, achieving both prompt fidelity and background preservation remains difficult.
Recent works have introduced score distillation techniques that leverage the rich generative prior of text-to-image diffusion models to solve this task without additional fine-tuning.
However, these methods often struggle with tasks such as object insertion.
Our investigation of these failures reveals significant variations in gradient magnitude and spatial distribution, making hyperparameter tuning highly input-specific or unsuccessful. To address this, we propose two simple yet effective modifications: attention-based spatial regularization and gradient filtering-normalization, both aimed at reducing these variations during gradient updates. Experimental results show our method outperforms state-of-the-art score distillation techniques in prompt fidelity, improving successful edits while preserving the background. Users also preferred our method over state-of-the-art techniques across three metrics, and by 58-64\% overall.

\end{abstract}
\renewcommand{\thefootnote}{\arabic{footnote}}
\vspace{-1em}
\section{Introduction}

\label{sec:intro}

In the problem of \emph{text-guided image editing}, we are given an image and a text prompt, and the goal is to modify the image to match the prompt. Unlike image generation, editing requires preserving elements of the input image, such as a fox's face when \emph{adding sunglasses}, a cat's outline when \emph{changing its color}, and the overall structure of an outdoor scene when \emph{transitioning it from summer to winter}.

Recent approaches leverage large-scale text-to-image diffusion models \cite{ho2020denoising, song2020denoising}, such as Stable Diffusion \cite{rombach2021highresolution}, to tackle the task.
Supervised methods \cite{brooks2023instructpix2pix, zhang2024hive, sheynin2023emuedit} fine-tune or train diffusion models on large-scale synthetic datasets conditioned on edit instructions, enabling intuitive user interaction via natural language prompts. 
For example, a user may provide an image and ask the system to ``\emph{add a cat on the sofa}''. 
In contrast, zero-shot methods \cite{meng2022sdedit, couairon2022diffedit, hertz2022prompt, parmar2023zeroshot, brack2023ledits} attempt to invert the diffusion process of the input image and \textit{regenerate} it with a new prompt.
Recently, Score Distillation Sampling (SDS) \cite{poole2022dreamfusion} has emerged as a promising alternative. Instead of relying on additional training data or diffusion inversion, SDS leverages the prior from pre-trained diffusion models to optimize the input image to align with the text prompt. 
DDS \cite{hertz2023delta} extends SDS by reducing noisy gradient directions, which helps preserve the original content and can be enhanced with additional regularization to better maintain structural information~\cite{nam2024contrastive}.

Despite many attempts, editing an image to match a prompt while preserving the background remains challenging. Supervised methods typically perform well only in limited scenarios, as they rely on small or synthetic training sets, which can be biased and fail to capture the diversity of real-world cases (Figure \ref{fig:teaser}).  
Notably, these methods also struggle to preserve the background---a limitation shared by most zero-shot methods as they often depend on inferred implicit binary masks that can be inaccurate.
While SDS-based methods can preserve the background better, they sometimes fail to insert objects altogether (see Figure \ref{fig:related_work_score_distillation}). Moreover, they only work within a narrow range of hyperparameters, requiring tuning for each input image. 
Object insertion, which involves deciding \emph{where} and \emph{how} to generate objects from scratch, poses even greater challenges for these methods, especially for unusual combinations, such as ``a Chinese lion statue wearing sunglasses'' (Figure \ref{fig:teaser}). 

In this paper, 
we investigate a solution based on score distillation and its associated challenges.
For such a method, an input image is gradually transformed through gradient updates derived from the denoising process of a text-conditioned diffusion model.
One difficulty lies in the extreme variations in gradient magnitudes, which make it difficult to determine the correct learning rate or apply a regularizer to preserve the background. These variations can come from simply changing the prompt or the input image. 
Moreover, even when the text prompt and image are fixed, the denoising process with different noise seeds still strongly influences the gradient magnitude and its spatial distributions \cite{liang2023luciddreamer}, leading to gradients in multiple locations counteracting each other's progress.



Our method, Localized Update Score Disilltation (LUSD), builds upon a score-distillation formulation~\cite{mcallister2024rethinking} with a simple $L_2$ regularizer that pulls updates toward the original image and incorporates two key ideas. First, to reduce variation in spatial distributions, we track the spatial locations of the edits made by SDS using attention-based features. By computing a moving average of these estimated locations during optimization and using it to modulate the gradients, updates progressively focus on narrower areas, increasing the rate at which new objects appear and allowing the background to be preserved better. Second, we implement a normalization and thresholding mechanism to filter out ``counterproductive'' gradients, identified by their low standard deviation. 




We evaluate our method against InstructPix2Pix \cite{brooks2022instructpix2pix}, HIVE \cite{zhang2024hive}, LEDITS++ \cite{brack2023ledits}, DDS \cite{hertz2023delta}, and CDS \cite{nam2024contrastive} on a standard image editing benchmark, MagicBrush \cite{zhang2023magicbrush}. Our method achieves a higher success rate in object addition and preserves the original background more effectively than the competitors. Additionally, it retains the general editing capabilities inherent to score distillation.

To summarize, our contributions are:
\begin{itemize}
    \item An analysis of gradient behavior, identifying key factors that hinder effective object insertion in score distillation.
    \item A novel attention-based spatial regularization and gradient normalization for mitigating \emph{bad gradients} effects.
\end{itemize}

\section{Related Work}
\label{sec:related_work}




\begin{figure}
    \centering
    \includegraphics[width=0.95\columnwidth]{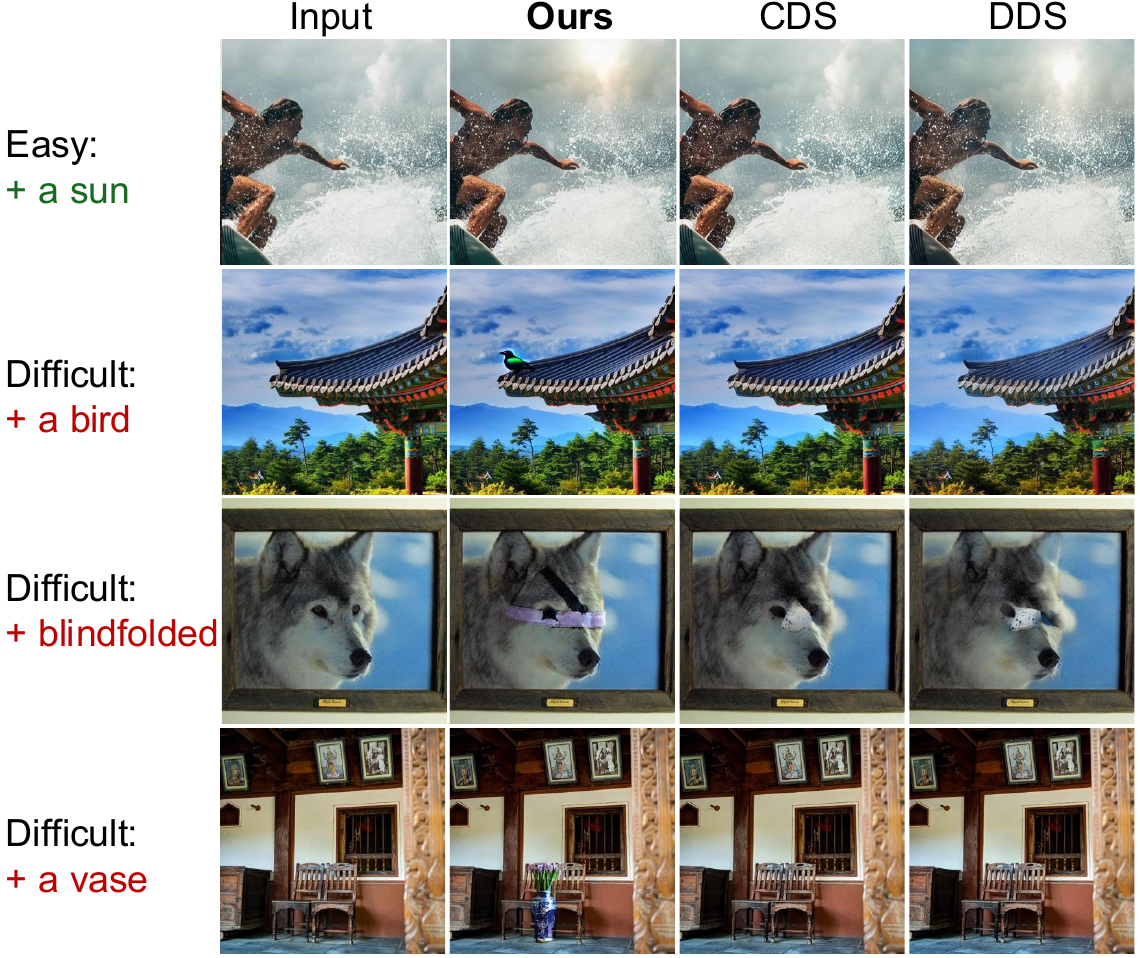}
    \captionsetup{skip=5pt}
    \caption{CDS \cite{nam2024contrastive} and DDS \cite{hertz2023delta} tend to generate incomplete objects, or completely fail to produce one. Our method mitigates these issues while preserving the background.}
    \label{fig:related_work_score_distillation}
    \vspace{-12pt}
\end{figure}

\myparagraph{Diffusion-based text-guided image editing.}
While there exist image editing techniques that require binary masks \cite{yang2022paint, avrahami2022blended, avrahami2023blended, xie2022smartbrush, wang2023imageneditor, zhuang2024task} or other conditions \cite{zhang2023adding}, we focus on approaches that do not require such explicit spatial conditions, categorized into zero-shot and supervised methods.


Most zero-shot methods first invert the input image along its diffusion trajectory conditioned on a caption and then denoise the result using a new target caption.
The inversion process can be achieved with DDIM \cite{song2020denoising}, DDPM \cite{ho2020denoising, huberman2024edit}, optimized text embeddings \cite{mokady2022null, miyake2023negative}, DPM-Solver++ \cite{brack2023ledits}, or simply adding noise as in SDEdit \cite{meng2022sdedit}. 
TiNO-Edit \cite{chen2024tino} proposes a strategy to determine the optimal timestep and noise required for this process.
To better preserve content of the input image, some methods derive an implicit mask to guide the editing process. 
DiffEdit \cite{couairon2022diffedit} infers such a mask by computing the noise difference when conditioning the model on the source and target texts. Other works \cite{hertz2022prompt, parmar2023zeroshot, tumanyan2022plugandplay, cao2023masactrl} extract attention features of the noised image during the inversion process, then inject them to preserve the original structure when denoising.
LEDITS++ \cite{brack2023ledits} combines both attention-based and noise-based methods to generate a more precise editing mask. 
Some works \cite{rout2025semantic, wang2024taming} focus on rectified flow models based on the MM-DiT architecture~\cite{esser2024scaling}; however, they are incompatible with diffusion models as they depend on rectified ODE properties \cite{rout2025semantic} or MM-DiT's multi-modal self-attention \cite{wang2024taming}.




Another line of work trains diffusion models conditioned on an edit instruction on large-scale synthetic datasets for general-purpose editing models.
InstructPix2Pix \cite{brooks2023instructpix2pix} synthesizes their training samples using Prompt-to-Prompt \cite{hertz2022prompt}, whereas Hive \cite{zhang2024hive} enhances this pipeline by fine-tuning with human feedback.
Emu Edit \cite{sheynin2023emuedit} improves data generation by localizing the editing area with a more accurate mask derived from large language models. It also introduces task-specific embeddings and trains the model on a wider range of tasks, resulting in better generalization.

Still, editing an image to align with a target text prompt without altering unrelated regions remains challenging.
Inversion-based methods contain no explicit constraints to preserve the background, except for an approximated mask, and
instruction-based supervised methods suffer from imperfect, synthetic datasets.
Our method is based on score distillation, an optimization-based approach that can easily incorporate a loss term to keep the background intact.




\begin{figure}
    \centering
    \includegraphics[width=1.0\columnwidth]{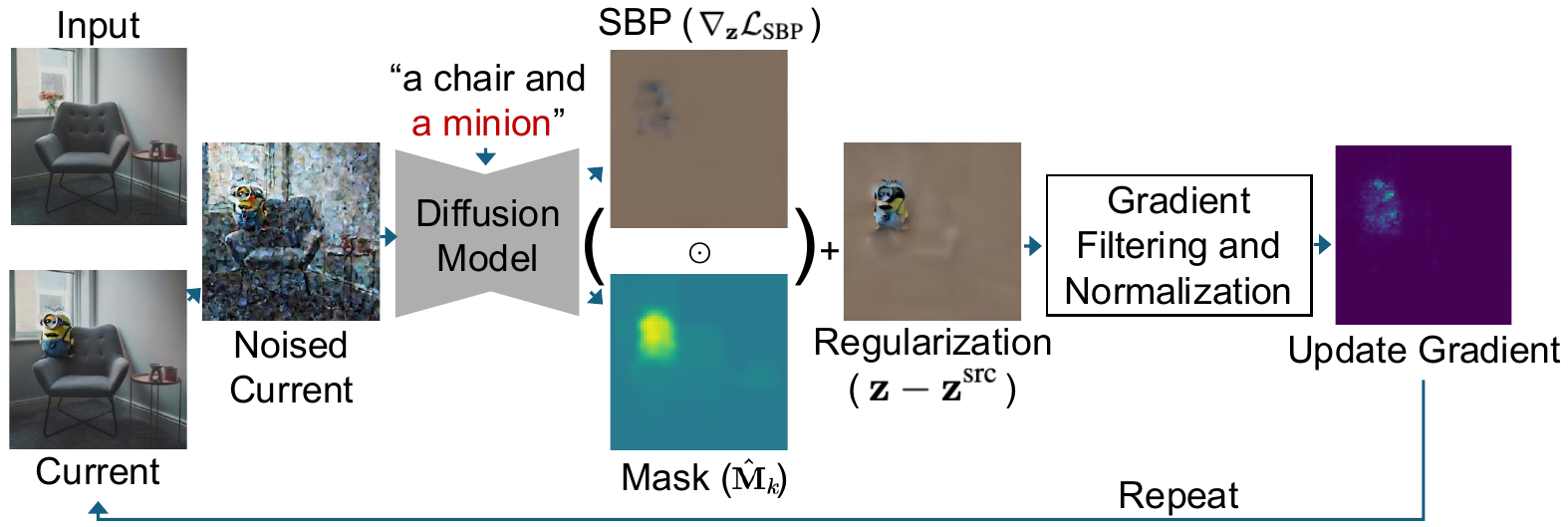}
    \caption{An overview of our method. Given an input image and a target prompt, we obtain gradient of the SBP loss \cite{mcallister2024rethinking} and an attention-based mask. With spatial regularization, gradient filtering and normalization, we modify the image to match the prompt.}
    \vspace{-12pt}
    
    \label{fig:overview}
\end{figure}

\myparagraph{Score distillation.} Score distillation technique has shown promising performance in text-to-3D generation, although with some flaws. 
The original formulation, SDS \cite{poole2022dreamfusion}, often produces blurry and over-saturated outputs due to its mode-seeking behavior and requires high classifier-free guidance (CFG) values \cite{ho2022classifier}. 
Subsequent formulations \cite{katzir2023noisefree, wang2023prolificdreamer, yu2023text, alldieck2024score, mcallister2024rethinking} addressed these issues to improve output quality. 
Recent works \cite{hertz2023delta, nam2024contrastive, koo2024posterior, burgert2023diffusion, kim2024dreamsampler} have also adapted these techniques to 2D image editing.
DDS \cite{hertz2023delta} leverages the source caption and the input image to reduce noisy gradients by using the difference between the gradients of the target pair and the source pair. CDS \cite{nam2024contrastive} builds upon DDS by regularizing structural changes with a CUT loss \cite{park2020cut} derived from the self-attention features of the diffusion model to better preserve the source image's structure. 
DreamSampler \cite{kim2024dreamsampler} explores data consistency terms controlled by a weighting hyperparameter $\lambda$ and improves noise schedules through reverse diffusion sampling. 

Although DDS and CDS excel at object replacement and global attribute manipulation (e.g., color, style), they struggle with inserting new objects (Figure \ref{fig:related_work_score_distillation}), as background preservation is not part of their optimization objectives. While DreamSampler explores regularizers, conditioning solely on $\lambda$ can be sensitive to variations in gradient magnitudes. Our method mitigates these variations, enhancing object insertion while preserving the input image.

\section{Approach}
\label{sec:method}

Given an input \emph{source image} and a \emph{target prompt} describing how the image should be modified, our goal is to modify the image to match the prompt. 
We focus on modifications that preserve parts of the original image, avoiding a complete transformation. For example, prompts may add an object, such as ``a cat \emph{wearing a hat}'', or adjust global features while retaining the image's structure, such as ``a city \emph{in winter}'' to add snow and make the sky cloudy.

Our method, LUSD, is based on score distillation sampling~\cite{poole2022dreamfusion}. We introduce regularization techniques to better preserve the background and a method to filter and normalize updates so that the optimization process become robust under diverse inputs without needing instance-specific hyperparameters. An overview is shown in Figure \ref{fig:overview}. 
We begin with a review of SDS and explain our choice of an SDS-based method we extend (Section \ref{sec:prelim}).
Next, we introduce the loss term for background preservation (Section~\ref{sec:method_challenges}) and then propose an attention-based spatial regularization (Section \ref{sec:method_attention}) and a gradient filtering-normalization technique (Section \ref{sec:method_gradients}) that improve the algorithm's reliability. 

\subsection{Preliminaries} \label{sec:prelim}


\myparagraph{Diffusion models} \cite{ho2020denoising} form a family of generative models that learn a target data distribution $p_{\mathrm{data}}$ by transforming samples from a noise distribution. 
A diffusion model $\epsilon_{\phi}$ is trained to predict noise $\epsilon \sim \mathcal{N}(0,I)$ that is used to generate a noisy sample $\vect{x}_t = \sqrt{\alpha_t} \vect{x} + \sqrt{1-\alpha_t}\epsilon$ where $\vect{x} \sim p_{\mathrm{data}}$ and $\alpha_t$ denotes the noise schedule of the model. The training loss is given by
\begin{equation} \label{eq:diffusion_objective}
    \mathcal{L} = \mathbb{E}_{\vect{x}, t, \vect{\epsilon}} \big[ \lVert \epsilon_{\vect{\theta}} ( \vect{x}_t, y, t) - \vect{\epsilon} \rVert_2^2 \big],
\end{equation}
where $y$ denotes conditioning signals such as text. In this paper, we use Stable Diffusion \cite{rombach2021highresolution}, which operates on noisy latent codes $\vect{z}_t = \sqrt{\alpha_t} \mathbf{z} + \sqrt{1 - \alpha_t} \epsilon$ where $\vect{z}$ is the (noise-free) latent code of a (noise-free) image $\vect{x}$. Note that $\vect{x}$ and $\vect{z}$ are related by $\vect{x} = \mathcal{D}(\vect{z})$ and $\vect{z} = \mathcal{E}(\vect{x})$ where $\mathcal{D}$ and $\mathcal{E}$ are the decoder and encoder of a variational autoencoder (VAE), respectively. Our method optimizes $\vect{z}$, which finally must be converted to an image with $\mathcal{D}$.

\myparagraph{Score distillation sampling} \cite{poole2022dreamfusion} uses $\epsilon_{\phi}$ to optimize a set of parameters $\theta$ that gets converted to a (noise-free) latent code $\vect{z}$ through a differentiable function $\vect{z} = g(\theta)$, given a text condition $y$. The loss function is defined implicitly by setting its gradient with respect to $\theta$ to be the gradient of \eqref{eq:diffusion_objective}, also with respect to $\theta$, without the Jacobian term $\partial \epsilon_{\phi}^{\omega} / \partial \vect{z}_t$:
\begin{align}\
    \nabla_{\theta} \mathcal{L}_{\text{SDS}} &= \mathbb{E}_{t,\epsilon} \Big [ \left( \epsilon_{\phi}^{\omega}(\vect{z}_t, y, t) - \epsilon \right) \frac{\partial \vect{z}_t}{\partial \theta} \Big], \label{eq:sds_loss} \\
    \epsilon_{\phi}^{\omega}(\vect{z}_t, y, t) &= (1 + \omega) \epsilon_{\phi}(\vect{z}_t, y, t) - \omega \epsilon_{\phi}(\vect{z}_t, t), \label{eq:cfg} \notag
\end{align}
where $\omega$ is the classifier-free guidance (CFG) \cite{ho2022classifier} scale.
DDS \cite{hertz2023delta} extends this formulation to transform the latent code of a source image $\vect{z}^{\text{src}} = \mathcal{E}(\vect{x}^{\mathrm{src}})$ into a target latent code $\vect{z}$, which aligns with a target text prompt $y^{\text{tgt}}$ while preserving the source content. 
DDS simplifies the problem by setting $\theta = \vect{z}$ and $g$ to the identity function. The DDS loss replaces the noise $\epsilon$ in \eqref{eq:sds_loss} with the noise predicted from $\vect{z}^{\text{src}}_t = \sqrt{\alpha_t} \vect{z}^{\text{src}} + \sqrt{1-\alpha_t}\epsilon$ and the source prompt $y^{\text{src}}$:
\begin{align*}
    \nabla_{\vect{z}} \mathcal{L}_{\text{DDS}} = \mathbb{E}_{t,\epsilon} \Big[ \left( \epsilon_{\phi}^{\omega}(\vect{z}_t, y^{\text{tgt}}, t) - \epsilon_{\phi}^{\omega}(\vect{z}^{\text{src}}_t, y^{\text{src}}, t) \right) \frac{\partial \vect{z}_t}{\partial \vect{z}} \Big].
\end{align*}
The source prompt can be either provided by the user or automatically generated by vision-language models such as BLIP \cite{li2022blip} or Chat-GPT~\cite{openai2023chatgpt}.

While DDS reduces noisy gradient directions, according to McAllister \etal~\cite{mcallister2024rethinking}, it does not achieve an accurate estimation of the source distribution because the $\epsilon_{\phi}^{\omega}$ term is not computed from $\vect{z}$, the latent code being optimized. As such, we adopt their SBP loss, which replaces $\vect{z}_t^{\mathrm{src}}$ with $\vect{z}_t$:
\begin{align*}
    \nabla_{\vect{z}} \mathcal{L}_{\text{SBP}} = \mathbb{E}_{t,\epsilon} \Big[ \left( \epsilon_{\phi}^{\omega}(\vect{z}_t, y^{\text{tgt}}, t) - \epsilon_{\phi}^{\omega}(\vect{z}_t, y^{\text{src}}, t) \right) \frac{\partial \vect{z}_t}{\partial \vect{z}}\Big].
\end{align*}

\subsection{Regularization for background preservation} \label{sec:method_challenges}

\begin{figure}
    \centering
    \includegraphics[width=1.0\columnwidth]{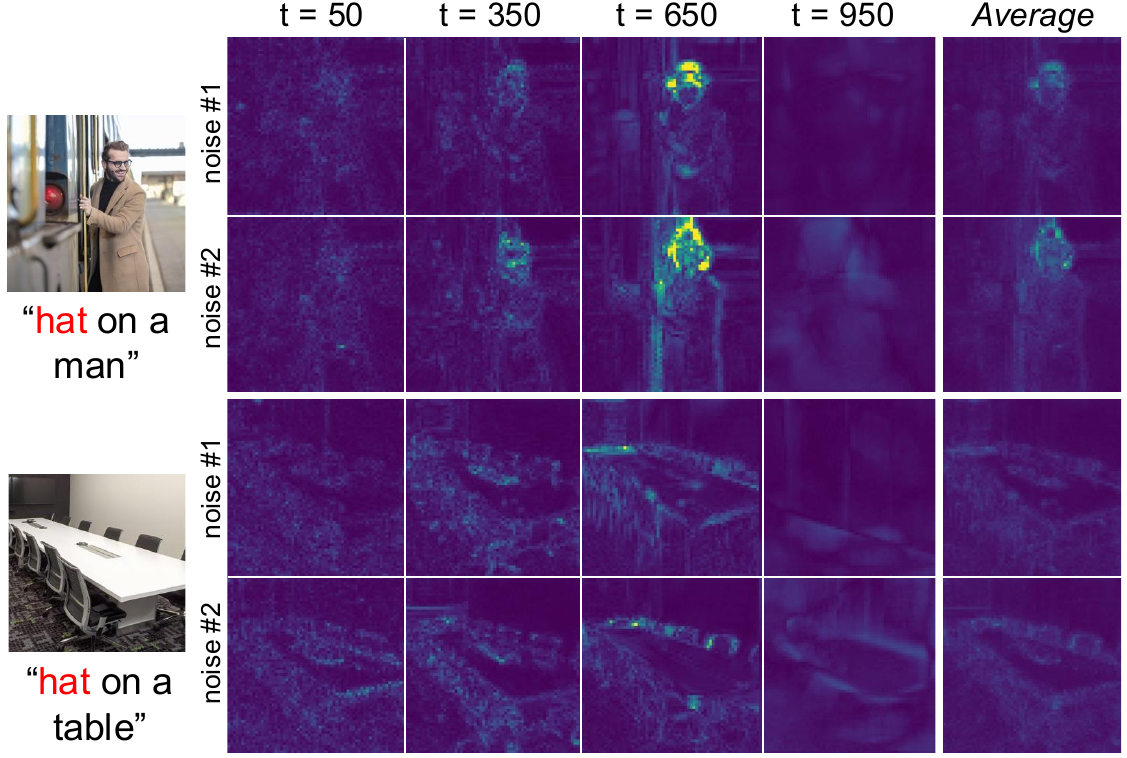}
    \captionsetup{skip=5pt}
    \caption{
    Gradients vary with different timesteps, noises, prompts, images, as well as the number of plausible placements for ``hat.''
    }
    \label{fig:difficulty}
    \vspace{-6pt}
\end{figure}

For many image editing tasks, such as inserting objects or transforming certain elements, it is essential to preserve background areas unrelated to the edit. However, previous SDS-based methods often lack explicit background constraints, leading to unintended changes. As previously explored in DDS \cite{hertz2023delta} and DreamSampler \cite{kim2024dreamsampler}, a straightforward solution is to add a regularizing term:
\begin{equation}
    \nabla_{\vect{z}} \mathcal{L}_{\text{SBP-reg}} = (1 - \lambda) \nabla_{\theta} \mathcal{L}_{\text{SBP}} + \lambda (\vect{z} - \vect{z}^{\text{src}}). \label{eq:sbp-with-reg}
\end{equation}
An issue with this formulation is that it is sensitive to the hyperparameter $\lambda$, whose optimal value can vary between inputs as different images and prompts naturally yield different gradients. An additional source of variability in tasks like object insertion is whether the object to be added has a single primary location, like a hat on a person, or multiple plausible locations, like a hat on a table. In this example, the gradient $\mathcal{L}_{\text{SBP}}$ \emph{averaged} over multiple optimization steps for a hat on a table will be much weaker than of a hat on the person's head (Figure \ref{fig:difficulty}).
Moreover, even with a fixed image and prompt, different sampled noises $\epsilon$ yield gradients with widely varying intensities and spatial distributions.



\subsection{Attention-based spatial regularization} \label{sec:method_attention}

Our idea is to use a fixed $\lambda$ value and to modulate the gradient $\nabla_{\vect{z}} \mathcal{L}_{\text{SBP}}$ in \eqref{eq:sbp-with-reg} so that it becomes low in areas that are likely to be the background. In such an area, the regularizing term will dominate, and the content there is thus encouraged to be the same as that in the source image. 

\begin{figure}[t]
    \centering
    \includegraphics[width=1\columnwidth]{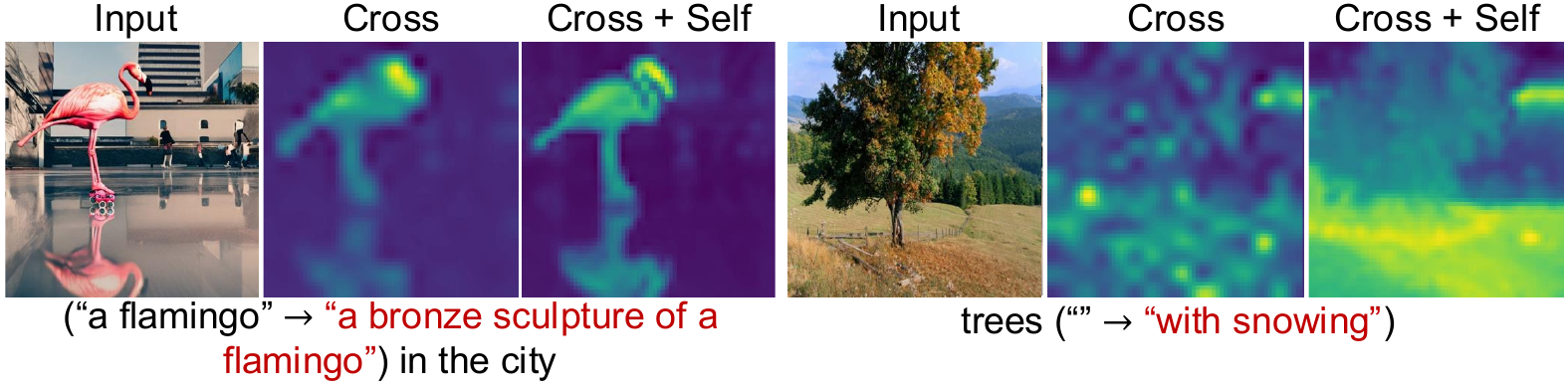}
    \captionsetup{skip=4pt}
    \caption{Refining cross-attention using self-attention \cite{quangtruong2023dataset} produces editing masks with higher contrast and sharper boundaries. 
    }
    
    \label{fig:compare_cross_self}
    \vspace{-0.6em}
\end{figure}

To modulate the gradient, we estimate a mask of the edited region (i.e., the inverse of background region) using a technique called self-attention exponentiation~\cite{quangtruong2023dataset}, where the mask is computed from attention maps inside the diffusion model's intermediate activations. Specifically, we assume that the diffusion model $\epsilon_\phi$ is a U-Net with $L$ self-attention layers, and each self-attention layer is immediately followed by a cross-attention layer \cite{rombach2021highresolution}. When we compute $\epsilon_{\phi}(\vect{z}_t, y^{\text{tgt}}, t)$ in each optimization step, we extract the $l^\text{th}$ self-attention maps $\vect{A}_{S}^{l, t}$ and a set of associated cross-attention maps $\{ \vect{A}_{C}^{l, t, \vect{e} } \}$ 
where $\vect{e}$ denotes a token in the text embedding of the target prompt $y^{\text{tgt}}$.
In our case, we extract such maps for all \emph{noun} tokens $\vect{e}$ associated with the edit, as in \cite{chefer2023attendandexcite}, which can be inferred by comparing the source and target prompts (see Appendix \ref{supp:implement}).
We then average these maps across all layers to obtain $\vect{A}_S^{t}$ and $\{ \vect{A}_C^{t,\vect{e}} \}$. Let $N$ denote the spatial size of the tensor from which the largest self-attention map is computed.\footnote{For Stable Diffusion, $N = 1024 = 32 \times 32$.} We may view $\vect{A}_S^{t}$ as an $N \times N$ matrix, and each $\vect{A}_C^{t,\vect{e}}$ as an $N \times 1$ vector. We then compute the enhanced cross-attention map $\hat{\vect{A}}_C^{t,\vect{e}} = \vect{A}_S^{t} \vect{A}_C^{t,\vect{e}}$ as a matrix-vector product. Finally, we average $\hat{\vect{A}}_C^{t,\vect{e}}$ across all noun tokens $\vect{e}$ to obtain $\hat{\vect{A}}_C^{t}$. As shown in Figure \ref{fig:compare_cross_self}, this method produces cross-attention maps with greater contrast and sharper boundaries, better highlighting edited areas.

In practice, the raw values of $\hat{\vect{A}}_C^{t}$ vary across different image-text pairs. 
To address this, following \cite{quangtruong2023dataset}, we apply min-max normalization to transform each pixel in $\hat{\vect{A}}_C^{t}$ to the range $[0,1]$, yielding a normalized mask $\vect{M}$.
During optimization, we compute a moving average of the mask as $\vect{M}_k = (1 - \alpha) \vect{M}_{k-1} + \alpha \vect{M}$, where $k$ is the optimization step and $\vect{M}_1 = \vect{M}$, to prevent sudden changes in the estimated mask. The gradient update at the $k$-th step is given by:
\begin{align*} 
    \nabla_{\vect{z}} \mathcal{L}_{\text{SBP-reg}} &= (1 - \lambda) (\hat{\vect{M}}_k \odot \nabla_{\vect{z}} \mathcal{L}_{\text{SBP}}) + \lambda (\vect{z} - \vect{z}^{\text{src}}), \\
    \hat{\vect{M}}_k &= \beta \vect{M}_k + (1 - \beta)\mathds{1},
\end{align*}
where $\odot$ denotes element-wise product, $\mathds{1}$ is an vector of ones with the same size as $\vect{M}$, and $\beta$ is a hyperparameter controlling the effect of $\vect{M}_k$. We found that linearly increasing $\beta$ from 0 to 1 during optimization generally suffices.

\subsection{Gradient filtering and normalization} \label{sec:method_gradients}
\begin{figure}
    \centering
    \includegraphics[width=0.9\columnwidth]{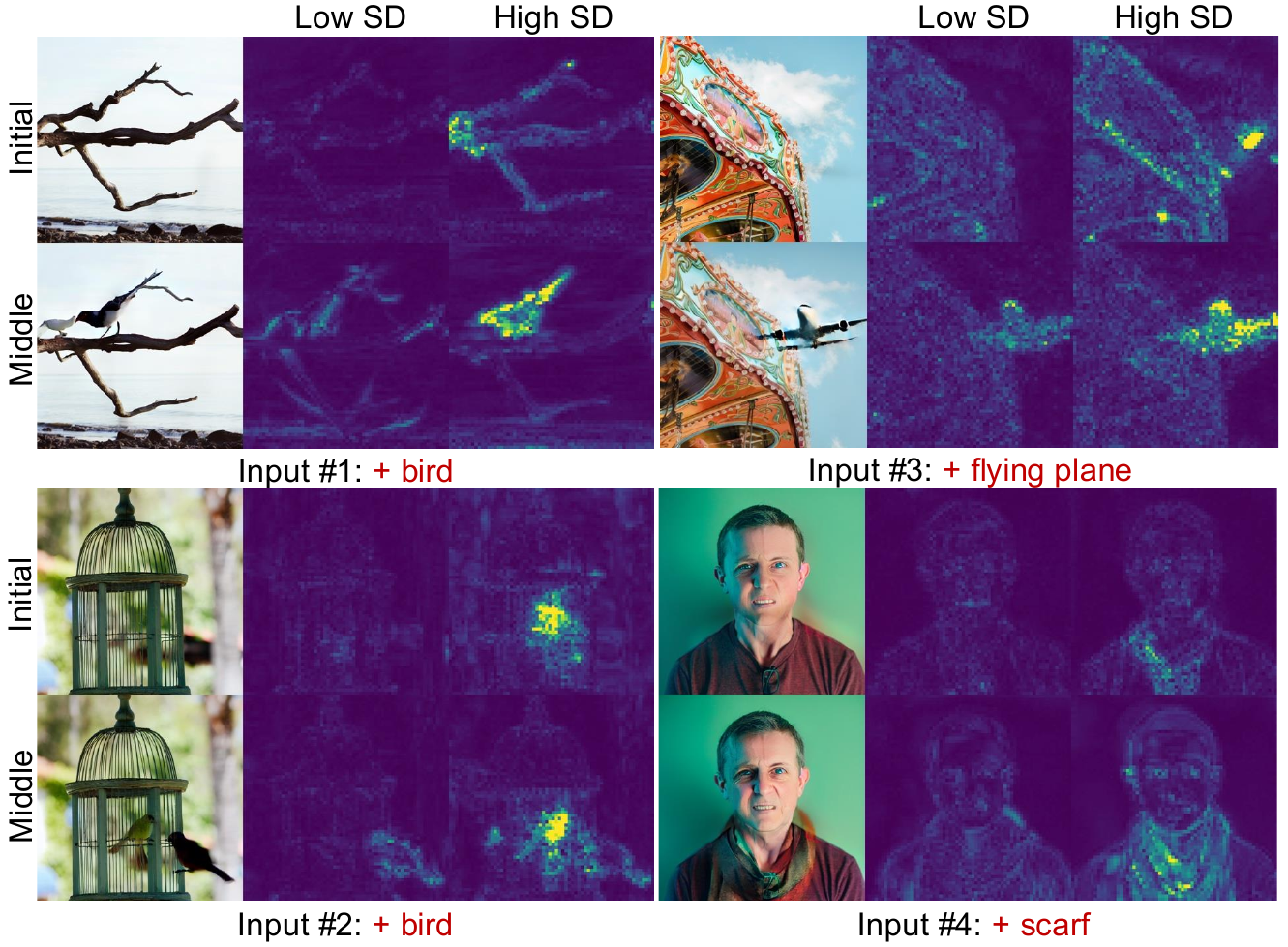}
    \captionsetup{skip=3pt}
    \caption{Example gradients with low or high standard deviations sampled from the beginning or middle of the optimization process.
    Low-SD gradients, which are less focused and counterproductive, are filtered out as they can revert the optimized image to the input.
    }
    \label{fig:grad_threshold}
    \vspace{-12pt}
\end{figure}

Another challenge we observe is that some noise samples $\epsilon$ produce gradients with very low magnitudes that scatter across the image and fail to drive meaningful progress. When combined with regularization, these weak gradients can even cause the optimized image to revert to the input, as the regularization overpowers $\nabla_{\vect{z}} \mathcal{L}_{\text{SBP}}$.
Such gradients tend to be less localized, and pixel values of $\nabla_{\vect{z}} \mathcal{L}_{\text{SBP}}$ tend to have a small standard deviation (Figure \ref{fig:grad_threshold}).


To prevent reversion, we detect the above ``bad'' gradient using a simple test: if the standard deviation of the pixel values of $\nabla_{\vect{z}} \mathcal{L}_{\text{SBP}}$ is below a certain threshold $\eta$, then the gradient is bad. When a bad gradient is found, we repeatedly sample a new noise $\epsilon$ while keeping the timestep $t$ constant until a ``good'' gradient is found for that step of the optimization process. While the range of standard deviations for good gradients can vary across images, our goal here is to avoid problematic ones and allow any sufficiently good gradients to make changes to the image. To ensure that progress is made even when $\eta$ is set too high, we begin the optimization process with an initial threshold $\eta_0$ and exponentially decay it whenever a gradient fails the test. Once a good gradient is found, we reset the threshold back to $\eta_0$.

Finally, to ensure a consistent optimization process and avoid issues with small gradients stalling progress or large gradients causing instability and highly saturated outputs, we normalize the gradients with its standard deviation:
\begin{equation}\label{eq:grad_norm}
    \nabla_{\vect{z}} \mathcal{L}_{\text{LUSD}} = \gamma \frac{\nabla_{\vect{z}} \mathcal{L}_{\text{SBP-reg}}}{\text{SD}(\nabla_{\vect{z}} \mathcal{L}_{\text{SBP-reg}})}.
\end{equation}
where $\gamma$ is a hyperparameter that enables annealing of the optimization process to gradually decrease fluctuation.

\section{Experiments}
\label{sec:experiment}

\myparagraph{Implementation details.} We initialize the latent code $\vect{z}$ with $\vect{z}^{\text{src}} = \mathcal{E}(\vect{x}^{\text{src}})$ and use our method to optimize it with the SGD optimizer (PyTorch's \texttt{torch.optim.SGD}) for 300 steps with a learning rate of 2000. Then, we decode $\vect{z}$ to the get output $\vect{x} = \mathcal{D}(\vect{z})$. In each optimization step, we sample timestep $t \sim U(50, 950)$ and set the CFG scale $\omega$ to $0$. To preserve the background (Section \ref{sec:method_attention}), we set $\lambda = 0.02$ and compute mask $\vect{M}_k$ with $\alpha = 0.1$. Following \cite{quangtruong2023dataset}, we extract cross- and self-attention maps from layers with the spatial resolution of 16 and 32, respectively. For gradient magnitude annealing (Section \ref{sec:method_gradients}), the values of $\gamma$ follow a reverse sigmoid schedule, transforming the range $[-5, 5]$ with a sigmoid function and scaling it to $[0.01, 0.15]$. We filter gradients with an initial threshold $\eta_0 = 0.01$, decaying it exponentially by 0.99 after each rejection.

\myparagraph{Baselines.} We compare our method to state-of-the-art diffusion-based ones from three different categories: (1) instruction-guided (InstructPix2Pix or IP2P \cite{brooks2022instructpix2pix} and Hive \cite{zhang2024hive}), (2) diffusion inversion-based (LEDITS++ \cite{brack2023ledits}),  and (3) score distillation (DDS \cite{hertz2023delta} and CDS \cite{nam2024contrastive}).

We provide reference comparisons with task-specific supervised object insertion methods \cite{wasserman2024paint, zhao2024diffree} and rectified-flow-based methods \cite{rout2025semantic, wang2024taming} in Appendix \ref{supp:object_insertion} and \ref{supp:rectifed_flow}, as these tackle related tasks but are orthogonal to our contributions of stabilizing score distillation for general image editing in diffusion models. 


\myparagraph{Dataset.} We use the MagicBrush test set \cite{zhang2023magicbrush}, a standard benchmark for image editing featuring diverse types of edits. 
It contains 1,053 examples, each with (1) a source image $\vect{x}^{\text{src}}$, (2) its global description $y^{\text{src}}$, (3) a ground-truth target image $\vect{x}^{\text{tgt}}$, (4) its global description $y^{\text{tgt}}$, (5) an edit instruction $y^{\text{edit}}$, and (6) a local description of the edited region $y^{\text{local}}$.
We use $y^{\text{edit}}$ as input for instruction-guided models and global descriptions $y^{\text{src}}$ and $y^{\text{tgt}}$, which describe the \textit{entire} image, to condition all other methods, including ours.

\subsection{Human evaluation} \label{sec:user_study} We conducted a user study to assess the quality of image editing using 200 randomly selected samples from the MagicBrush test set~\cite{zhang2023magicbrush}. For each sample, we presented a side-by-side comparison between our editing results and those of a state-of-the-art competitor. Each comparison was shown to 5 workers on Amazon Mechanical Turk, who were asked to choose their preferred image based on four criteria: (1) background preservation, (2) prompt fidelity, (3) quality of edited elements, and (4) overall preference.
These criteria are similar to those proposed by EditVal \cite{basu2023editval}.
As shown in Table \ref{tab:main_user_study}, our method outperforms both instruction-based and global description-based competitors across all metrics.
See Appendix \ref{supp:user_study} for details on the study design.

\vspace{-4pt}
\begin{table}[!h]
\centering
\small


\begin{tabular}{
    l@{\hspace{5pt}}
    c@{\hspace{5pt}}
    c@{\hspace{5pt}}
    c@{\hspace{5pt}}
    c
}
\toprule
\textbf{Method} & \textbf{Background} & \textbf{Prompt} & \textbf{Quality} & \textbf{Overall} \\
\midrule
IP2P \cite{brooks2022instructpix2pix} & 33.5\% & 40.0\% & 36.5\% & 36.0\% \\
HIVE \cite{zhang2024hive}           & 47.0\% & 40.5\% & 45.0\% & 39.0\% \\
LEDITS++ \cite{brack2023ledits}        & 35.5\% & 33.0\% & 37.0\% & 35.0\% \\
DDS \cite{hertz2023delta}            & 43.5\% & 37.0\% & 38.0\% & 38.5\% \\
CDS \cite{nam2024contrastive}            & 44.5\% & 40.0\% & 43.0\% & 42.0\% \\
SBP \cite{mcallister2024rethinking}            & 61.7\% & 59.8\% & 61.5\% & 57.7\% \\
\bottomrule
\end{tabular}
\vspace{-0.5em}
\caption{Percentage of times users preferred other methods over ours in 1-on-1 comparisons across different criteria. }
\label{tab:main_user_study}
\vspace{-12pt}

\end{table}

\subsection{Quantitative evaluation} \label{sec:quantitative_eval}

\myparagraph{Metrics.} We evaluate two main aspects with five metrics: prompt fidelity (CLIP-T) and background preservation (CLIP-R, CLIP-AUC, L1$^*$, and CLIP-I$^*$). For prompt fidelity, we use CLIP-T, following \cite{zhang2023magicbrush, sheynin2023emuedit}, which calculates the cosine similarity between the CLIP embeddings of the edited image $\vect{x}$ and the text prompt $y^{\text{local}}$, describing the local changes. 
For background preservation, our study (Appendix \ref{supp:score_dump}) shows that the L1, CLIP-I, DINO metrics used in prior work~\cite{zhang2023magicbrush, sheynin2023emuedit}, which are computed between the output and the single ground-truth edited image, are inherently biased: a method that does nothing to the input ranks first across the board, leading to misleading interpretations. To address this, we propose CLIP-R, CLIP-AUC, and improved versions of L1$^*$ and CLIP-I$^*$. CLIP-R is defined as:

\begin{align}
\operatorname{CLIP-R} = \frac{\operatorname{CosineSim}(\operatorname{CLIP}(\vect{x}), \operatorname{CLIP}(y^{\text{tgt}}))}{\operatorname{CosineSim}(\operatorname{CLIP}(\vect{x}^{\text{src}}), \operatorname{CLIP}(y^{\text{tgt}}))},
\end{align}
which quantifies how much more the edited image $\vect{x}$ conforms to the target prompt $y^{\text{tgt}}$ than the input image $\vect{x}^{\text{src}}$ does. 
Since $y^{\text{tgt}}$ describes entire images, methods are penalized for altering background elements specified in $y^{\text{tgt}}$. 
Because different inputs require varying degrees of modification to match $y^{\text{tgt}}$,
we plot the ratio of edits whose $\text{CLIP-R} > k$ for various thresholds $k \geq 1$ and compute the area under this curve as another metric, CLIP-AUC.

To address the shortcomings of L1 and CLIP-I for background preservation, we first ensure that edits from each method reach the same degree before computing scores. This is done by plotting the mean L1 and CLIP-I scores for edits with $\text{CLIP-R} > k$ for multiple $k$ values and computing the area under curves ($L_1^*$ and $\text{CLIP-I}^*$). The integrals are computed for $k \in [1.0, 1.22]$, which excludes failed edits ($\text{CLIP-R} < 1$) and extends to the largest $k$ that still produces at least 30 examples in any method for statistical analysis.

\begin{figure}[h]
    \centering
    \includegraphics[width=0.9\columnwidth]{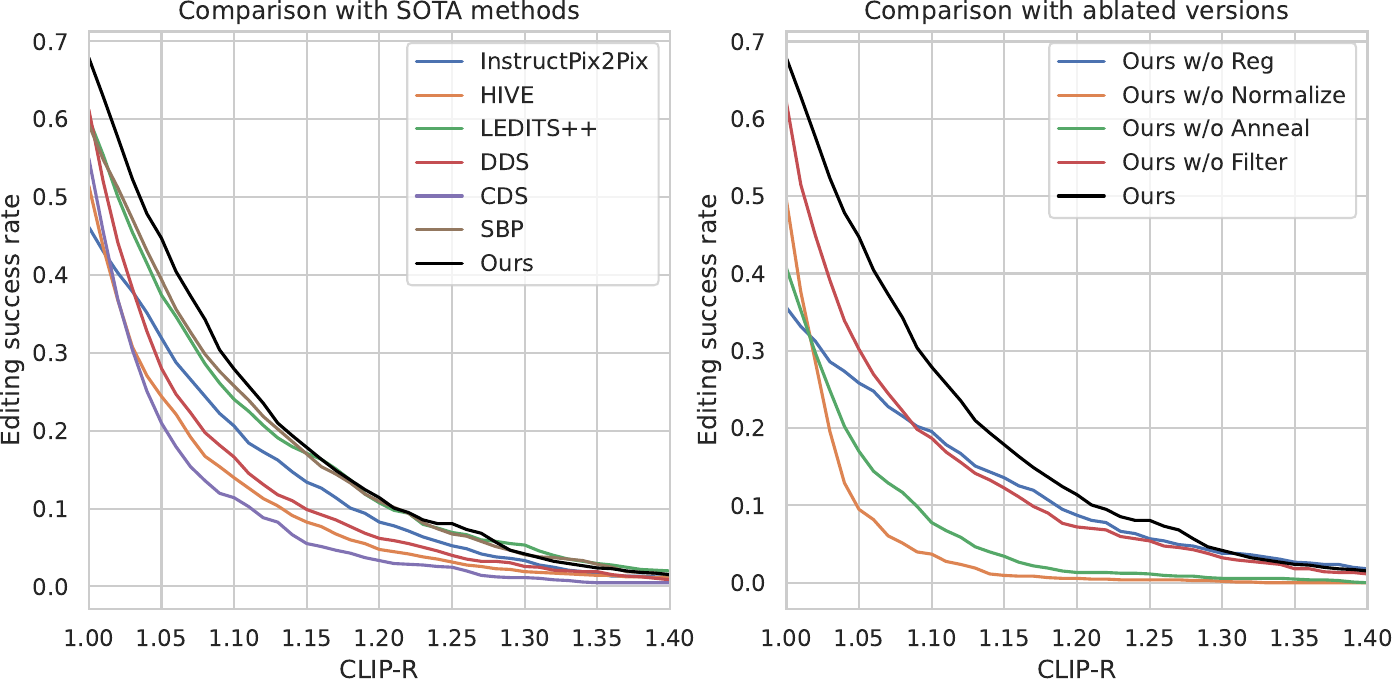}
    \captionsetup{skip=3pt}
    \caption{Editing success rate, defined as the ratio of edits with CLIP-R $> k$ at various thresholds $k \geq 1$. Our method achieves the most successful edits on MagicBrush \cite{zhang2023magicbrush}.}
    
    \vspace{-13pt}
    \label{fig:success_rate}
\end{figure}

\myparagraph{Results.} As shown in Table \ref{tab:main_compare_sota}, our method outperforms both existing zero-shot and supervised methods across most metrics, except for $\text{CLIP-I}^*$. 
Since DDS and CDS struggle with object insertion, our improvements in Table \ref{tab:main_compare_sota_object} are more pronounced when evaluating only on such examples (see Appendix \ref{supp:magicbrush_subset} for how we classify examples).
Figure \ref{fig:success_rate} indicates that at the same success rate, our results best align with the target caption $y^{\text{tgt}}$, which requires both strong textual alignment and background preservation. Our AUCs are also the largest, suggesting a better trade-off between the two aspects among all methods. 
This finding also concurs with the user study in Table \ref{tab:main_user_study}.








\begin{table}[!h]
\centering
\small
\resizebox{1\columnwidth}{!}{%
\setlength{\tabcolsep}{3pt}
\begin{tabular}{
    l@{\hspace{2pt}}
    c@{\hspace{2pt}}
    c@{\hspace{2pt}}
    c@{\hspace{2pt}}
    c@{\hspace{2pt}}
    c@{\hspace{2pt}}
}
\toprule
\textbf{Method} & \textbf{Time (mins)} & \textbf{CLIP-T} $\uparrow$   & \textbf{CLIP-AUC} $\uparrow$  & $\textbf{L1}^*$ $\downarrow$         & $\textbf{CLIP-I}^*$ $\uparrow$     \\
\hline
\multicolumn{6}{c}{\cellcolor[HTML]{EFEFEF} \rule{0pt}{2.3ex}\textbf{Instruction-guided methods}} \\
IP2P~\cite{brooks2022instructpix2pix}\rule{0pt}{2.3ex} & 0.06 & 0.275 & 0.053 & 0.029 & 0.180 \\

HIVE \cite{zhang2024hive} & 0.13  & 0.272 & 0.040 & 0.024 & 0.189 \\
\hline
\multicolumn{6}{c}{\cellcolor[HTML]{EFEFEF} \rule{0pt}{2.3ex} \textbf{Global description-guided methods}} \\
LEDITS++~\cite{brack2023ledits}\rule{0pt}{2.3ex} & 0.13  & 0.279 & 0.067 & 0.022 & 0.182 \\
DDS \cite{hertz2023delta} & 0.22   & 0.277 & 0.048 & 0.017 & \colorbox{tabsecond}{0.195} \\
CDS \cite{nam2024contrastive} & 0.62     & 0.272 & 0.034 & \colorbox{tabsecond}{0.016} & \colorbox{tabfirst}{0.197} \\
SBP \cite{mcallister2024rethinking} & 0.30 & \colorbox{tabsecond}{0.285} & \colorbox{tabsecond}{0.068} & 0.024 & 0.174 \\
\textbf{Ours}  & 1.79  & \colorbox{tabfirst}{0.287} & \colorbox{tabfirst}{0.074} & \colorbox{tabfirst}{0.015} & 0.192 \\
\bottomrule
\end{tabular}
}
\vspace{-0.5em}

\caption{
Scores on MagicBrush of state-of-the-art methods and our method. 
The \colorbox{tabfirst}{best} and \colorbox{tabsecond}{second-best} scores are color-coded.
}

\label{tab:main_compare_sota}
\vspace{-12pt}

\end{table}
\begin{table}[!h]
\centering
\small
\resizebox{1\columnwidth}{!}{%
\setlength{\tabcolsep}{4pt}
\begin{tabular}{lcccc
}
\toprule
\multirow{2}{*}{\textbf{Method}} & \multicolumn{2}{c}{\textbf{CLIP-T} $\uparrow$} & \multicolumn{2}{c}{\textbf{CLIP-AUC} $\uparrow$} \\
& All & Add & All & Add \\
\midrule
DDS~\cite{hertz2023delta}    & 0.277{\scriptsize \red{(-3.6\%)}} & 0.266{\scriptsize \red{(-6.0\%)}} & 0.048{\scriptsize \red{(-35.1\%)}} & 0.043{\scriptsize \red{(-47.2\%)}} \\

CDS~\cite{nam2024contrastive}    & 0.272{\scriptsize \red{(-5.1\%)}} & 0.262{\scriptsize \red{(-7.7\%)}} & 0.034{\scriptsize \red{(-53.8\%)}} & 0.029{\scriptsize \red{(-64.4\%)}} \\

SBP~\cite{mcallister2024rethinking}    & 0.285{\scriptsize \red{(-0.6\%)}} & 0.281{\scriptsize \red{(-0.8\%)}} & 0.068{\scriptsize \red{(-8.6\%)}} & 0.069{\scriptsize \red{(-14.7\%)}} \\

\textbf{Ours}      & \colorbox{tabfirst}{0.287} & \colorbox{tabfirst}{0.283} & \colorbox{tabfirst}{0.074} & \colorbox{tabfirst}{0.080} \\
\bottomrule
\end{tabular}
}
\captionsetup{skip=3pt}
\caption{
Our method outperforms DDS, CDS, and SBP, with larger gains in object insertion tasks (Add), compared to all tasks (All).
}
\label{tab:main_compare_sota_object}
\vspace{-12pt}

\end{table}

\begin{figure}[!htbp]
    \centering
    \includegraphics[width=0.92\columnwidth]{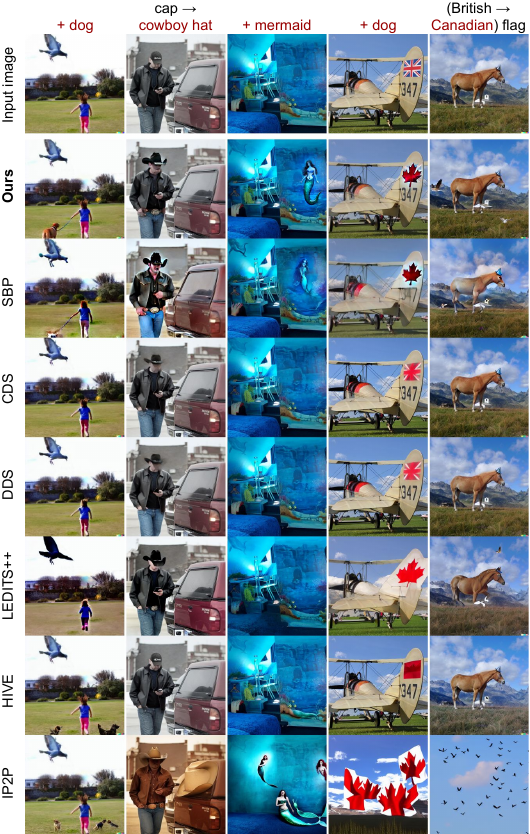}
    \captionsetup{skip=3pt}
    \caption{Qualitative results on MagicBrush dataset \cite{zhang2023magicbrush}.}
    \label{fig:qual_magicbrush_main}
    \vspace{-6pt}
\end{figure}

\subsection{Qualitative results} 
We present qualitative results of MagicBrush test inputs (Figure \ref{fig:qual_magicbrush_main} and Appendix \ref{supp:qual_magicbrush}) and in-the-wild inputs (Figure~\ref{fig:teaser}, Figure~\ref{fig:qual_results}, and Appendix~\ref{supp:qual_other}).
Compared to other general image editing techniques, our method better preserves the source background while more faithfully reflecting the target prompt for both large edits (e.g., dragon, pizza toppings, lava) and subtle ones (e.g., cat’s eyes, sunglasses).




\subsection{Ablation studies} 
We perform an ablation study on the MagicBrush dataset by removing (1) attention-based spatial regularization (setting $\lambda = 0$ and the mask $\vect{M}_k = \mathds{1}$), (2) normalization (setting the denominator in Equation \ref{eq:grad_norm} to 1), (3) annealing (setting $\gamma$ in Equation \ref{eq:grad_norm} to 1), and (4) filtering (setting the threshold for filtering gradients $\eta_0 = 0$).

Table \ref{tab:main_compare_ablation} shows that our full method outperforms all ablated versions on CLIP-T. 
Without gradient filtering and normalization, our method often fails to insert objects or adds incomplete ones, resulting in a lower CLIP-T score.
Using a constant $\gamma = 1$ instead of annealing leads to unstable optimization, over-saturated outputs, and worse CLIP-T. 
Studies on the moving average in attention masks and other hyperparameters are in Appendices \ref{supp:moving_average_mask} and \ref{supp:hyperparams}, respectively.


\begin{table}[!h]
\centering
\small
\vspace{-3pt}
\begin{tabular}{
    l@{\hspace{1pt}}
    c@{\hspace{1pt}}
    c@{\hspace{1pt}}
    c@{\hspace{1pt}}
    c@{\hspace{1pt}}
}
\toprule
\textbf{Method} & \textbf{CLIP-T} $\uparrow$   & \textbf{CLIP-AUC} $\uparrow$  & $\textbf{L1}^*$ $\downarrow$         & $\textbf{CLIP-I}^*$ $\uparrow$     $\uparrow$  \\
\midrule
w/o Spatial Reg.   & 0.277 & 0.049 & 0.057 & 0.137 \\
w/o Normalize    & 0.268 & 0.017 & \colorbox{tabfirst}{0.013} & \colorbox{tabfirst}{0.200} \\
w/o Anneal ($\gamma = 1$)   & 0.265 & 0.025 & 0.021 & 0.185 \\
w/o Filtering               & \colorbox{tabsecond}{0.279} & \colorbox{tabsecond}{0.053} & \colorbox{tabsecond}{0.015} & \colorbox{tabsecond}{0.195} \\
\textbf{Ours}            & \colorbox{tabfirst}{0.287} & \colorbox{tabfirst}{0.074} & \colorbox{tabsecond}{0.015} & 0.192 \\
\bottomrule
\end{tabular}
\captionsetup{skip=3pt}
\caption{
Ablations on MagicBrush.
Our method best balances prompt fidelity and background preservation (CLIP-AUC).
}
\label{tab:main_compare_ablation}
\vspace{-12pt}

\end{table}

\begin{figure}[!htbp]
    \centering
    \vspace{-0.5em}
    \includegraphics[width=0.95\columnwidth]{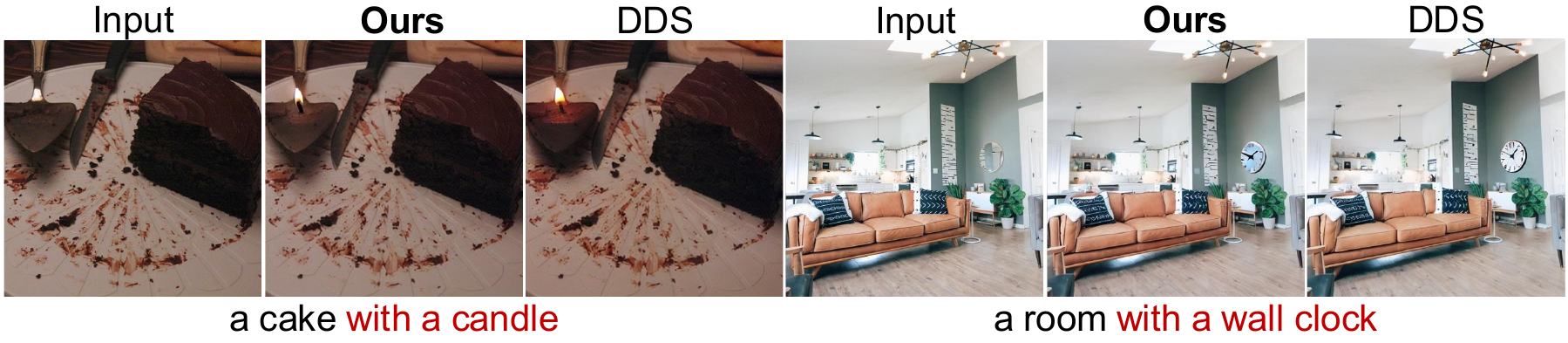}
    \captionsetup{skip=3pt}
    \caption{Score distillation tends to be biased towards regions with existing visual cues, as they require less \textit{effort} to modify.}
    \vspace{-1.1em}
    \label{fig:minimal_effort_main}
\end{figure}



\subsection{Limitations and discussion}
\label{sec:limit}
We highlight interesting failure cases of our method, and potentially score distillation in general, in Figure \ref{fig:minimal_effort_main}.
These methods tend to favor minimal-effort regions, where visual cues for object formation already exist, leading to unnatural placements in some cases (see Appendix \ref{supp:minimal_effort}).


Additionally, our technique can be slow with gradient filtering, especially for challenging prompts that produce many problematic gradients. It also struggles with certain edits due to limited language understanding of diffusion models. However, using larger models with improved text understanding \cite{podell2023sdxl} can directly improve its performance. 
We also show in Appendix \ref{supp:extension_dds} that our proposed techniques can function as plug-and-play components and improve other distillation algorithms, such as DDS \cite{hertz2023delta}.


\begin{figure*}[h]
    \centering
    \includegraphics[width=0.95\textwidth]{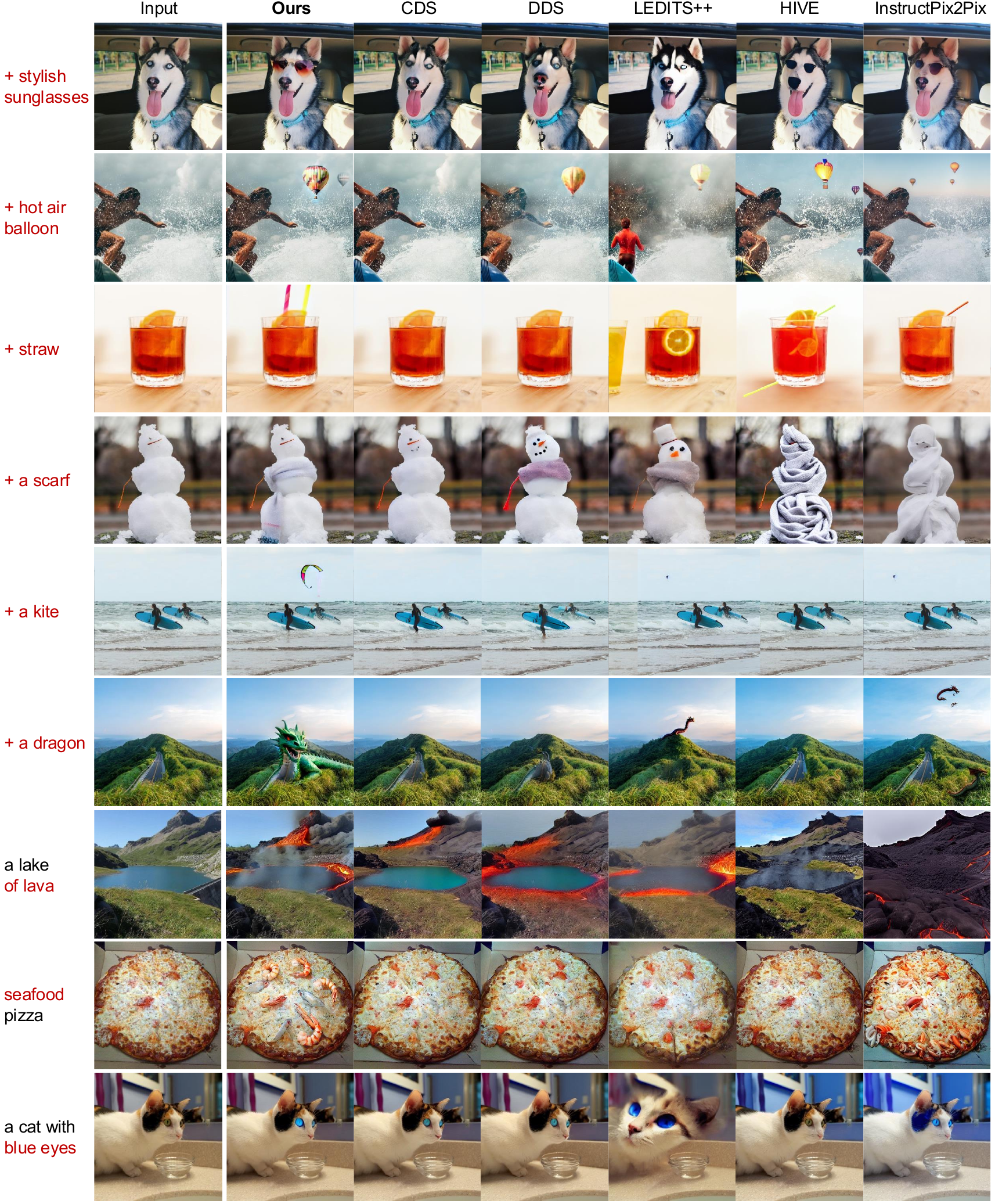}
    \caption{Qualitative results on various in-the-wild editing tasks. While state-of-the-art methods require instance-level hyperparameter tuning, our method successfully performs edits with a higher success rate using a \textit{single} configuration. Hypertuning grids for results from state-of-the-art methods are provided in Appendix \ref{supp:tuning_grid}.}
    \label{fig:qual_results}
\end{figure*}




\section{Conclusion}
\label{sec:conclusion}

We introduce LUSD, an SDS-based method for text-guided image editing with an emphasis on object insertion. It features two new techniques that improve reliability in the face of input variation and randomness inherent to the optimization process: (1) attention-based spatial regularization to modulate gradients for background preservation, and (2) gradient filtering and normalization to mitigate counterproductive gradients. It outperforms other SOTA methods in prompt fidelity and has a higher rate of editing success while using a \textit{single} set of hyperparameters for all inputs.


\textbf{Acknowledgement}: This research was supported by Research Fellowships from VISTEC, SCB public company limited, and PTT public company limited.

{
    \small
    \bibliographystyle{ieeenat_fullname}

\begin{thebibliography}{53}
\providecommand{\natexlab}[1]{#1}
\providecommand{\url}[1]{\texttt{#1}}
\expandafter\ifx\csname urlstyle\endcsname\relax
  \providecommand{\doi}[1]{doi: #1}\else
  \providecommand{\doi}{doi: \begingroup \urlstyle{rm}\Url}\fi

\bibitem[Alldieck et~al.(2024)Alldieck, Kolotouros, and
  Sminchisescu]{alldieck2024score}
Thiemo Alldieck, Nikos Kolotouros, and Cristian Sminchisescu.
\newblock Score {{Distillation Sampling}} with {{Learned Manifold Corrective}},
  2024.

\bibitem[Avrahami et~al.(2022)Avrahami, Lischinski, and
  Fried]{avrahami2022blended}
Omri Avrahami, Dani Lischinski, and Ohad Fried.
\newblock Blended diffusion for text-driven editing of natural images.
\newblock In \emph{2022 IEEE/CVF Conference on Computer Vision and Pattern
  Recognition (CVPR)}. IEEE, 2022.

\bibitem[Avrahami et~al.(2023)Avrahami, Fried, and
  Lischinski]{avrahami2023blended}
Omri Avrahami, Ohad Fried, and Dani Lischinski.
\newblock Blended latent diffusion.
\newblock \emph{ACM Transactions on Graphics}, 42\penalty0 (4):\penalty0
  1–11, 2023.

\bibitem[Basu et~al.(2023)Basu, Saberi, Bhardwaj, Chegini, Massiceti, Sanjabi,
  Hu, and Feizi]{basu2023editval}
Samyadeep Basu, Mehrdad Saberi, Shweta Bhardwaj, Atoosa~Malemir Chegini,
  Daniela Massiceti, Maziar Sanjabi, Shell~Xu Hu, and Soheil Feizi.
\newblock Editval: Benchmarking diffusion based text-guided image editing
  methods, 2023.

\bibitem[Brack et~al.(2023)Brack, Friedrich, Kornmeier, Tsaban, Schramowski,
  Kersting, and Passos]{brack2023ledits}
Manuel Brack, Felix Friedrich, Katharina Kornmeier, Linoy Tsaban, Patrick
  Schramowski, Kristian Kersting, and Apolinário Passos.
\newblock Ledits++: Limitless image editing using text-to-image models.
\newblock 2023.

\bibitem[Brooks et~al.(2022)Brooks, Holynski, and
  Efros]{brooks2022instructpix2pix}
Tim Brooks, Aleksander Holynski, and Alexei~A Efros.
\newblock Instructpix2pix: Learning to follow image editing instructions.
\newblock \emph{arXiv preprint arXiv:2211.09800}, 2022.

\bibitem[Brooks et~al.(2023)Brooks, Holynski, and
  Efros]{brooks2023instructpix2pix}
Tim Brooks, Aleksander Holynski, and Alexei~A. Efros.
\newblock Instructpix2pix: Learning to follow image editing instructions, 2023.

\bibitem[Burgert et~al.(2023)Burgert, Li, Leite, Ranasinghe, and
  Ryoo]{burgert2023diffusion}
Ryan Burgert, Xiang Li, Abe Leite, Kanchana Ranasinghe, and Michael~S. Ryoo.
\newblock Diffusion illusions: Hiding images in plain sight, 2023.

\bibitem[Cao et~al.(2023)Cao, Wang, Qi, Shan, Qie, and Zheng]{cao2023masactrl}
Mingdeng Cao, Xintao Wang, Zhongang Qi, Ying Shan, Xiaohu Qie, and Yinqiang
  Zheng.
\newblock Masactrl: Tuning-free mutual self-attention control for consistent
  image synthesis and editing.
\newblock In \emph{Proceedings of the IEEE/CVF International Conference on
  Computer Vision (ICCV)}, pages 22560--22570, 2023.

\bibitem[Caron et~al.(2021)Caron, Touvron, Misra, Jégou, Mairal, Bojanowski,
  and Joulin]{caron2021emerging}
Mathilde Caron, Hugo Touvron, Ishan Misra, Hervé Jégou, Julien Mairal, Piotr
  Bojanowski, and Armand Joulin.
\newblock Emerging properties in self-supervised vision transformers, 2021.

\bibitem[Chefer et~al.(2023)Chefer, Alaluf, Vinker, Wolf, and
  Cohen-Or]{chefer2023attendandexcite}
Hila Chefer, Yuval Alaluf, Yael Vinker, Lior Wolf, and Daniel Cohen-Or.
\newblock Attend-and-excite: Attention-based semantic guidance for
  text-to-image diffusion models, 2023.

\bibitem[Chen et~al.(2024)Chen, Vaxman, Ben~Baruch, Asulin, Moreshet, Lien,
  Sra, and Sen]{chen2024tino}
Sherry~X Chen, Yaron Vaxman, Elad Ben~Baruch, David Asulin, Aviad Moreshet,
  Kuo-Chin Lien, Misha Sra, and Pradeep Sen.
\newblock Tino-edit: Timestep and noise optimization for robust diffusion-based
  image editing.
\newblock In \emph{Proceedings of the IEEE/CVF Conference on Computer Vision
  and Pattern Recognition}, pages 6337--6346, 2024.

\bibitem[Couairon et~al.(2022)Couairon, Verbeek, Schwenk, and
  Cord]{couairon2022diffedit}
Guillaume Couairon, Jakob Verbeek, Holger Schwenk, and Matthieu Cord.
\newblock Diffedit: Diffusion-based semantic image editing with mask guidance,
  2022.

\bibitem[Esser et~al.(2024)Esser, Kulal, Blattmann, Entezari, Müller, Saini,
  Levi, Lorenz, Sauer, Boesel, Podell, Dockhorn, English, Lacey, Goodwin,
  Marek, and Rombach]{esser2024scaling}
Patrick Esser, Sumith Kulal, Andreas Blattmann, Rahim Entezari, Jonas Müller,
  Harry Saini, Yam Levi, Dominik Lorenz, Axel Sauer, Frederic Boesel, Dustin
  Podell, Tim Dockhorn, Zion English, Kyle Lacey, Alex Goodwin, Yannik Marek,
  and Robin Rombach.
\newblock Scaling rectified flow transformers for high-resolution image
  synthesis, 2024.

\bibitem[Hertz et~al.(2022)Hertz, Mokady, Tenenbaum, Aberman, Pritch, and
  Cohen-Or]{hertz2022prompt}
Amir Hertz, Ron Mokady, Jay Tenenbaum, Kfir Aberman, Yael Pritch, and Daniel
  Cohen-Or.
\newblock Prompt-to-prompt image editing with cross attention control.
\newblock 2022.

\bibitem[Hertz et~al.(2023)Hertz, Aberman, and Cohen-Or]{hertz2023delta}
Amir Hertz, Kfir Aberman, and Daniel Cohen-Or.
\newblock Delta denoising score.
\newblock In \emph{Proceedings of the IEEE/CVF International Conference on
  Computer Vision (ICCV)}, pages 2328--2337, 2023.

\bibitem[Ho and Salimans(2022)]{ho2022classifier}
Jonathan Ho and Tim Salimans.
\newblock Classifier-free diffusion guidance, 2022.

\bibitem[Ho et~al.(2020)Ho, Jain, and Abbeel]{ho2020denoising}
Jonathan Ho, Ajay Jain, and Pieter Abbeel.
\newblock Denoising diffusion probabilistic models.
\newblock In \emph{Proceedings of the 34th International Conference on Neural
  Information Processing Systems}, pages 6840--6851, 2020.

\bibitem[Huberman-Spiegelglas et~al.(2024)Huberman-Spiegelglas, Kulikov, and
  Michaeli]{huberman2024edit}
Inbar Huberman-Spiegelglas, Vladimir Kulikov, and Tomer Michaeli.
\newblock An edit friendly {DDPM} noise space: Inversion and manipulations.
\newblock In \emph{Proceedings of the IEEE/CVF Conference on Computer Vision
  and Pattern Recognition}, pages 12469--12478, 2024.

\bibitem[Katzir et~al.(2023)Katzir, Patashnik, Cohen-Or, and
  Lischinski]{katzir2023noisefree}
Oren Katzir, Or Patashnik, Daniel Cohen-Or, and Dani Lischinski.
\newblock Noise-free score distillation, 2023.

\bibitem[Kim et~al.(2024)Kim, Park, and Ye]{kim2024dreamsampler}
Jeongsol Kim, Geon~Yeong Park, and Jong~Chul Ye.
\newblock Dreamsampler: Unifying diffusion sampling and score distillation for
  image manipulation.
\newblock \emph{arXiv preprint arXiv:2403.11415}, 2024.

\bibitem[Koo et~al.(2024)Koo, Park, and Sung]{koo2024posterior}
Juil Koo, Chanho Park, and Minhyuk Sung.
\newblock Posterior distillation sampling.
\newblock In \emph{CVPR}, 2024.

\bibitem[Li et~al.(2022)Li, Li, Xiong, and Hoi]{li2022blip}
Junnan Li, Dongxu Li, Caiming Xiong, and Steven Hoi.
\newblock Blip: Bootstrapping language-image pre-training for unified
  vision-language understanding and generation, 2022.

\bibitem[Liang et~al.(2023)Liang, Yang, Lin, Li, Xu, and
  Chen]{liang2023luciddreamer}
Yixun Liang, Xin Yang, Jiantao Lin, Haodong Li, Xiaogang Xu, and Yingcong Chen.
\newblock Luciddreamer: Towards high-fidelity text-to-3d generation via
  interval score matching, 2023.

\bibitem[Lin et~al.(2014)Lin, Maire, Belongie, Hays, Perona, Ramanan,
  Doll{\'a}r, and Zitnick]{lin2014microsoft}
Tsung-Yi Lin, Michael Maire, Serge Belongie, James Hays, Pietro Perona, Deva
  Ramanan, Piotr Doll{\'a}r, and C~Lawrence Zitnick.
\newblock Microsoft \text{COCO}: Common objects in context.
\newblock In \emph{Computer Vision--ECCV 2014: 13th European Conference,
  Zurich, Switzerland, September 6-12, 2014, Proceedings, Part V 13}, pages
  740--755. Springer, 2014.

\bibitem[McAllister et~al.(2024)McAllister, Ge, Huang, Jacobs, Efros, Holynski,
  and Kanazawa]{mcallister2024rethinking}
David McAllister, Songwei Ge, Jia-Bin Huang, David~W. Jacobs, Alexei~A. Efros,
  Aleksander Holynski, and Angjoo Kanazawa.
\newblock Rethinking {{Score Distillation}} as a {{Bridge Between Image
  Distributions}}, 2024.

\bibitem[Meng et~al.(2022)Meng, He, Song, Song, Wu, Zhu, and
  Ermon]{meng2022sdedit}
Chenlin Meng, Yutong He, Yang Song, Jiaming Song, Jiajun Wu, Jun-Yan Zhu, and
  Stefano Ermon.
\newblock {SDE}dit: Guided image synthesis and editing with stochastic
  differential equations.
\newblock In \emph{International Conference on Learning Representations}, 2022.

\bibitem[Miyake et~al.(2023)Miyake, Iohara, Saito, and
  Tanaka]{miyake2023negative}
Daiki Miyake, Akihiro Iohara, Yu Saito, and Toshiyuki Tanaka.
\newblock Negative-prompt inversion: Fast image inversion for editing with
  text-guided diffusion models, 2023.

\bibitem[Mokady et~al.(2022)Mokady, Hertz, Aberman, Pritch, and
  Cohen-Or]{mokady2022null}
Ron Mokady, Amir Hertz, Kfir Aberman, Yael Pritch, and Daniel Cohen-Or.
\newblock Null-text inversion for editing real images using guided diffusion
  models, 2022.

\bibitem[Nam et~al.(2024)Nam, Kwon, Park, and Ye]{nam2024contrastive}
Hyelin Nam, Gihyun Kwon, Geon~Yeong Park, and Jong~Chul Ye.
\newblock Contrastive denoising score for text-guided latent diffusion image
  editing, 2024.

\bibitem[Nguyen et~al.(2023)Nguyen, Vu, Tran, and
  Nguyen]{quangtruong2023dataset}
Quang~Ho Nguyen, Truong Vu, Anh Tran, and Khoi Nguyen.
\newblock Dataset diffusion: Diffusion-based synthetic dataset generation for
  pixel-level semantic segmentation.
\newblock In \emph{Proceedings of the Thirty-Seventh Conference on Neural
  Information Processing Systems (NeurIPS)}, 2023.

\bibitem[OpenAI(2023)]{openai2023chatgpt}
OpenAI.
\newblock Chatgpt: Language model for conversational ai.
\newblock \url{https://chat.openai.com}, 2023.
\newblock Accessed: 2024-11-15.

\bibitem[Park et~al.(2020)Park, Efros, Zhang, and Zhu]{park2020cut}
Taesung Park, Alexei~A. Efros, Richard Zhang, and Jun-Yan Zhu.
\newblock Contrastive learning for unpaired image-to-image translation.
\newblock In \emph{European Conference on Computer Vision}, 2020.

\bibitem[Parmar et~al.(2023)Parmar, Singh, Zhang, Li, Lu, and
  Zhu]{parmar2023zeroshot}
Gaurav Parmar, Krishna~Kumar Singh, Richard Zhang, Yijun Li, Jingwan Lu, and
  Jun-Yan Zhu.
\newblock Zero-shot image-to-image translation, 2023.

\bibitem[Podell et~al.(2023)Podell, English, Lacey, Blattmann, Dockhorn,
  Müller, Penna, and Rombach]{podell2023sdxl}
Dustin Podell, Zion English, Kyle Lacey, Andreas Blattmann, Tim Dockhorn, Jonas
  Müller, Joe Penna, and Robin Rombach.
\newblock Sdxl: Improving latent diffusion models for high-resolution image
  synthesis, 2023.

\bibitem[Poole et~al.(2022)Poole, Jain, Barron, and
  Mildenhall]{poole2022dreamfusion}
Ben Poole, Ajay Jain, Jonathan~T. Barron, and Ben Mildenhall.
\newblock Dreamfusion: Text-to-3d using 2d diffusion.
\newblock \emph{arXiv}, 2022.

\bibitem[Rombach et~al.(2021)Rombach, Blattmann, Lorenz, Esser, and
  Ommer]{rombach2021highresolution}
Robin Rombach, Andreas Blattmann, Dominik Lorenz, Patrick Esser, and Björn
  Ommer.
\newblock High-resolution image synthesis with latent diffusion models, 2021.

\bibitem[Rout et~al.(2025)Rout, Chen, Ruiz, Caramanis, Shakkottai, and
  Chu]{rout2025semantic}
L Rout, Y Chen, N Ruiz, C Caramanis, S Shakkottai, and W Chu.
\newblock Semantic image inversion and editing using rectified stochastic
  differential equations.
\newblock In \emph{The Thirteenth International Conference on Learning
  Representations}, 2025.

\bibitem[Sheynin et~al.(2023)Sheynin, Polyak, Singer, Kirstain, Zohar, Ashual,
  Parikh, and Taigman]{sheynin2023emuedit}
Shelly Sheynin, Adam Polyak, Uriel Singer, Yuval Kirstain, Amit Zohar, Oron
  Ashual, Devi Parikh, and Yaniv Taigman.
\newblock Emu edit: Precise image editing via recognition and generation tasks,
  2023.

\bibitem[Song et~al.(2020)Song, Meng, and Ermon]{song2020denoising}
Jiaming Song, Chenlin Meng, and Stefano Ermon.
\newblock Denoising diffusion implicit models.
\newblock 2020.

\bibitem[Tumanyan et~al.(2022)Tumanyan, Geyer, Bagon, and
  Dekel]{tumanyan2022plugandplay}
Narek Tumanyan, Michal Geyer, Shai Bagon, and Tali Dekel.
\newblock Plug-and-play diffusion features for text-driven image-to-image
  translation, 2022.

\bibitem[Wang et~al.(2024)Wang, Pu, Qi, Guo, Ma, Huang, Chen, Li, and
  Shan]{wang2024taming}
Jiangshan Wang, Junfu Pu, Zhongang Qi, Jiayi Guo, Yue Ma, Nisha Huang, Yuxin
  Chen, Xiu Li, and Ying Shan.
\newblock Taming rectified flow for inversion and editing.
\newblock \emph{arXiv preprint arXiv:2411.04746}, 2024.

\bibitem[Wang et~al.(2023{\natexlab{a}})Wang, Saharia, Montgomery, Pont-Tuset,
  Noy, Pellegrini, Onoe, Laszlo, Fleet, Soricut, Baldridge, Norouzi, Anderson,
  and Chan]{wang2023imageneditor}
Su Wang, Chitwan Saharia, Ceslee Montgomery, Jordi Pont-Tuset, Shai Noy,
  Stefano Pellegrini, Yasumasa Onoe, Sarah Laszlo, David~J. Fleet, Radu
  Soricut, Jason Baldridge, Mohammad Norouzi, Peter Anderson, and William Chan.
\newblock Imagen editor and editbench: Advancing and evaluating text-guided
  image inpainting, 2023{\natexlab{a}}.

\bibitem[Wang et~al.(2023{\natexlab{b}})Wang, Lu, Wang, Bao, Li, Su, and
  Zhu]{wang2023prolificdreamer}
Zhengyi Wang, Cheng Lu, Yikai Wang, Fan Bao, Chongxuan Li, Hang Su, and Jun
  Zhu.
\newblock {{ProlificDreamer}}: {{High-Fidelity}} and {{Diverse Text-to-3D
  Generation}} with {{Variational Score Distillation}}, 2023{\natexlab{b}}.

\bibitem[Wasserman et~al.(2024)Wasserman, Rotstein, Ganz, and
  Kimmel]{wasserman2024paint}
Navve Wasserman, Noam Rotstein, Roy Ganz, and Ron Kimmel.
\newblock Paint by inpaint: Learning to add image objects by removing them
  first.
\newblock \emph{arXiv preprint arXiv:2404.18212}, 2024.

\bibitem[Xie et~al.(2022)Xie, Zhang, Lin, Hinz, and Zhang]{xie2022smartbrush}
Shaoan Xie, Zhifei Zhang, Zhe Lin, Tobias Hinz, and Kun Zhang.
\newblock Smartbrush: Text and shape guided object inpainting with diffusion
  model, 2022.

\bibitem[Yang et~al.(2022)Yang, Gu, Zhang, Zhang, Chen, Sun, Chen, and
  Wen]{yang2022paint}
Binxin Yang, Shuyang Gu, Bo Zhang, Ting Zhang, Xuejin Chen, Xiaoyan Sun, Dong
  Chen, and Fang Wen.
\newblock Paint by example: Exemplar-based image editing with diffusion models.
\newblock \emph{arXiv preprint arXiv:2211.13227}, 2022.

\bibitem[Yu et~al.(2023)Yu, Guo, Li, Liang, Zhang, and Qi]{yu2023text}
Xin Yu, Yuan-Chen Guo, Yangguang Li, Ding Liang, Song-Hai Zhang, and Xiaojuan
  Qi.
\newblock Text-to-3d with classifier score distillation.
\newblock \emph{arXiv preprint arXiv:2310.19415}, 2023.

\bibitem[Zhang et~al.(2023{\natexlab{a}})Zhang, Mo, Chen, Sun, and
  Su]{zhang2023magicbrush}
Kai Zhang, Lingbo Mo, Wenhu Chen, Huan Sun, and Yu Su.
\newblock Magicbrush: A manually annotated dataset for instruction-guided image
  editing.
\newblock In \emph{Advances in Neural Information Processing Systems},
  2023{\natexlab{a}}.

\bibitem[Zhang et~al.(2023{\natexlab{b}})Zhang, Rao, and
  Agrawala]{zhang2023adding}
Lvmin Zhang, Anyi Rao, and Maneesh Agrawala.
\newblock Adding conditional control to text-to-image diffusion models,
  2023{\natexlab{b}}.

\bibitem[Zhang et~al.(2024)Zhang, Yang, Feng, Qin, Chen, Yu, Chen, Wang,
  Savarese, Ermon, Xiong, and Xu]{zhang2024hive}
Shu Zhang, Xinyi Yang, Yihao Feng, Can Qin, Chia-Chih Chen, Ning Yu, Zeyuan
  Chen, Huan Wang, Silvio Savarese, Stefano Ermon, Caiming Xiong, and Ran Xu.
\newblock Hive: Harnessing human feedback for instructional visual editing,
  2024.

\bibitem[Zhao et~al.(2024)Zhao, Yang, Shao, Zhang, Qiao, Luo, Zhang, and
  Ji]{zhao2024diffree}
Lirui Zhao, Tianshuo Yang, Wenqi Shao, Yuxin Zhang, Yu Qiao, Ping Luo, Kaipeng
  Zhang, and Rongrong Ji.
\newblock Diffree: Text-guided shape free object inpainting with diffusion
  model.
\newblock \emph{arXiv preprint arXiv:2407.16982}, 2024.

\bibitem[Zhuang et~al.(2024)Zhuang, Zeng, Liu, Yuan, and Chen]{zhuang2024task}
Junhao Zhuang, Yanhong Zeng, Wenran Liu, Chun Yuan, and Kai Chen.
\newblock A task is worth one word: Learning with task prompts for high-quality
  versatile image inpainting, 2024.

\end{thebibliography}

}
\clearpage
\setcounter{page}{1}
\maketitlesupplementary

\section{Implementation Details} \label{supp:implement}

\subsection{Identifying noun tokens} \label{supp:noun_tokens}

As mentioned in Section \ref{sec:method_attention}, we extract cross-attention maps for all noun tokens related to an edit, which can be inferred by comparing the source and target prompts. We assume that the target prompt is a modified version of the source prompt that either (1) expands on the source prompt or (2) alters specific details within it. Such modifications can appear in various forms, for example:
\begin{itemize}
    \item a waterfall \textit{\underline{with a small boat floating near it}}.
    \item a girl \textit{\underline{wearing glasses}} sitting in front of a mirror.
    \item \textit{\underline{a bird on}} a roof.
    \item a cup of (``coffee'' $\rightarrow$ \textit{\underline{``matcha''}}).
\end{itemize}
We refer to the modified portion as the \textit{differing substring}, which represents the edit. To identify the differing substring, we first remove the longest common suffix and prefix from both prompts, then extract nouns from the remaining target prompt using Part-of-Speech (POS) tags\footnote{We use Natural Language Toolkit's \texttt{nltk.tag.pos\_tag} and select tokens tagged as \texttt{NN} or \texttt{NNS}.}. If the last word of the substring is not (1) a noun, (2) an article, or (3) a preposition, we expand the substring by appending additional words from the target prompt until a noun is included. This step ensures that the extracted segment captures complete noun phrases.

This simple rule-based approach relies on the accuracy of the POS tagger and may not work for all prompt pairs. However, we employ this algorithm to ensure a consistent methodology for both qualitative and quantitative comparisons. In practice, the differing substring can be specified by the user. 


\subsection{LUSD algorithm} \label{supp:algo_pseudocode}

The pseudocode of our LUSD described in Section \ref{sec:method} is given in Algorithm \ref{algo:lusd} and \ref{algo:attn}. Our implementation uses $N = 300$, $\eta_0 = 0.01$, $\alpha = 0.1$, $\lambda = 0.02$, $lr = 2000$, and a reverse sigmoid schedule $\gamma$.


\newcommand\mycommfont[1]{\footnotesize\ttfamily\textcolor{blue}{#1}}
\SetCommentSty{mycommfont}

\begin{algorithm}[h!]
\caption{Image Editing with LUSD}\label{algo:lusd}
\DontPrintSemicolon 
\SetNoFillComment

\SetKwFunction{AttentionMask}{AttentionMask}
\SetKwFunction{Decode}{Decode}

\KwIn{$\vect{z}^{\text{src}}$: latent code of input image \\
      \quad \quad $y^{\text{src}}, y^{\text{tgt}}$: source and target prompts \\
      \quad \quad $lr, \lambda, N, \eta_0$: hyperparameters
      }
\KwOut{Edited image}

$\vect{z} \gets \vect{z}^{\text{src}}$ \;

\For{$k \gets 1$ \textbf{to} $N$} {
    $\eta \gets \eta_0$ \;
    $t \sim \mathcal{U}(50, 950)$ \;

    \While{True} {
        $\vect{\epsilon} \sim \mathcal{N}(\vect{0}, \vect{I})$ \;
        $\vect{z}_t \gets \sqrt{\alpha_t}\vect{z} + \sqrt{1-\alpha_t} \vect{\epsilon}$ \;
        $\vect{\epsilon}^{\text{tgt}}, \vect{\epsilon}^{\text{src}} \gets \epsilon_{\vect{\phi}}(\vect{z}_t, t, (y^{\text{tgt}}, y^{\text{src}}))$ \;
        $\nabla_{\vect{z}} \mathcal{L}_{\text{SBP}} \gets \vect{\epsilon}^{\text{tgt}} - \vect{\epsilon}^{\text{src}}$ \;

        \eIf{$\text{SD}(\nabla_{\vect{z}} \mathcal{L}_{\text{SBP}}) \geq \eta$} {
            
            $\hat{\vect{M}}_k \gets $\AttentionMask{$\epsilon_{\vect{\phi}}, \vect{E}, k, \alpha$} \; \tcp*{Algorithm \ref{algo:attn}}
            $\nabla_{\vect{z}} \mathcal{L}_{\text{SBP-reg}} \gets (1 - \lambda) (\hat{\vect{M}}_k \odot \nabla_{\vect{z}} \mathcal{L}_{\text{SBP}}) + \lambda (\vect{z} - \vect{z}^{\text{src}})$ \;
            $\nabla_{\vect{z}} \mathcal{L}_{\text{LUSD}} \gets \gamma \frac{\nabla_{\vect{z}} \mathcal{L}_{\text{SBP-reg}}}{\text{SD}(\nabla_{\vect{z}} \mathcal{L}_{\text{SBP-reg}})}$ \;
            $\vect{z} \gets \vect{z} - lr \cdot \nabla_{\vect{z}} \mathcal{L}_{\text{LUSD}}$ \;
            \textbf{break} \;
        }{
            $\eta \gets 0.99 \eta$ \;
        }
    }
}
\Return \Decode{$\vect{z}$} \;
\end{algorithm}

\begin{algorithm}[h!]
\caption{AttentionMask}\label{algo:attn}
\DontPrintSemicolon
\SetKwFunction{getself}{get\_self}
\SetKwFunction{getcross}{get\_cross}
\KwIn{ $\epsilon_{\phi}$: diffusion model \\
    \quad \quad $\vect{E}$: Set of target noun tokens \\
    \quad \quad $k$: Current optimization step \\
    \quad \quad $\alpha$: Moving average parameter}
\KwOut{Attention-based mask $\hat{\vect{M}}_k$}

\For{$l \gets 1$ \textbf{to} $L$}{
    $\vect{A}_S^{l, t} \gets $\getself{$\epsilon_{\phi}, l$}\;
    $\vect{A}_C^{l, t, \vect{e}} \gets $\getcross{$\epsilon_{\phi}, l, \vect{e}$}, $\forall \vect{e} \in \vect{E}$\;
}

$\vect{A}_S^{t} \gets \frac{1}{L} \sum_{l=1}^{L} \vect{A}_S^{l, t}$\;
$\vect{A}_C^{t, \vect{e}} \gets \frac{1}{L} \sum_{l=1}^{L} \vect{A}_C^{l, t, \vect{e}}, \forall \vect{e} \in \vect{E}$\;

$\hat{\vect{A}}_C^t \gets \vect{A}_S^{t} \cdot \parens*{\frac{1}{\abs{\vect{E}}} \sum_{\vect{e} \in \vect{E}} \vect{A}_C^{t, \vect{e}}}$\;
$\vect{M} \gets \frac{\hat{\vect{A}}_C^t - \min(\hat{\vect{A}}_C^t)}{\max(\hat{\vect{A}}_C^t) - \min(\hat{\vect{A}}_C^t)}$\;

\eIf{$k = 1$}{
    $\vect{M}_k \gets \vect{M}$\;
}{
    $\vect{M}_k \gets (1 - \alpha) \vect{M}_{k-1} + \alpha \vect{M}$\;
}

$\beta \gets k / N$\;
$\hat{\vect{M}}_k \gets \beta \vect{M}_k + (1 - \beta)\mathds{1}$\;

\Return $\hat{\vect{M}}_k$\;

\end{algorithm}

\section{Study on Moving Average in Attention Mask} \label{supp:moving_average_mask}

\begin{figure}[ht]
    \centering
    \includegraphics[width=0.9\columnwidth, trim=0 10cm 16.5cm 0, clip]{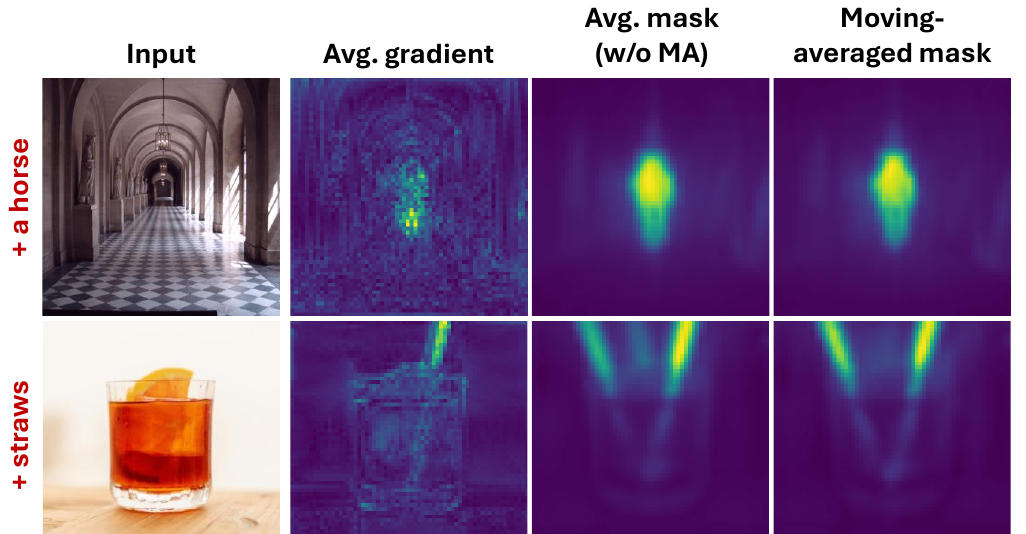}
        \vskip-5pt
        \caption{Attention masks are more localized than SBP gradients.}
        \label{fig:average_gradient_mask}

    \vspace{-15pt}
\end{figure}

As discussed in Section~\ref{sec:method_attention}, spatial regularization is introduced to modulate SBP gradients, which may be averaged out over multiple optimization steps (see Figure~\ref{fig:difficulty}). By estimating the editing region using attention features, our method produces more localized masks than the naive SBP gradients, even without using a moving average (see Figure~\ref{fig:average_gradient_mask}). Nonetheless, we observe that attention masks with a moving average consistently outperform those without it across all metrics (see Table~\ref{tab:mask_moving_average}).

\begin{table}[!h]
\centering
\small

\begin{tabular}{
    l@{\hspace{2pt}}
    c@{\hspace{2pt}}
    c@{\hspace{2pt}}
    c@{\hspace{2pt}}
    c@{\hspace{2pt}}
}
 \hline
\textbf{Moving average} & \textbf{CLIP-T} $\uparrow$   & \textbf{CLIP-AUC} $\uparrow$  & $\textbf{L1}^*$ $\downarrow$         & $\textbf{CLIP-I}^*$ $\uparrow$ \\
\hline
Without & 0.286 & 0.071 & 0.0148 & 0.1921 \\
\textbf{With (Ours)} & \colorbox{tabfirst}{0.287} & \colorbox{tabfirst}{0.074} & \colorbox{tabfirst}{0.0146} & \colorbox{tabfirst}{0.1923} \\
\hline
\end{tabular}

\captionsetup{skip=5pt}
\caption{Applying moving average when computing attention mask yields better results on MagicBrush across all metrics.
}
\vspace{-6pt}
\label{tab:mask_moving_average}
\end{table}

\section{Effects of Hyperparameters} \label{supp:hyperparams}

This section discusses how hyperparameters influence background preservation, gradient filtering, and detail editing. Our default configuration of the regularizer ($\lambda$), filtering threshold ($\eta_0$), and timestep range ($t_{\text{min}}, t_{\text{max}}$) aims to ensure the right extent of image modification, robustness against bad gradients from uncommon concepts, and the ability to alter both low- and high-frequency image features.



\begin{figure}[h]
    \centering
    \includegraphics[width=1.0\columnwidth]{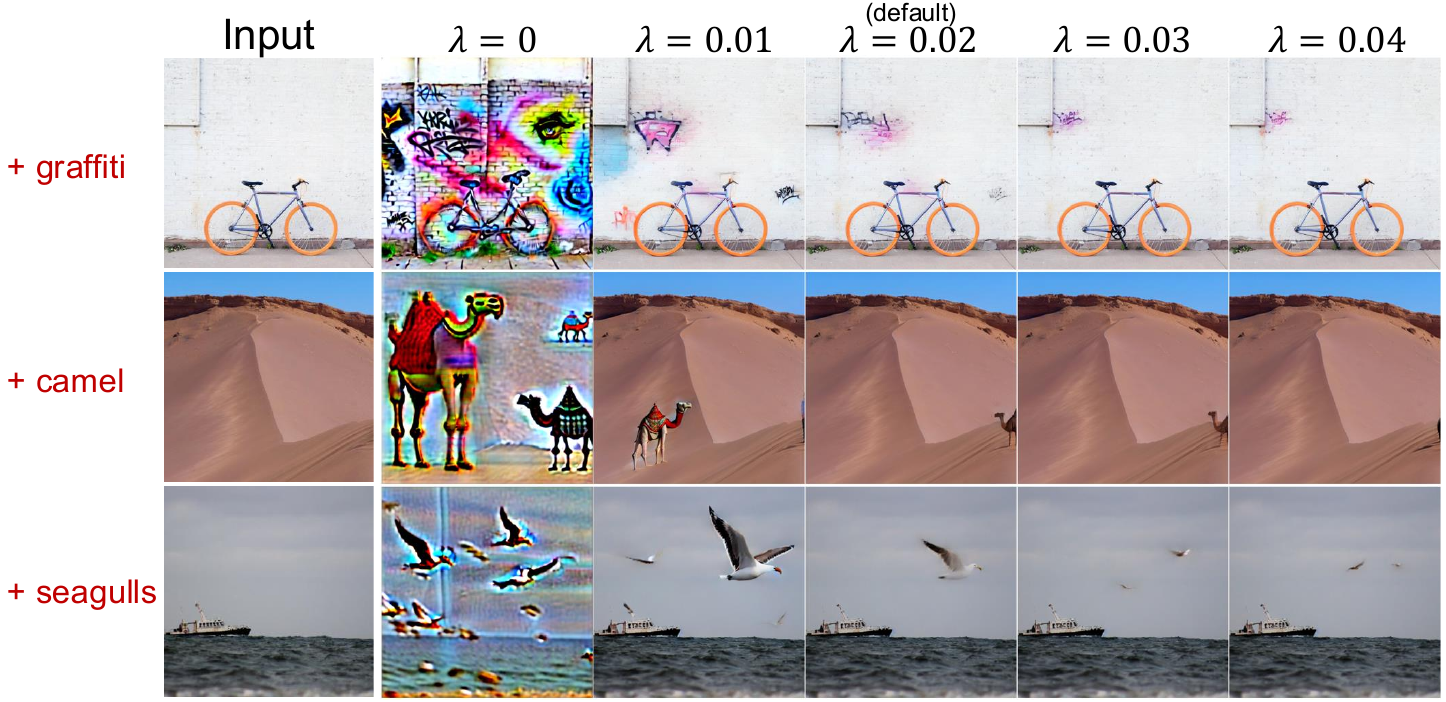}
    \caption{Regularizer $\lambda$ is necessary for background preservation; however, a higher $\lambda$ may restrict the size of the edited region.}
    \vspace{-0.8em}
    \label{fig:hyper_reg}
\end{figure}

\myparagraph{Regularizer ($\lambda$).} The regularizer, as used in Equation \ref{eq:sbp-with-reg}, is crucial for preserving the background during edits. Without the regularizer ($\lambda=0$), the method modifies the entire image to match the prompt. Conversely, increasing $\lambda$ limits the extent of the edited region. An overly high $\lambda$ can prematurely eliminate essential visual cues before larger objects form during the optimization process and thus worsen the quality of the results. Figure \ref{fig:hyper_reg} illustrates how varying $\lambda$ affects outcomes.



\begin{figure}[h]
    \centering
    \includegraphics[width=1.0\columnwidth]{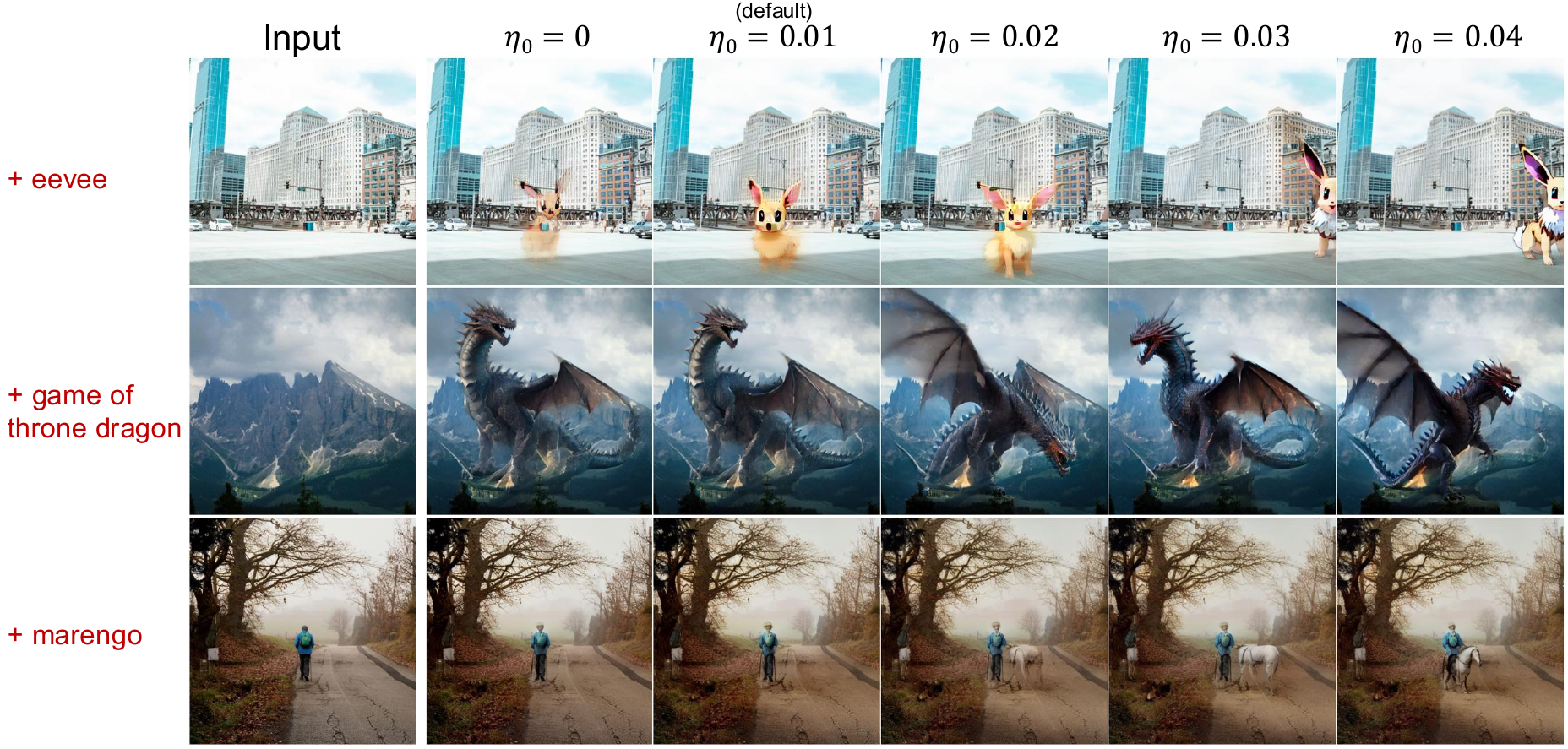}
    \captionsetup{skip=3pt}
    \caption{Higher filtering threshold ($\eta_0$) mitigates the \textit{bad} gradient issue with less known concepts such as ``Marengo'' (the war horse of Napoleon), albeit requiring more optimization time.}
    \vspace{-0.8em}
    \label{fig:hyper_eta}
\end{figure}

\myparagraph{Filtering threshold ($\eta_0$).} The filtering threshold $\eta_0$ helps prevent edit reversion caused by applying \textit{bad} gradients (Section \ref{sec:method_gradients}). Its necessity varies based on input concepts due to the differing prior knowledge encoded in Stable Diffusion \cite{rombach2021highresolution}. For instance, less recognizable concepts like ``Marengo'' (the war horse of Napoleon) has higher chances of encountering bad gradients compared to more common ones like ``Eevee'' (the Pokémon), necessitating a higher $\eta_0$. The right value of $\eta_0$ also depends on the input image and the composition of the input prompt. For instance, a prompt such as ``Game of Thrones dragon'' would already yield a high editing success rate without gradient filtering ($\eta_0=0$) because it includes the common term ``dragon.'' Effects of various $\eta_0$ values are shown in Figure \ref{fig:hyper_eta}. Lastly, the value of $\eta_0$ affects our method's speed because a higher $\eta_0$ requires more optimization time as more gradients are filtered.

\begin{figure}[h]
    \centering
    \includegraphics[width=1.0\columnwidth]{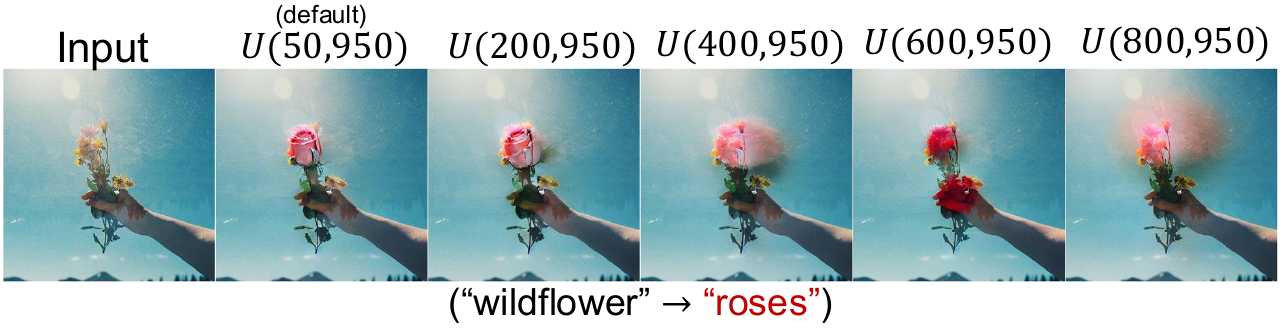}
    \captionsetup{skip=3pt}
    \caption{Low timestep range's lower bound ($t_{\text{min}}$) is necessary for editing high-frequency details, such as the texture of ``roses.''}
    \vspace{-0.8em}
    \label{fig:hyper_tmin}
\end{figure}

\begin{figure}[h]
    \centering
    \includegraphics[width=1.0\columnwidth]{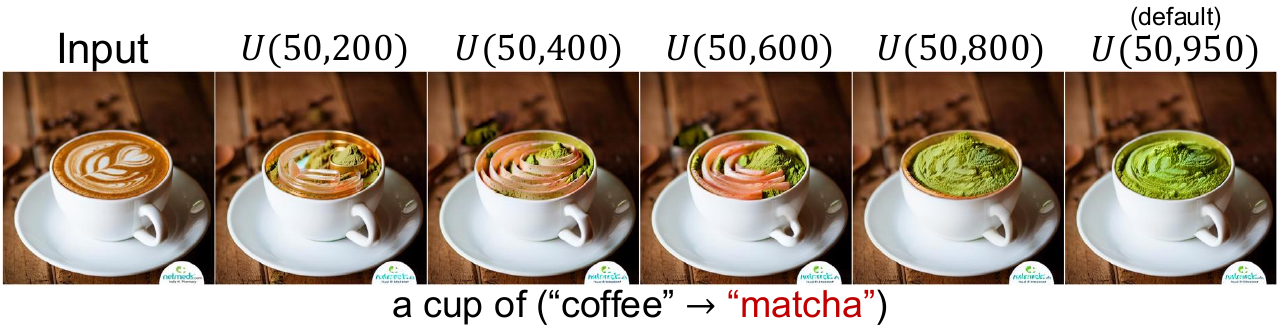}
    \captionsetup{skip=3pt}
    \caption{High timestep range's upper bound ($t_{\text{max}}$) is necessary for editing low-frequency details, such as the color of ``matcha.''}
    \vspace{-0.8em}
    \label{fig:hyper_tmax}
\end{figure}

\myparagraph{Timestep range ($t_{\text{min}}, t_{\text{max}}$).} The default configuration samples diffusion timesteps $t \sim U(t_{\text{min}}, t_{\text{max}})$, with $t_{\text{min}}=50$ and $t_{\text{max}}=950$. 
Lower timesteps allow the method to better resolve high-frequency details such as texture, which is essential for tasks like transforming ``wildflower'' into ``roses'' (Figure \ref{fig:hyper_tmin}).
Higher timesteps, by contrast, focus on low-frequency details like color, which is crucial for edits such as altering ``coffee'' to ``matcha'' (Figure \ref{fig:hyper_tmax}).




\begin{figure*}[h]
    \centering
    \includegraphics[width=0.94\textwidth]{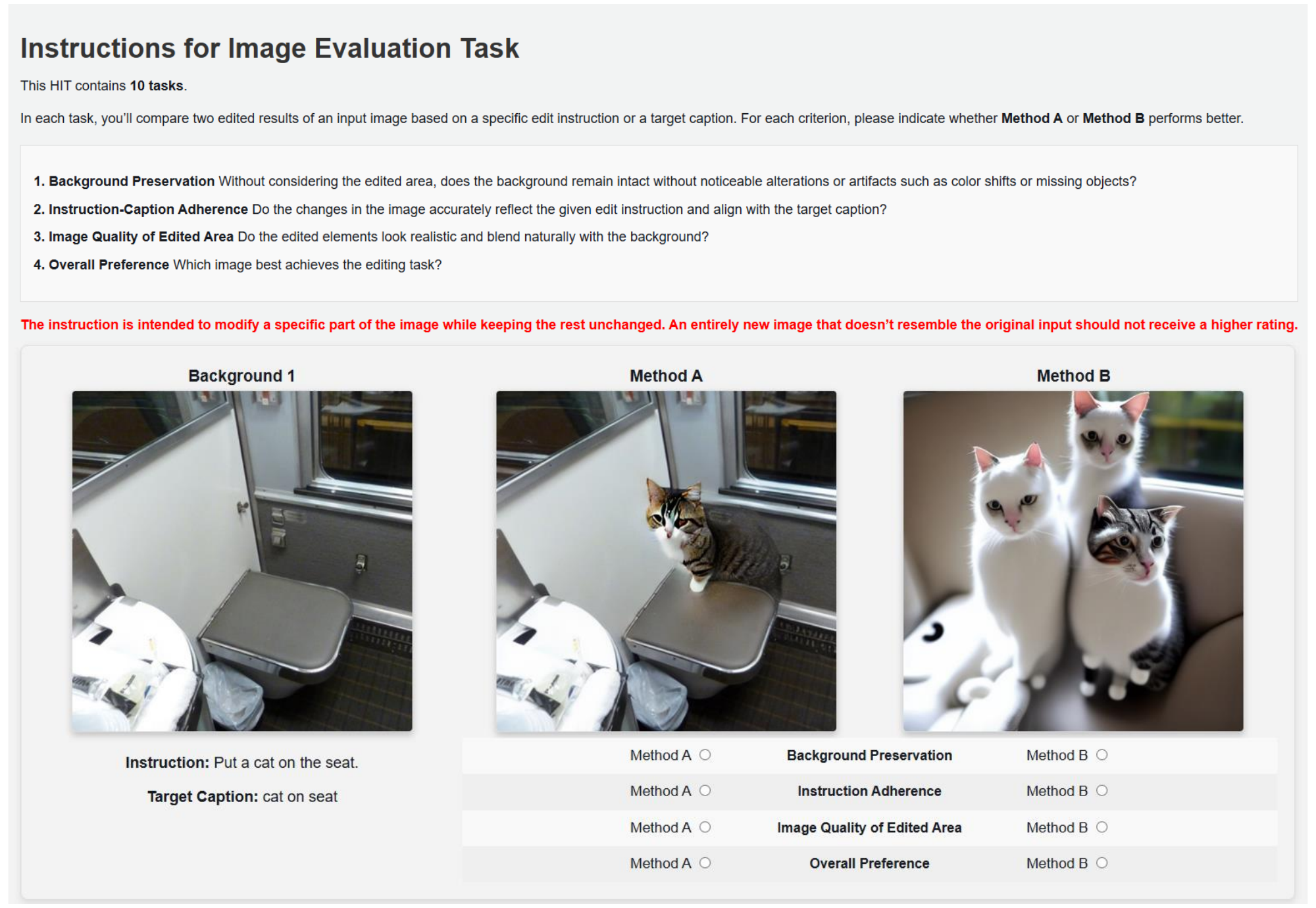}
    \captionsetup{skip=5pt}
    \caption{User study interface.}
    \label{fig:user_study_ui}
    \vspace{-0.8em}
\end{figure*}

\section{Additional Experimental Details}

\subsection{MagicBrush classification} \label{supp:magicbrush_subset}

In Table \ref{tab:main_compare_sota_object} of the main paper, we report scores for examples from the MagicBrush test set \cite{zhang2023magicbrush} involving object insertion. To identify these examples, we first compile a list of keywords for each editing task. 
For object insertion, we use \textit{add}, \textit{put}, and \textit{let there be}. 
For other tasks, we use \textit{remove}, \textit{erase}, \textit{delete}, \textit{replace}, \textit{swap}, \textit{make}, \textit{change}, \textit{turn}, \textit{smaller}, \textit{bigger}, \textit{larger}, \textit{smile}, \textit{cry}, and \textit{look}. Instructions containing these keywords are automatically categorized accordingly, and the rest of the instructions, approximately 35\%,  are classified manually by the authors.

\subsection{Human evaluation} \label{supp:user_study}
As mentioned in Section \ref{sec:user_study}, the user study evaluated 200 samples from the MagicBrush test set. Of these, 100 were randomly selected from object insertion tasks and the rest from other tasks. We compared our method against five state-of-the-art competitors in a one-on-one setup, which results in $200 \times 5 = 1000$ sample-competitor pairs.

Each worker was presented with multiple sample-competitor pairs. For each pair, they saw the input image, the edit instruction, the target caption, and the outputs of our method and the competitor. The worker was not informed of the task type the sample belongs to, and the outputs were presented side-by-side in a randomized order to prevent positional bias. They were asked to choose between our method and the competitor as the better method based on 4 criteria: (1) background preservation, (2) prompt fidelity, (3) quality of edited elements, and (4) overall preference. The user interface and detailed instructions are shown in Figure \ref{fig:user_study_ui}.

A total of 350 unique workers participated via Amazon Mechanical Turk\footnote{https://www.mturk.com/}. We designed the study so that five different workers evaluated each sample-competitor pair. For each sample, the better method in a one-on-one comparison was determined by majority vote (i.e., at least 3 out of 5 workers selected it). As discussed in Section \ref{sec:user_study}, our method outperforms all state-of-the-art approaches when considering all tasks.

\begin{figure}[ht]
    \centering
    \includegraphics[width=0.9\columnwidth]{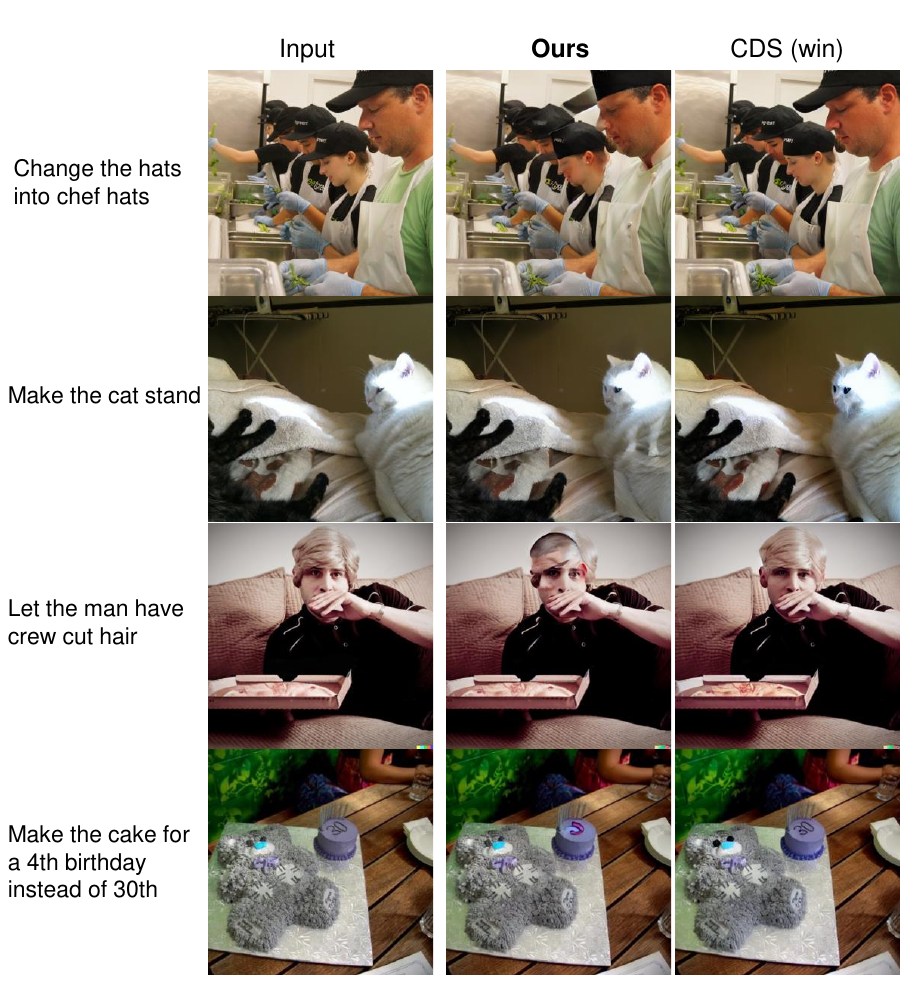}
    \captionsetup{skip=3pt}
    \caption{For complex edits, both CDS and our method fail to match the target prompt. However, CDS typically returns almost unchanged results, whereas our method may introduce artifacts.}
    \vspace{-0.8em}
    \label{fig:user_study_cds}
\end{figure}

Table \ref{tab:aba_user_study} presents the scores separately for object insertion (Add) and other tasks (Other). Our method achieves higher overall preference scores in both task categories, except when compared to CDS in the ``Other'' category. Upon examination, we found that this outcome stems from examples involving complex edits where both methods struggle to match the target prompt, such as those illustrated in Figure \ref{fig:user_study_cds}. In such cases, our method attempts modifications, sometimes introducing slight visual artifacts or corrupted elements. In contrast, CDS makes almost no changes to the input image. While this conservative behavior in CDS does not adhere to the target prompt, it avoids introducing errors, leading to higher scores in these specific cases.

\begin{table*}[ht]
\centering

\begin{tabular}{lcccccccc}
\toprule
\multirow{2}{*}{\textbf{Method}} & \multicolumn{2}{c}{\textbf{Background}} & \multicolumn{2}{c}{\textbf{Prompt}} & \multicolumn{2}{c}{\textbf{Quality}} & \multicolumn{2}{c}{\textbf{Overall}}\\
                                 & Add                & Other              & Add             & Other           & Add               & Other            & Add               & Other            \\
\midrule
InstructPix2Pix \cite{brooks2022instructpix2pix}  & 30.0\%            & 37.0\%            & 39.0\%         & 41.0\%         & 35.0\%           & 38.0\%          & 36.0\%           & 36.0\%          \\
HIVE \cite{zhang2024hive}                          & 42.0\%            & 52.0\%            & 34.0\%         & 47.0\%         & 42.0\%           & 48.0\%          & 34.0\%           & 44.0\%          \\
LEDITS++ \cite{brack2023ledits}                    & 31.0\%            & 40.0\%            & 27.0\%         & 39.0\%         & 27.0\%           & 47.0\%          & 29.0\%           & 41.0\%          \\
DDS \cite{hertz2023delta}                          & 39.0\%            & 48.0\%            & 31.0\%         & 43.0\%         & 34.0\%           & 42.0\%          & 35.0\%           & 42.0\%          \\
CDS \cite{nam2024contrastive}                      & 28.0\%            & 61.0\%            & 25.0\%         & 55.0\%         & 31.0\%           & 55.0\%          & 29.0\%           & 55.0\%      \\
\bottomrule
\end{tabular}

\captionsetup{skip=5pt}

\caption{
Percentage of times users preferred other methods over ours in 1-on-1 comparisons. We present the scores separately for samples involving object insertion (Add) and other tasks (Other). Please refer to Section~\ref{sec:user_study}.
}
\label{tab:aba_user_study}
\vspace{-0.8em}
\end{table*}







\subsection{Inherent biases in commonly used background preservation metrics} \label{supp:score_dump}

Metrics commonly used to assess background preservation in previous works \cite{zhang2023magicbrush, sheynin2023emuedit} are L1, CLIP-I, and DINO, all computed on the MagicBrush test set \cite{zhang2023magicbrush}. L1 is defined as the $L_1$-norm between the edited and reference images, while CLIP-I measures the cosine similarity between their CLIP embeddings. Similarly, DINO computes the cosine similarity between their DINO \cite{caron2021emerging} embeddings, making it highly correlated with CLIP-I.

In MagicBrush~\cite{zhang2023magicbrush}, the reference, or ``ground-truth'' images, were created by workers on Amazon Mechanical Turk with a mask-based inpainting model DALL-E 2\footnote{\url{https://openai.com/index/dall-e-2/}}. While this process yields high-quality edited images verified by humans, each input has only one ``correct'' ground-truth image.
As a result, the metrics may penalize good results where changes are made in a perfectly valid location but not in the single ground-truth location specified by workers during dataset creation. 

We illustrate this on the MagicBrush test set. Specifically, we compute pixel-wise differences between the input and ground-truth reference image to infer a ground-truth edit region $\vect{B}_1$. We then use Paint-by-Example \cite{yang2022paint} to insert \textit{same-sized objects} (sourced from Unsplash) into both $\vect{B}_1$ and other plausible regions $\vect{B}_2$. As shown in Table \ref{tab:supp_metric_location_bias}, these metrics disfavor editing made in $\vect{B}_2$.

\begin{table}[!h]
\centering
\small

\begin{tabular}{
    l@{\hspace{6pt}}
    c@{\hspace{6pt}}
    c@{\hspace{6pt}}
    c@{\hspace{6pt}}
}
 \hline
\textbf{Locations} & \textbf{L1} $\downarrow$ & \textbf{CLIP-I} $\uparrow$ & \textbf{DINO} $\uparrow$ \\
\hline
Same ($\vect{B}_1$) & 0.048 & 0.911 & 0.876 \\
Different ($\vect{B}_2$) & 0.057 & 0.899 & 0.839 \\
p-value & 2.33e-9 & 1.88e-2 & 1.28e-3 \\
\hline
\end{tabular}

\captionsetup{skip=5pt}
\caption{
Editing the same region as the reference images yields statistically better scores ($N = 70$). We restrict our test to cases where the ground-truth region $\vect{B}_1$ is sufficiently small, allowing us to select a non-overlapping region of the same size $\vect{B}_2$ for inpainting using Paint-by-Example.}
\vspace{-6pt}
\label{tab:supp_metric_location_bias}


\end{table}

Moreover, these metrics can produce misleading rankings by favoring unchanged outputs over valid edits that deviate from the ground truth (see Table \ref{tab:supp_metric_flaw}). Additionally, as DINO is trained via self-supervised training to capture differences between objects of the same class, the DINO metric may penalize valid edits that produce the correct object but with an appearance different from the one in the ground-truth image.

For these reasons, we exclude these metrics from Table \ref{tab:main_compare_sota}, propose four new metrics (Section~\ref{sec:quantitative_eval}) and assess visual quality with a user study (Section~\ref{sec:user_study}).

\begin{table}[!h]
\centering
\small
\begin{tabular}{
    l@{\hspace{6pt}}
    c@{\hspace{6pt}}
    c@{\hspace{6pt}}
    c@{\hspace{6pt}}
    c@{\hspace{6pt}}
    c@{\hspace{6pt}}
    c@{\hspace{6pt}}
    c@{\hspace{6pt}}
    c@{\hspace{6pt}}
}
\toprule
\textbf{Method} & $\textbf{L1}$ $\downarrow$ & $\textbf{CLIP-I}$ $\uparrow$      & $\textbf{DINO}$ $\uparrow$   \\
\midrule
Do Nothing & \colorbox{tabfirst}{0.037} & \colorbox{tabfirst}{0.943} & \colorbox{tabfirst}{0.917} \\
InstructPix2Pix \cite{brooks2022instructpix2pix} & 0.147 & 0.782 & 0.607 \\

HIVE \cite{zhang2024hive}   & 0.090 & 0.893 & 0.824 \\
LEDITS++ \cite{brack2023ledits}  & 0.097 & 0.864 & 0.775 \\
DDS \cite{hertz2023delta}    & 0.066 & 0.920 & 0.886 \\
CDS \cite{nam2024contrastive}     & \colorbox{tabsecond}{0.061} & \colorbox{tabsecond}{0.931} & \colorbox{tabsecond}{0.902} \\
SBP \cite{mcallister2024rethinking} & 0.095 & 0.825 & 0.752 \\
\textbf{Ours}       & 0.063 & 0.900 & 0.853 \\
\bottomrule
\end{tabular}

\captionsetup{skip=5pt}
\caption{The \colorbox{tabfirst}{best} and \colorbox{tabsecond}{second-best} scores are color-coded. We observe that the commonly used L1, CLIP-I, and DINO metrics for this task are biased toward unchanged results, with a method that does nothing to the input (Do Nothing) ranking first across the board. As a result, comparisons based on these scores can be misleading. We discuss this limitation in Section~\ref{supp:score_dump} and propose less biased evaluation in Section~\ref{sec:experiment}.
}
\vspace{-3pt}
\label{tab:supp_metric_flaw}


\end{table}

\section{Additional Qualitative Results}

\subsection{Benchmark dataset} \label{supp:qual_magicbrush}

\begin{figure*}
    \centering
    \includegraphics[width=1.0\textwidth]{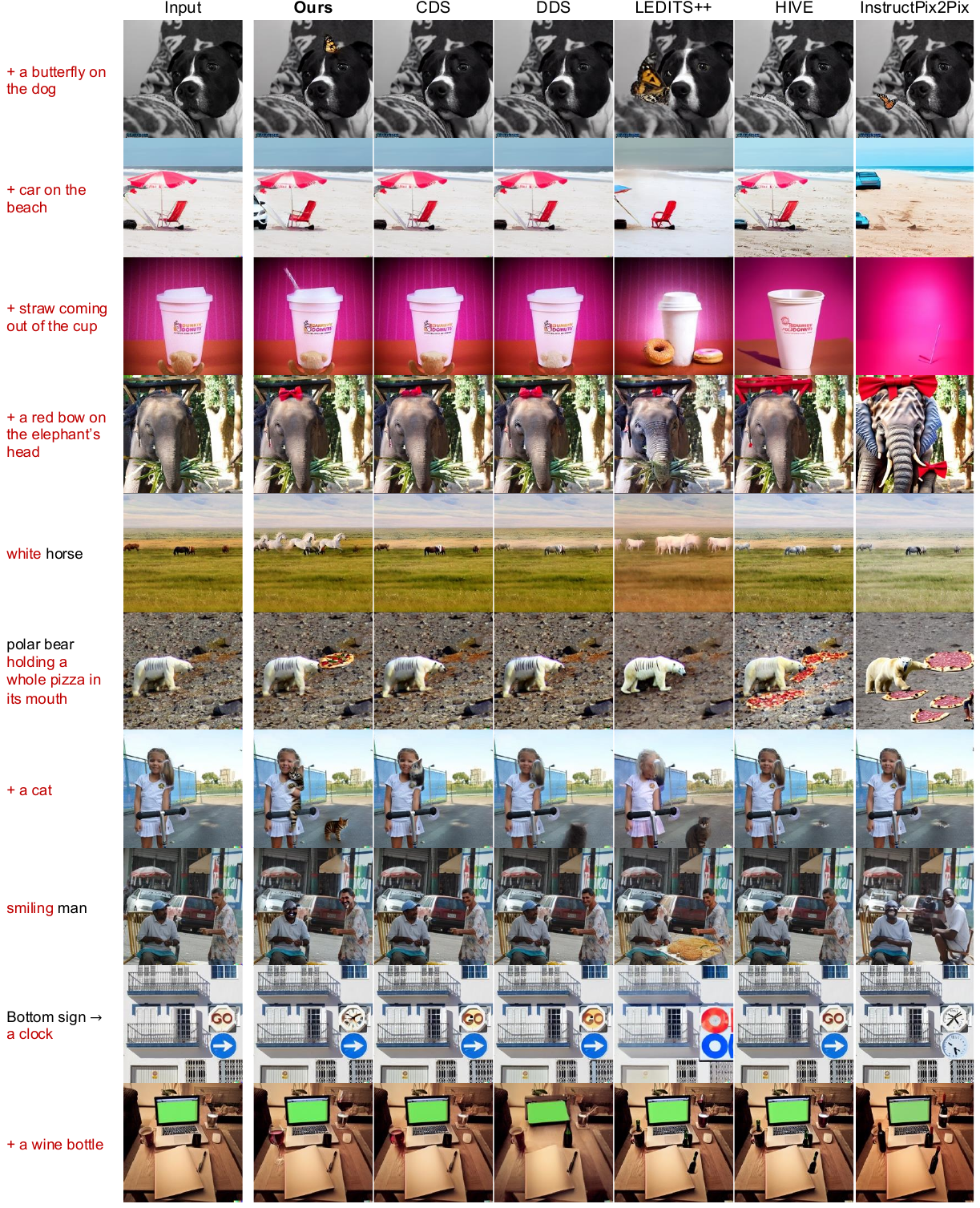}
    \captionsetup{skip=5pt}
    \caption{Qualitative results on MagicBrush dataset \cite{zhang2023magicbrush} between our method and other state-of-the-art methods}
    \vspace{-12pt}
    \label{fig:aba_magicbrush_sota1}
\end{figure*}

\begin{figure*}
    \centering
    \includegraphics[width=1.0\textwidth]{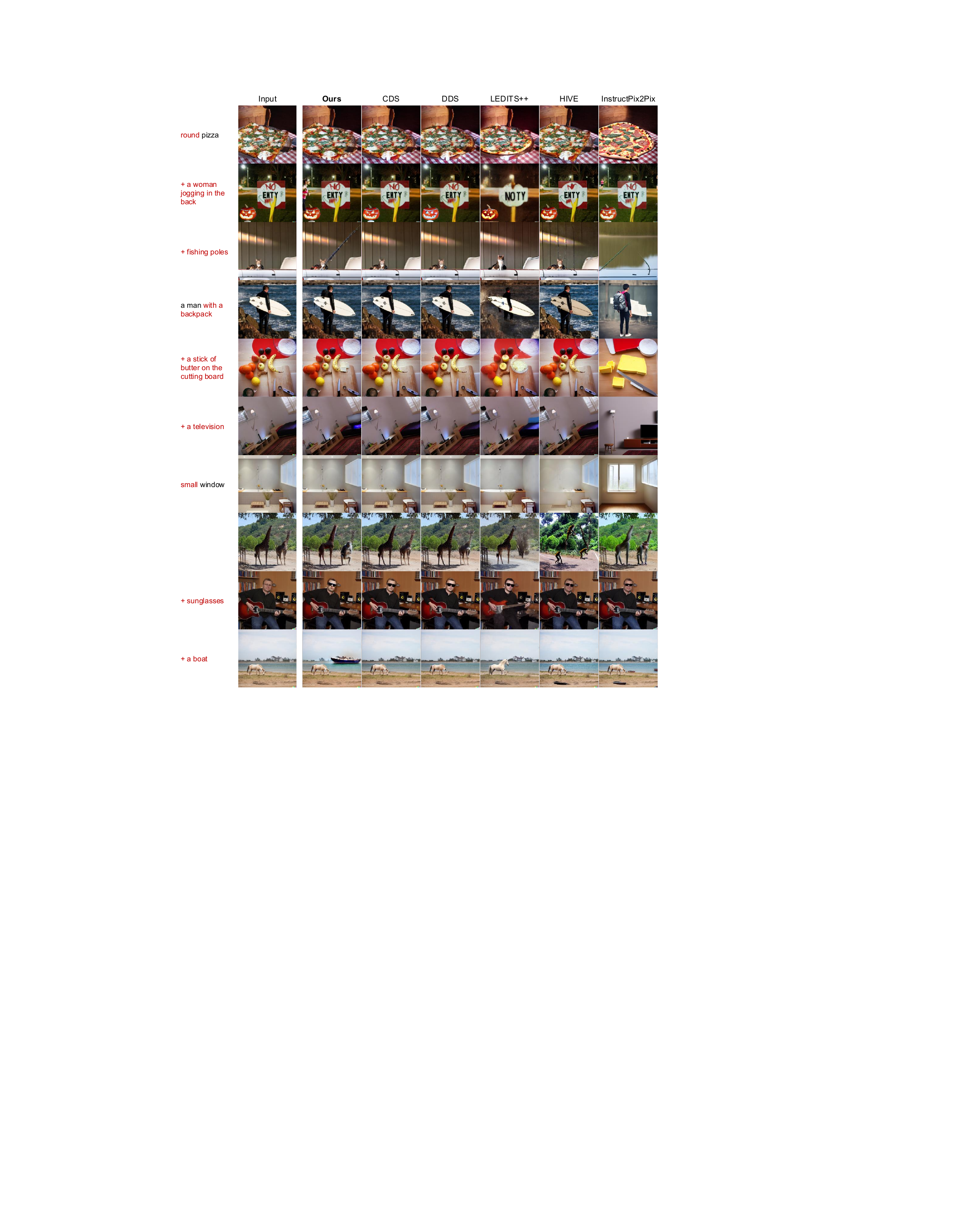}
    \captionsetup{skip=5pt}
    \caption{Qualitative results on MagicBrush dataset \cite{zhang2023magicbrush} between our method and other state-of-the-art methods}
    \vspace{-12pt}
    \label{fig:aba_magicbrush_sota2}
\end{figure*}

This section provides qualitative results for the experiment on MagicBrush test set \cite{zhang2023magicbrush} in Section \ref{sec:experiment} of the main paper. We show editing results from our LUSD method alongside other state-of-the-art approaches in Figures \ref{fig:aba_magicbrush_sota1} and \ref{fig:aba_magicbrush_sota2}. 
While our method may occasionally make incorrect edits (e.g., the bottom two examples in Figure \ref{fig:aba_magicbrush_sota1}) due to the inherently limited language understanding of Stable Diffusion, it generally offers a good balance between prompt fidelity and background preservation.

\begin{figure*}
    \centering
    \includegraphics[width=0.9\textwidth]{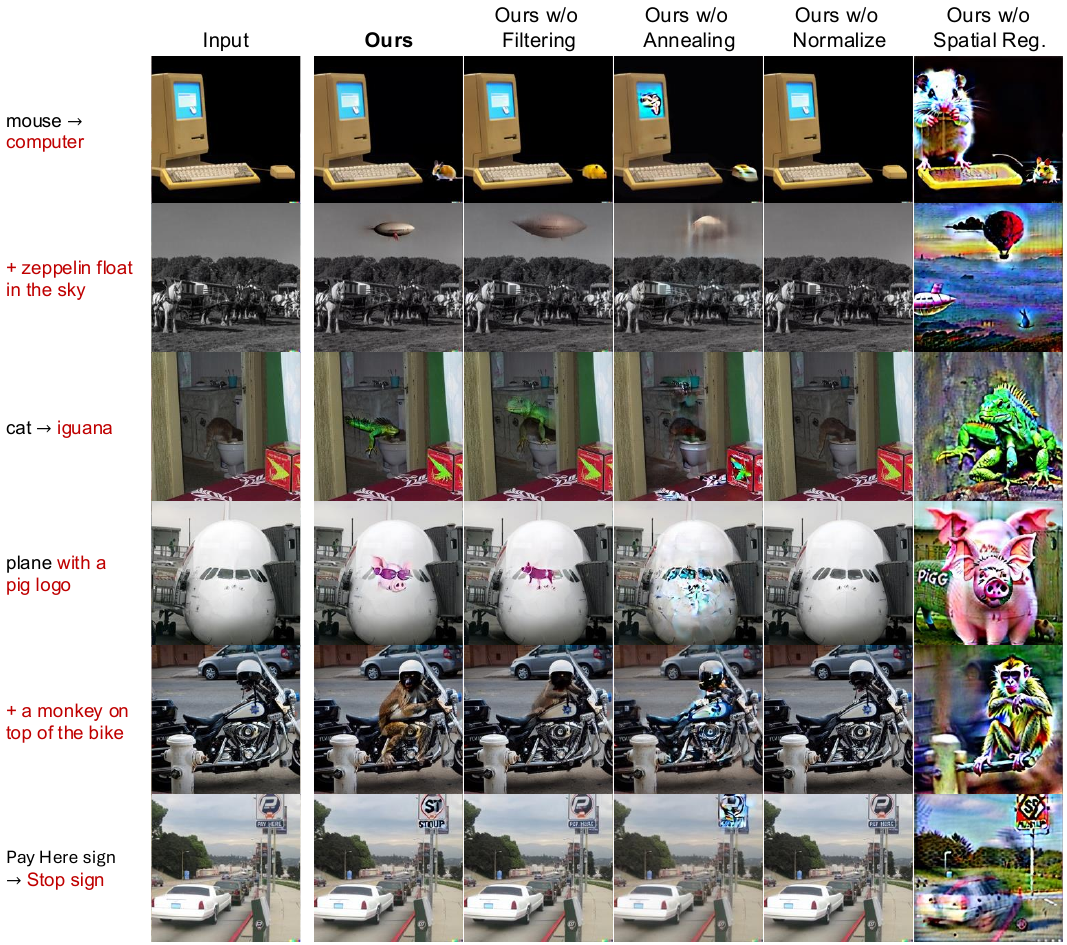}
    \captionsetup{skip=5pt}
    \caption{Qualitative results on MagicBrush dataset \cite{zhang2023magicbrush} between our full method and its ablated versions.}
    \vspace{-12pt}
    \label{fig:aba_magicbrush_aba}
\end{figure*}

Additionally, Figure \ref{fig:aba_magicbrush_aba} compares the performance of our full method against its ablated versions. Excluding spatial regularization results in entirely new images. Not annealing normalized gradients magnitude via $\gamma$ produces visual artifacts due to unstable optimization. Without gradient filtering and normalization, our method often struggles to insert objects correctly or produces incomplete additions.


\subsection{In-the-wild images} \label{supp:qual_other}
As images in MagicBrush \cite{zhang2023magicbrush} are curated from MS COCO dataset \cite{lin2014microsoft} only, we present additional qualitative results for diverse images under CC4.0 license from Unsplash\footnote{https://unsplash.com/} and other websites, using multiple random seeds in Figures \ref{fig:supp_seed1} to \ref{fig:supp_seed3}. 
We input source prompts and target prompts directly into LEDITS++ \cite{brack2023ledits}, DDS \cite{hertz2023delta}, CDS \cite{nam2024contrastive}, and our method.
For InstructPix2Pix \cite{brooks2023instructpix2pix} and HIVE \cite{zhang2024hive}, we use edit instructions generated by ChatGPT, as these models are trained on edit instructions. To generate these inputs, given a source prompt and a target prompt from MagicBrush, we ask chatgpt to generate an edit instruction, prepping it with a short prompt that contains a couple of examples of desired text transformation. Note that this approach is similar to the procedure used in MagicBrush, where a global description (i.e., a target prompt) is inferred from a source prompt and an edit instruction using ChatGPT.

Unlike other methods, which require adjusting hyperparameters for each image to achieve good editing results, LUSD achieves competitive performance---or even better in challenging cases involving object insertion---using a single configuration. It also works across diverse scenarios, such as adding a Google logo to a t-shirt, adding a party hat to a cat, and replacing meatballs with chrome balls. Refer to Appendix \ref{supp:tuning_grid} for hyperparameter tuning grids.

\subsection{Comparison with object insertion works} 
\label{supp:object_insertion}

\begin{figure*}
    \centering
    \includegraphics[width=1.0\textwidth]{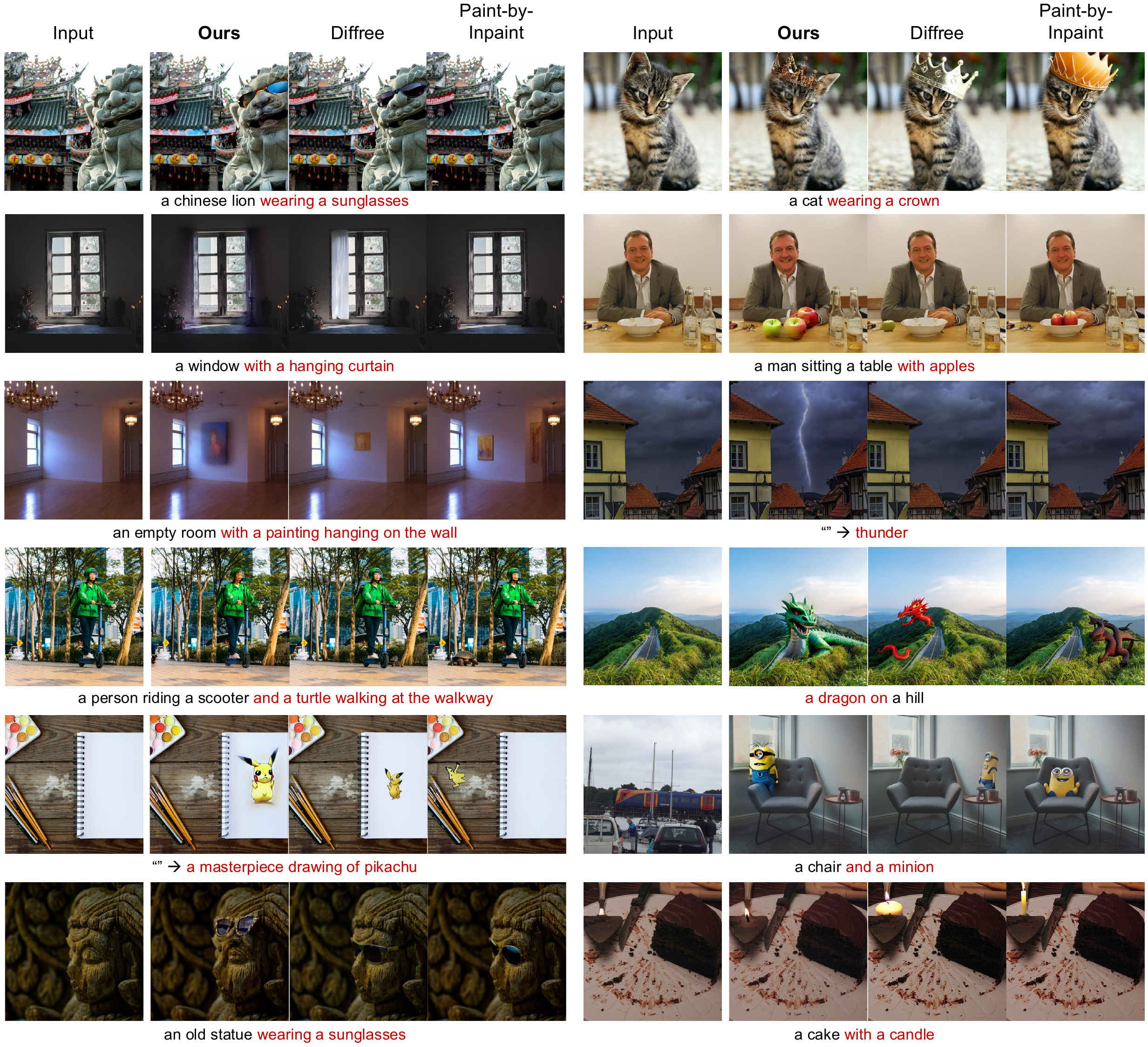}
    \captionsetup{skip=5pt}
    \caption{Comparison of our method with other supervised object insertion methods. While task-specific approaches perform better on common objects, they struggle with objects outside their training classes.}
    \vspace{-12pt}
    \label{fig:aba_object_insertion}
\end{figure*}

Our work focuses on stabilizing score distillation, which enables general image editing using diffusion priors. This differs from object insertion techniques that specifically tackle object insertion with supervised fine-tuning on datasets such as Paint-by-Inpaint \cite{wasserman2024paint} and Diffree \cite{zhao2024diffree}. While supervised approaches generally perform better for common objects (e.g., curtain, apple, turtle), they can produce qualitatively worse results for objects outside their training classes (e.g., dragon, Pikachu, Minion), as shown in Figure \ref{fig:aba_object_insertion}. Interestingly, even fine-tuned models exhibit the minimal-effort issue (Section~\ref{supp:minimal_effort}), albeit to a lesser extent (e.g., sunglasses on a statue, candle). Bridging the gap between these two approaches remains an interesting research direction.

\subsection{Comparison with rectified flow models} \label{supp:rectifed_flow}

In Section \ref{sec:experiment}, we limit our comparison to methods applicable to Stable Diffusion \cite{rombach2021highresolution} and those fine-tuned on it to ensure a fair evaluation, as models vary in their prior knowledge and language understanding. Nonetheless, we also include comparisons with RF-Inversion \cite{rout2025semantic} and RF-Edit \cite{wang2024taming}, both zero-shot methods designed for rectified flow models. For implementation, we use FLUX.1-dev \footnote{\url{https://huggingface.co/black-forest-labs/FLUX.1-dev}} with Diffusers' implementation for RF-Inversion and the official implementation for RF-Edit.
Following the paper's recommendation, we set the inversion prompt in RF-Inversion to an empty string and limit the number of feature-sharing steps in RF-edit to 5, with other hyperparameters set to default values. Note that the number of parameters in Stable Diffusion and FLUX.1-dev are 1.3 billion and 12 billion, respectively.

\begin{figure*}
    \centering
    \includegraphics[width=1.0\textwidth]{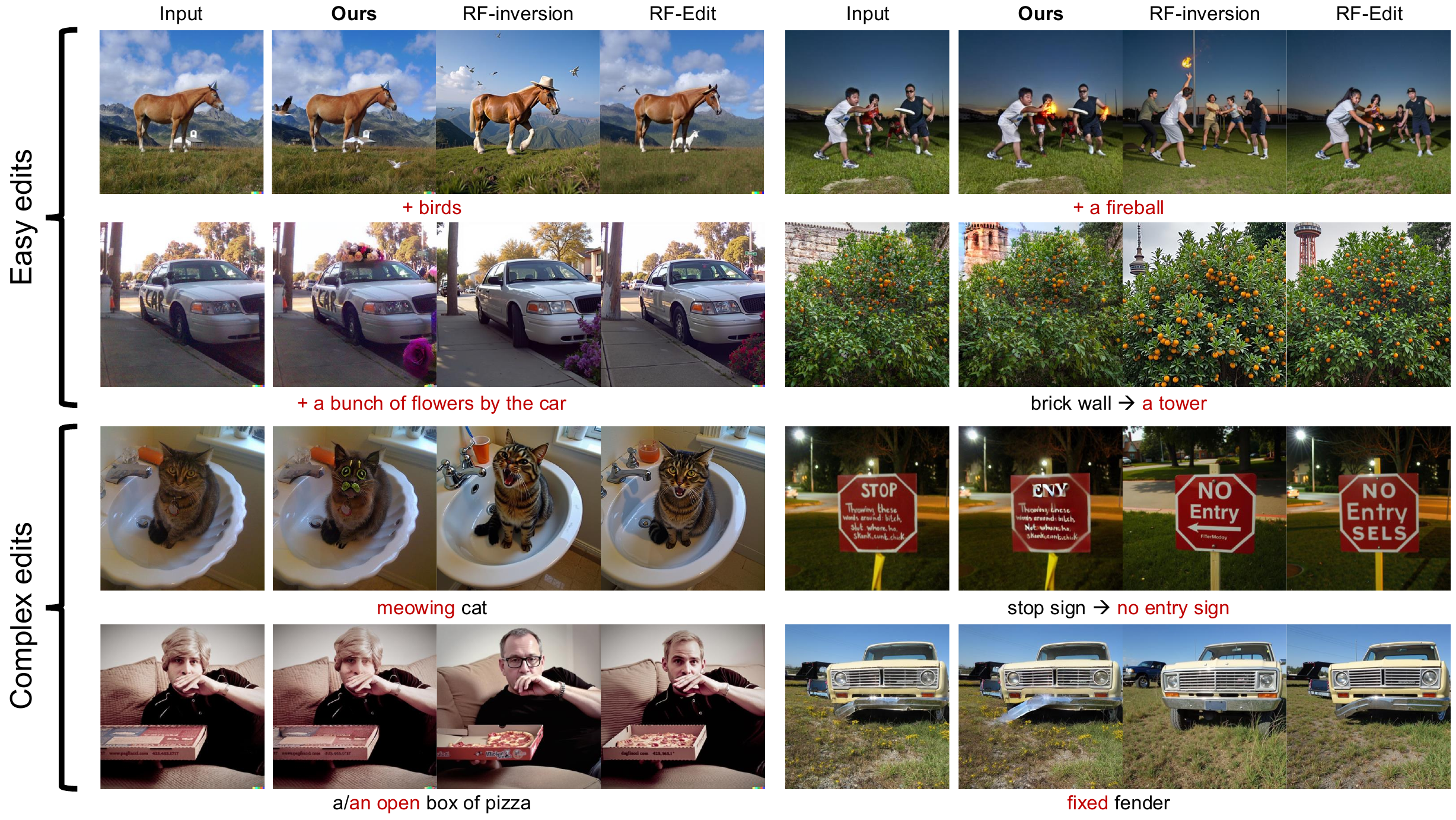}
    \captionsetup{skip=5pt}
    \caption{
    Comparison of our method with zero-shot rectified-flow-based approaches. For simple edits, all methods can follow the text prompt, but our approach better preserves background elements, such as the horse’s hat, people's poses, the graffiti on the car, and the distribution of fruits on the plant. However, RF-Inversion and RF-Edit can handle more complex edits by leveraging the richer prior and better language understanding of the larger base model (FLUX.1-dev).
    }
    \vspace{-12pt}
    \label{fig:aba_rectified_flow}
\end{figure*}

As shown in Table \ref{tab:aba_compare_rf}, our method outperforms RF-Edit and is competitive with RF-Inversion in CLIP-T on the MagicBrush \cite{zhang2023magicbrush} test set. However, RF-Inversion outperforms our method in CLIP-AUC. This improvement can be due to RF-Inversion and RF-Edit's ability to handle more complex edits (making a cat meowing, altering texts, and opening a pizza box) by leveraging the richer prior and better language understanding of the larger FLUX.1-dev (Figure \ref{fig:aba_rectified_flow}). Nonetheless, these methods still struggle with background preservation, which is the central challenge addressed by our work.

\begin{table}[!h]
\centering
\small
\begin{tabular}{
    l@{\hspace{2pt}}
    c@{\hspace{2pt}}
    c@{\hspace{2pt}}
    c@{\hspace{2pt}}
    c@{\hspace{2pt}}
}
\toprule
\textbf{Method} & \textbf{CLIP-T} $\uparrow$   & \textbf{CLIP-AUC} $\uparrow$  & $\textbf{L1}^*$ $\downarrow$         & $\textbf{CLIP-I}^*$ $\uparrow$  \\
\midrule
RF-Inversion   & \colorbox{tabfirst}{0.287} & \colorbox{tabfirst}{0.096} & 0.026 & 0.171 \\
RF-Edit    & 0.279 & 0.068 & 0.016 & 0.182 \\
\textbf{Ours}  & \colorbox{tabfirst}{0.287} & 0.074 & \colorbox{tabfirst}{0.015} & \colorbox{tabfirst}{0.192} \\
\bottomrule
\end{tabular}
\vspace{-0.5em}
\caption{
Comparison on MagicBrush between rectified-flow-based methods and our method.}
\label{tab:aba_compare_rf}
\vspace{-12pt}

\end{table}





\section{Additional Failure Cases} \label{supp:minimal_effort}
\begin{figure*}[h]
    \centering
    \includegraphics[width=0.85\textwidth]{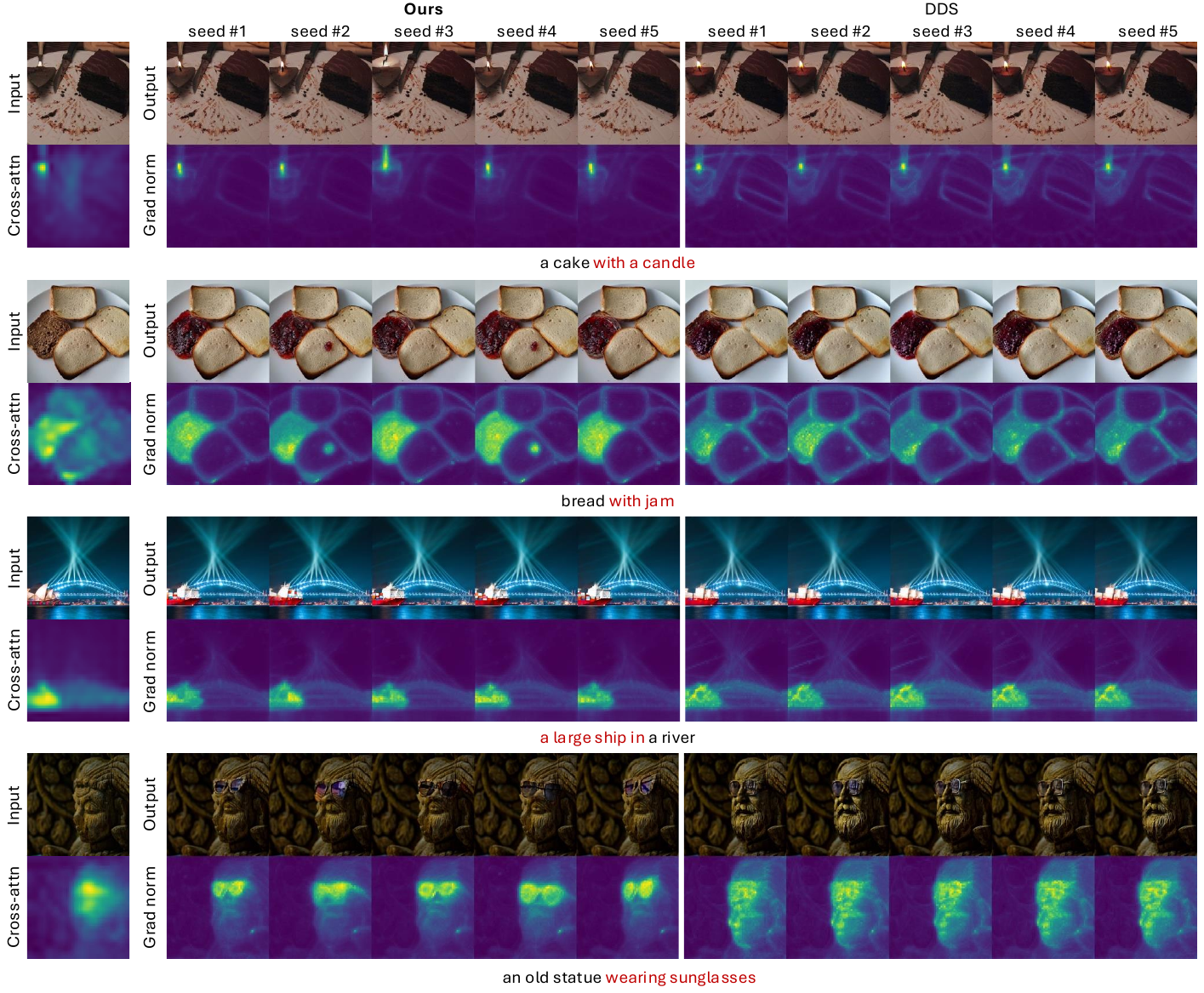}
    \captionsetup{skip=5pt}
    \caption{Failure mode: Our method and other SDS-based methods (e.g., DDS \cite{hertz2023delta}) favor minimal-effort regions, where the visual cues needed for object formation are already present. This bias may lead to unnatural object placements or limited diversity in image edits.}
    \vspace{-12pt}
    \label{fig:aba_limitation}
\end{figure*}

Our technique successfully improves the success rate of SDS-based image editing, particularly for object insertion. However, it remains susceptible to \textit{minimal-effort regions}, where the visual cues needed for object formation are already present, leading our method to only add objects there. As shown in Figure \ref{fig:aba_limitation}, these cues can manifest as intensity (e.g., a candle), color (e.g., bread), or shape (e.g., a ship or sunglasses). We observed that such regions are associated with unusually high values in the cross-attention map $\vect{A}_{C}^{l, t, \vect{e}}$, averaged across layers $l$, timesteps $t$, and target noun tokens $\vect{e}$ (see Section \ref{sec:method_attention} and Appendix \ref{supp:noun_tokens}). Since the magnitude of averaged gradients correlates with the spatial location of these bright spots, SDS-based methods that derive gradient updates directly from model predictions are inherently vulnerable to this issue. To address this spatial bias, a potential solution might be reweighting attention features. This problem is an interesting area for future work.




\section{More Comparison with SOTA Image Editing Methods}
\label{supp:tuning_grid}

\begin{figure*}
    \centering
    \begin{subfigure}[b]{0.99\textwidth}
        \centering
        \includegraphics[width=\textwidth]{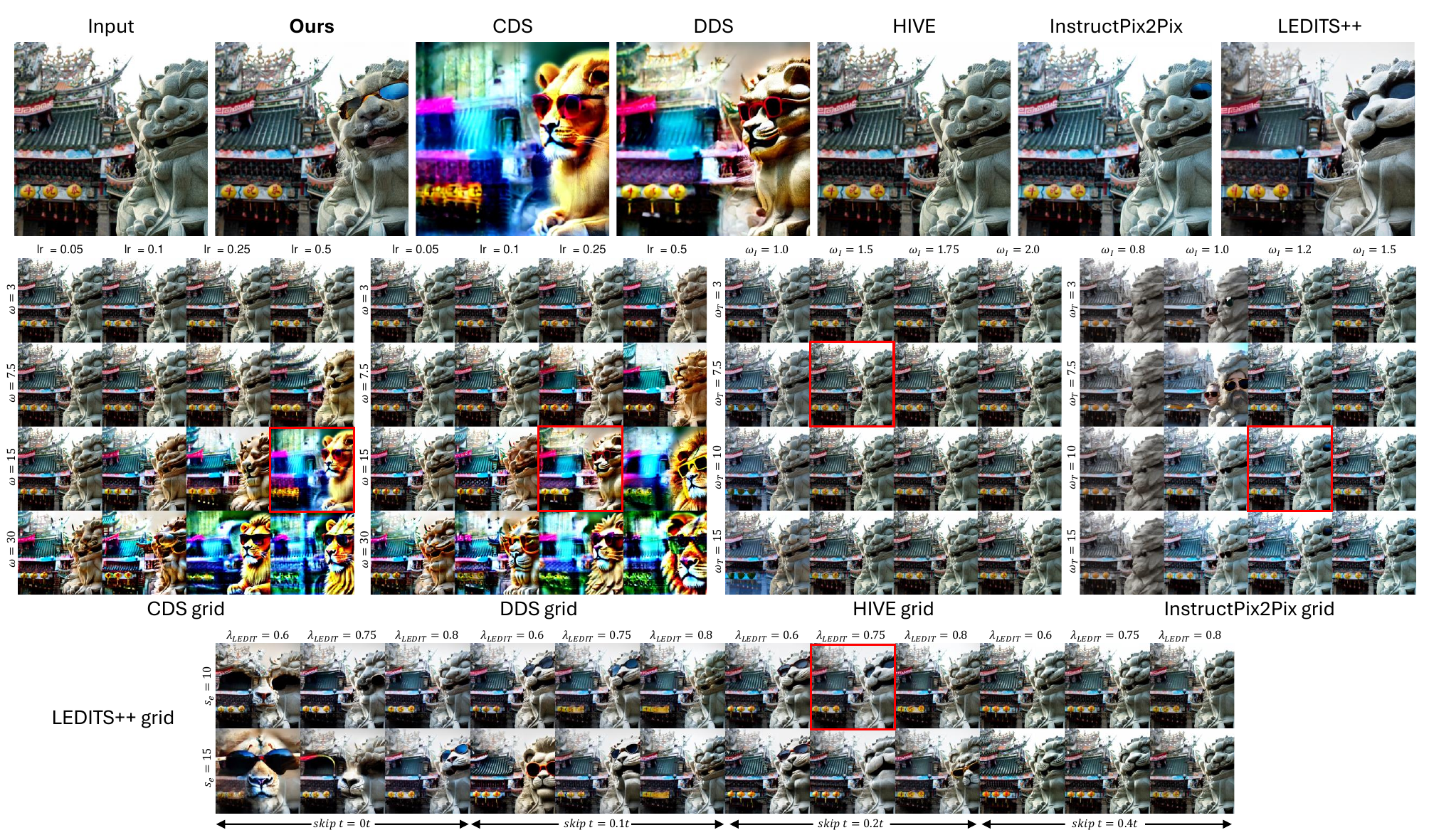}
        \caption{Prompt: a Chinese lion wearing \textcolor{red}{a sunglasses}}
        
    \end{subfigure}
    
    \begin{subfigure}[b]{0.99\textwidth}
        \centering
        \includegraphics[width=\textwidth]{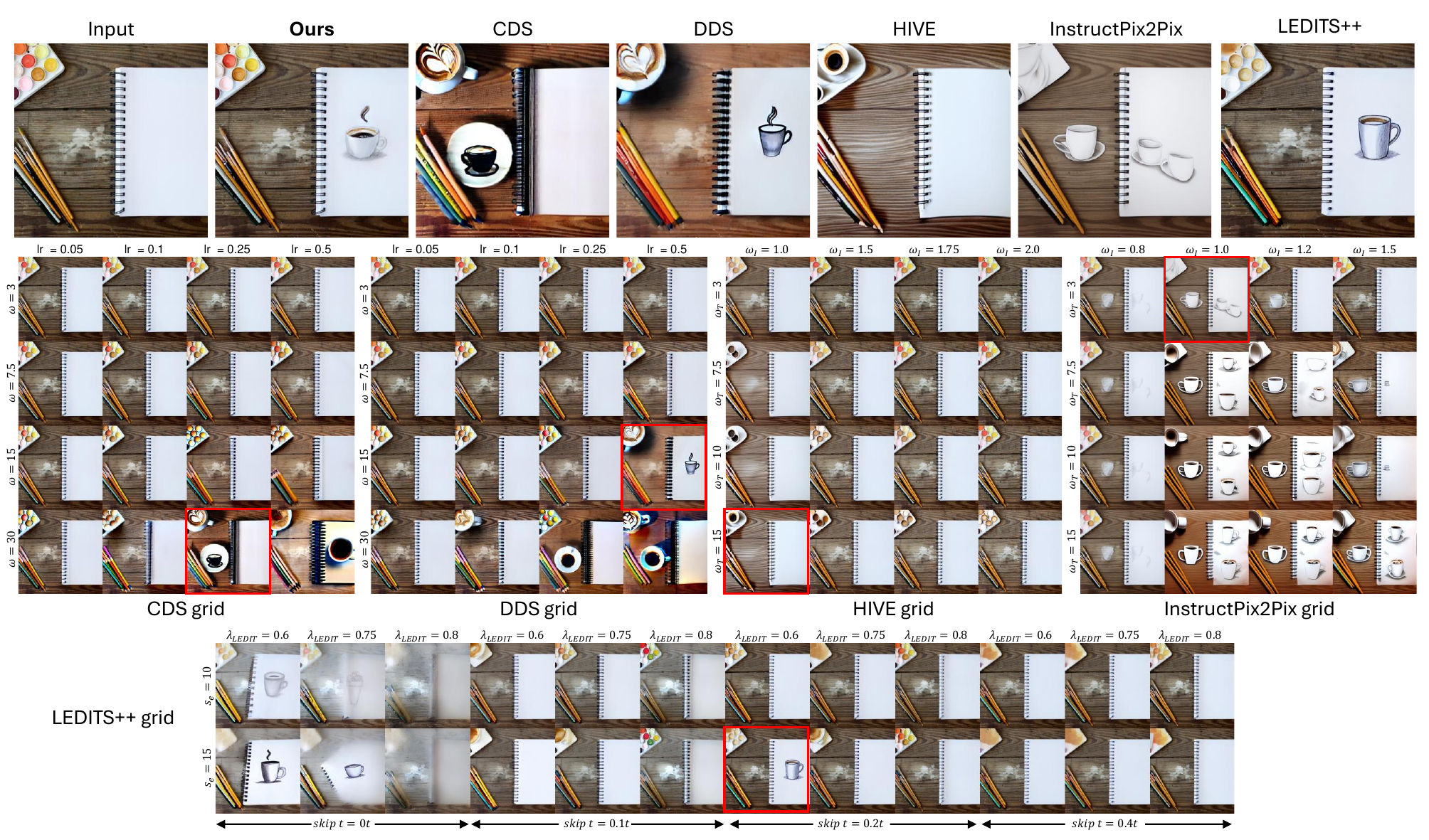}
        \caption{Prompt: a notebook \textcolor{red}{with a drawing of a coffee cup}}
        
    \end{subfigure}
    
    \caption{Hypertuning grids on various in-the-wild editing tasks. We compare the best results (prioritizing object appearance) from state-of-the-art methods, optimized through hyperparameter tuning (highlighted by red boxes), with our results all generated using a single configuration. When several configurations perform equally, we choose the one that best preserves the background.}
    \vspace{-12pt}
    \label{fig:supp_grid1}
\end{figure*}

\begin{figure*}
    \centering
    \begin{subfigure}[b]{0.99\textwidth}
        \centering
        \includegraphics[width=\textwidth]{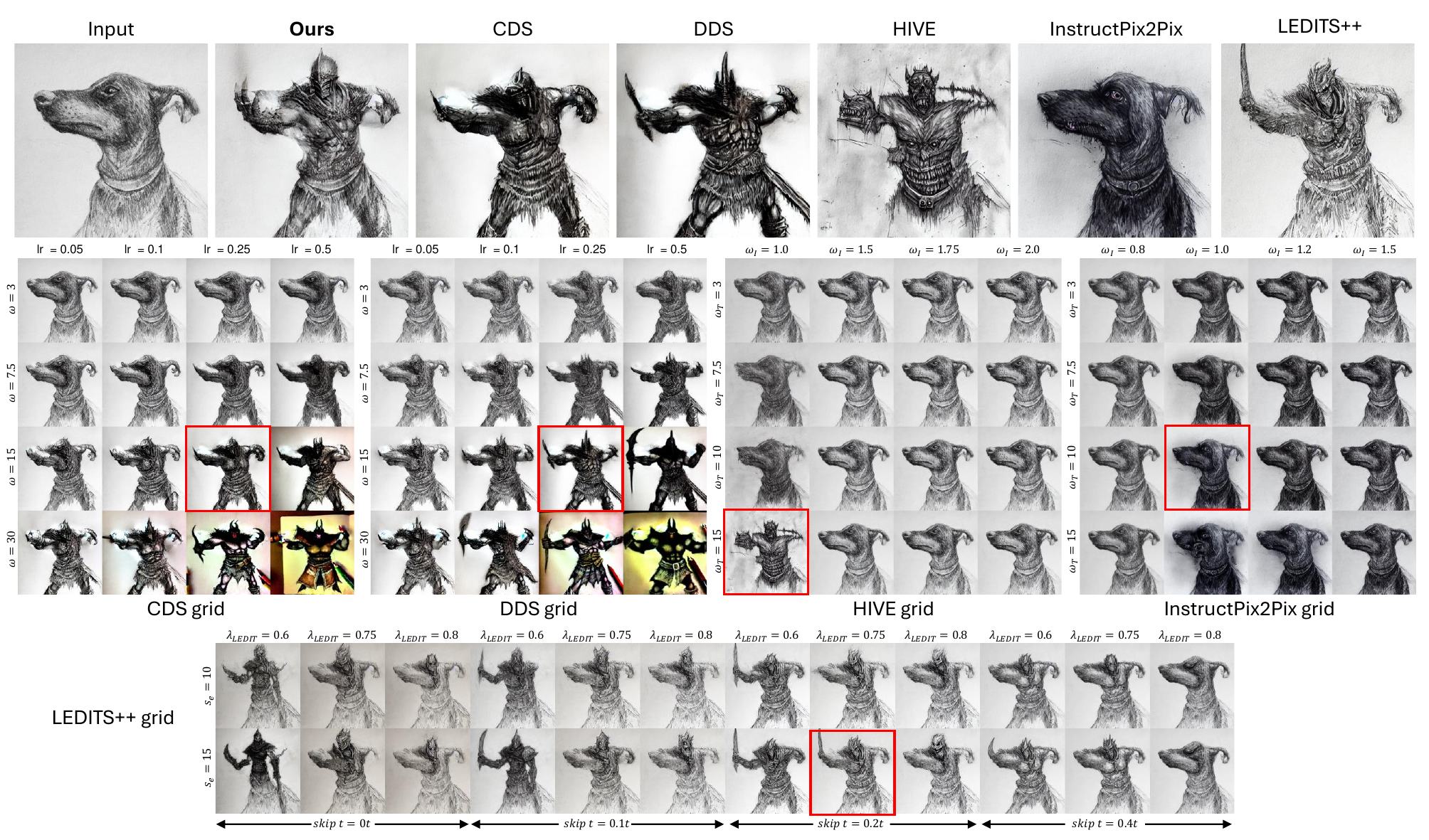}
        \caption{Prompt: a drawing of (``a dog'' $\rightarrow$ \textcolor{red}{``a boss in dark soul''})}
        
    \end{subfigure}
    
    \begin{subfigure}[b]{0.99\textwidth}
        \centering
        \includegraphics[width=\textwidth]{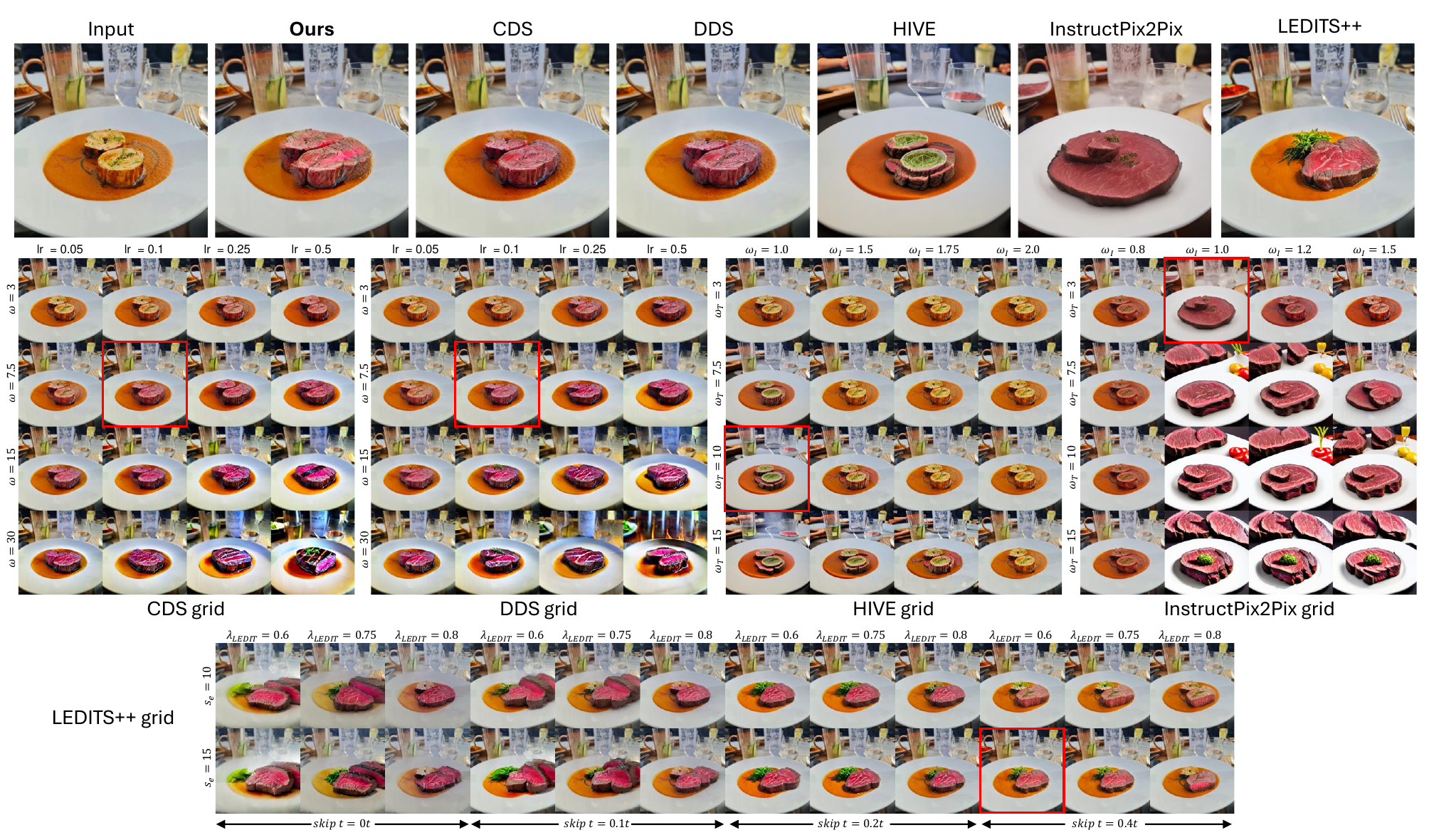}
        \caption{Prompt: (``food'' $\rightarrow$ \textcolor{red}{``wagyu steak''})}
        
    \end{subfigure}
    
    \caption{Hypertuning grids on various in-the-wild editing tasks. We compare the best results (prioritizing object appearance) from state-of-the-art methods, optimized through hyperparameter tuning (highlighted by red boxes), with our results all generated using a single configuration. When several configurations perform equally, we choose the one that best preserves the background.}
    \vspace{-12pt}
    \label{fig:supp_grid2}
\end{figure*}

\begin{figure*}
    \centering
    \begin{subfigure}[b]{0.99\textwidth}
        \centering
        \includegraphics[width=\textwidth]{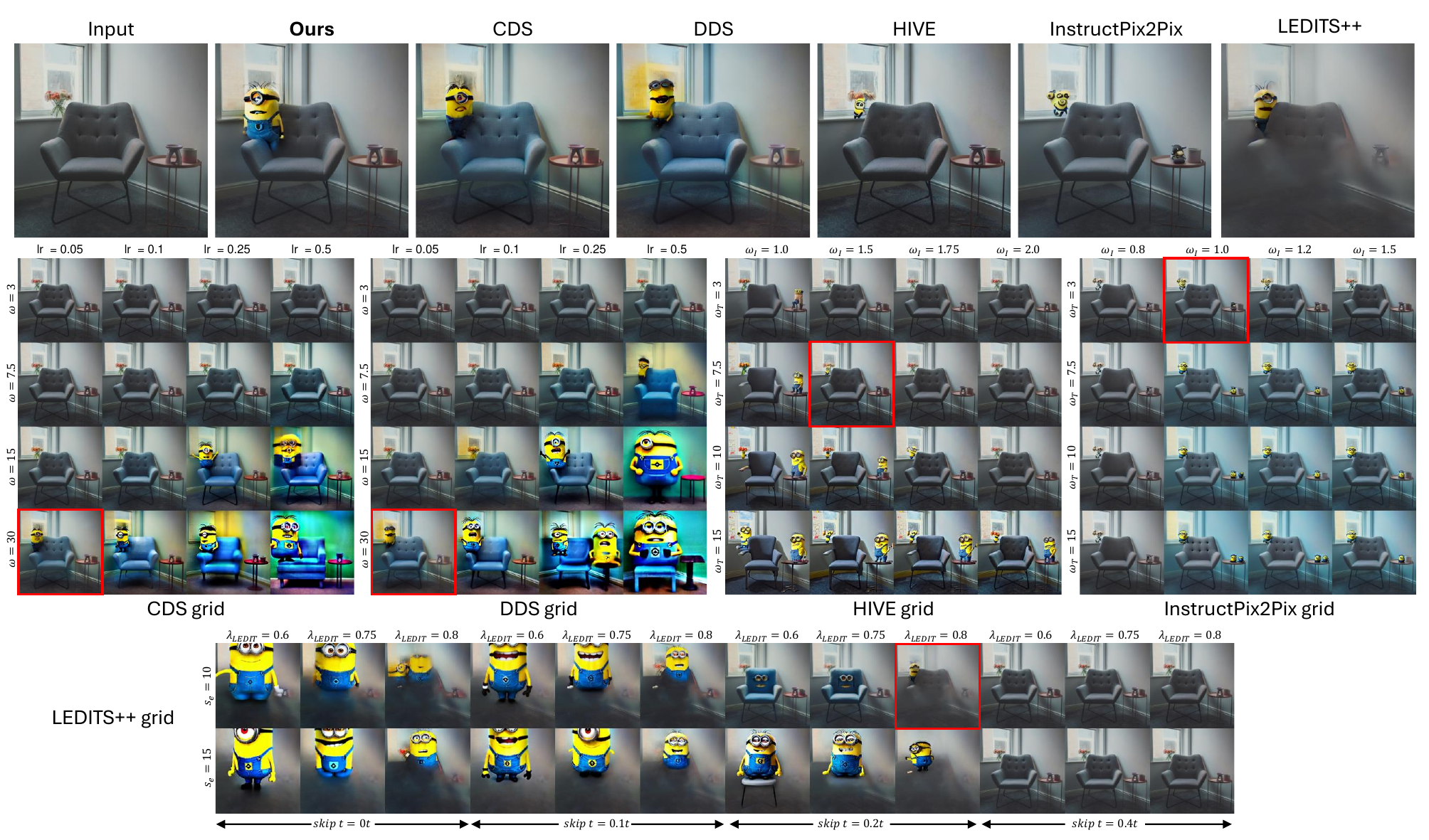}
        \caption{Prompt: a chair \textcolor{red}{and a minion}}
        
    \end{subfigure}
    
    \begin{subfigure}[b]{0.99\textwidth}
        \centering
        \includegraphics[width=\textwidth]{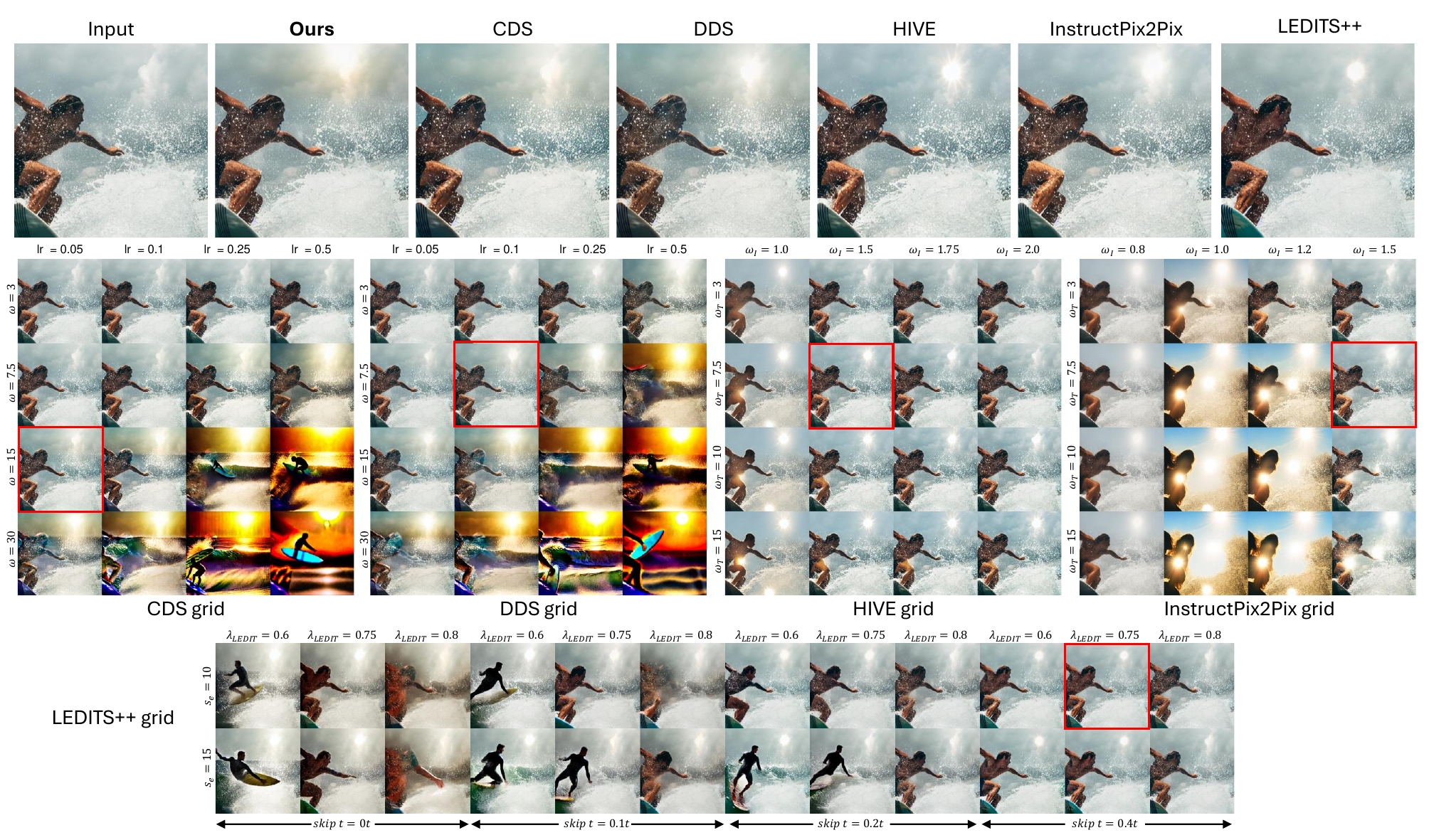}
        \caption{Prompt: a man surfing \textcolor{red}{and a sun}}
        
    \end{subfigure}
    
    \caption{Hypertuning grids on various in-the-wild editing tasks. We compare the best results (prioritizing object appearance) from state-of-the-art methods, optimized through hyperparameter tuning (highlighted by red boxes), with our results all generated using a single configuration. When several configurations perform equally, we choose the one that best preserves the background.}
    \vspace{-12pt}
    \label{fig:supp_grid3}
\end{figure*}

\begin{figure*}
    \centering
    \begin{subfigure}[b]{0.99\textwidth}
        \centering
        \includegraphics[width=\textwidth]{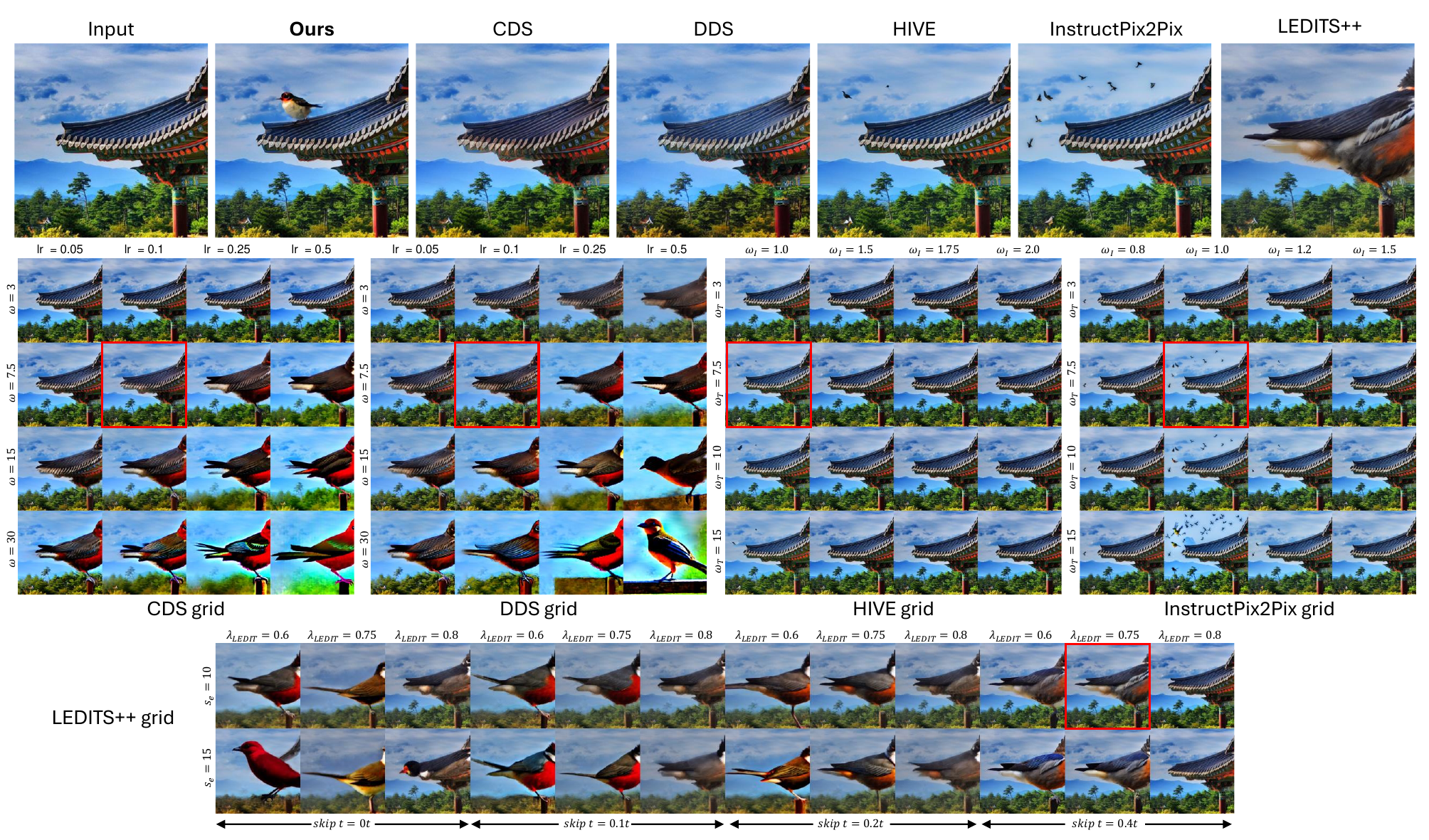}
        \caption{Prompt: \textcolor{red}{a bird on} a roof}
        
    \end{subfigure}
    
    \begin{subfigure}[b]{0.99\textwidth}
        \centering
        \includegraphics[width=\textwidth]{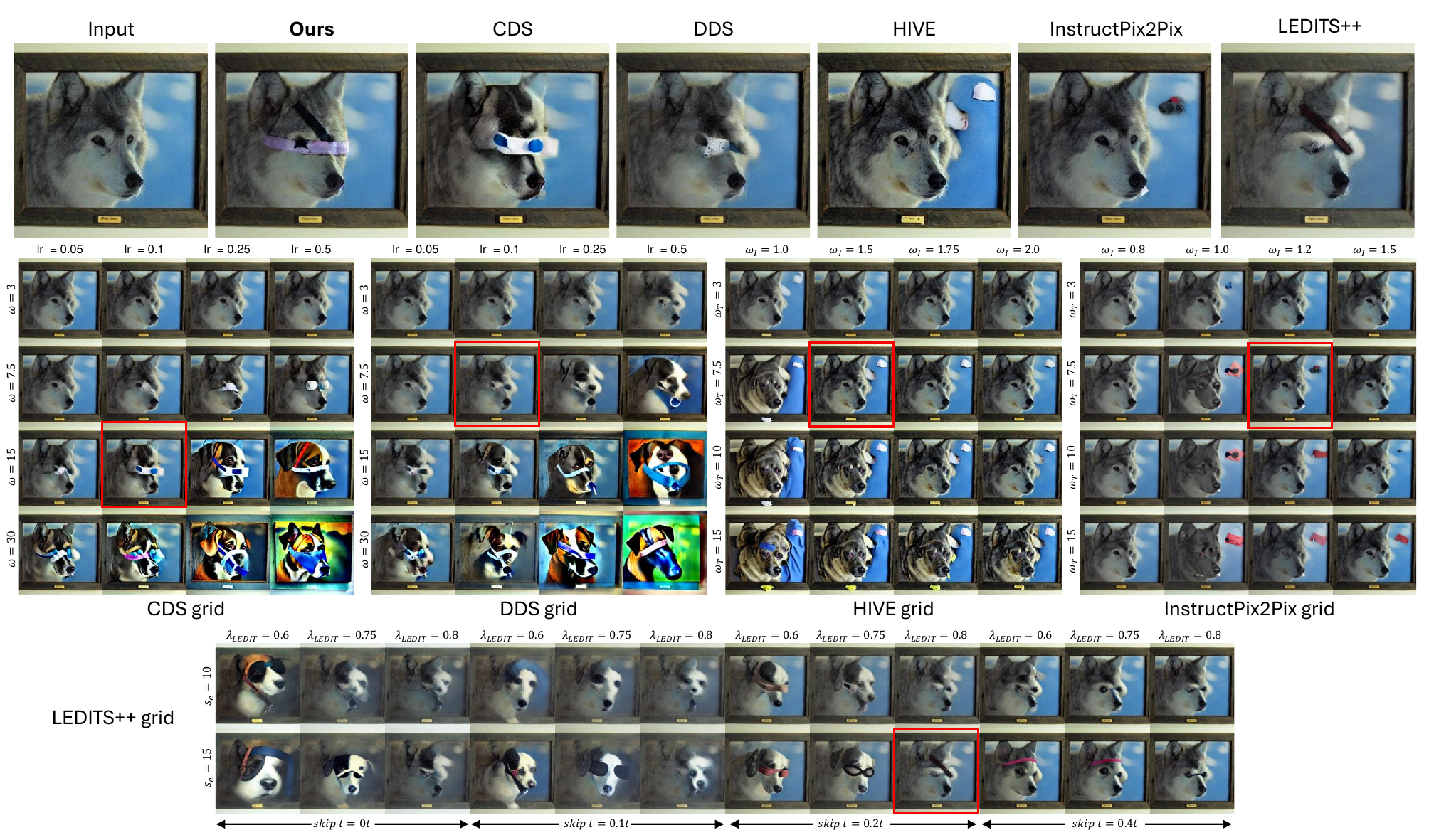}
        \caption{Prompt: a painting of a dog \textcolor{red}{wearing a blindfold over its eyes}}
        
    \end{subfigure}
    
    \caption{Hypertuning grids on various in-the-wild editing tasks. We compare the best results (prioritizing object appearance) from state-of-the-art methods, optimized through hyperparameter tuning (highlighted by red boxes), with our results all generated using a single configuration. When several configurations perform equally, we choose the one that best preserves the background.}
    \vspace{-12pt}
    \label{fig:supp_grid4}
\end{figure*}

\begin{figure*}
    \centering
    \begin{subfigure}[b]{0.99\textwidth}
        \centering
        \includegraphics[width=\textwidth]{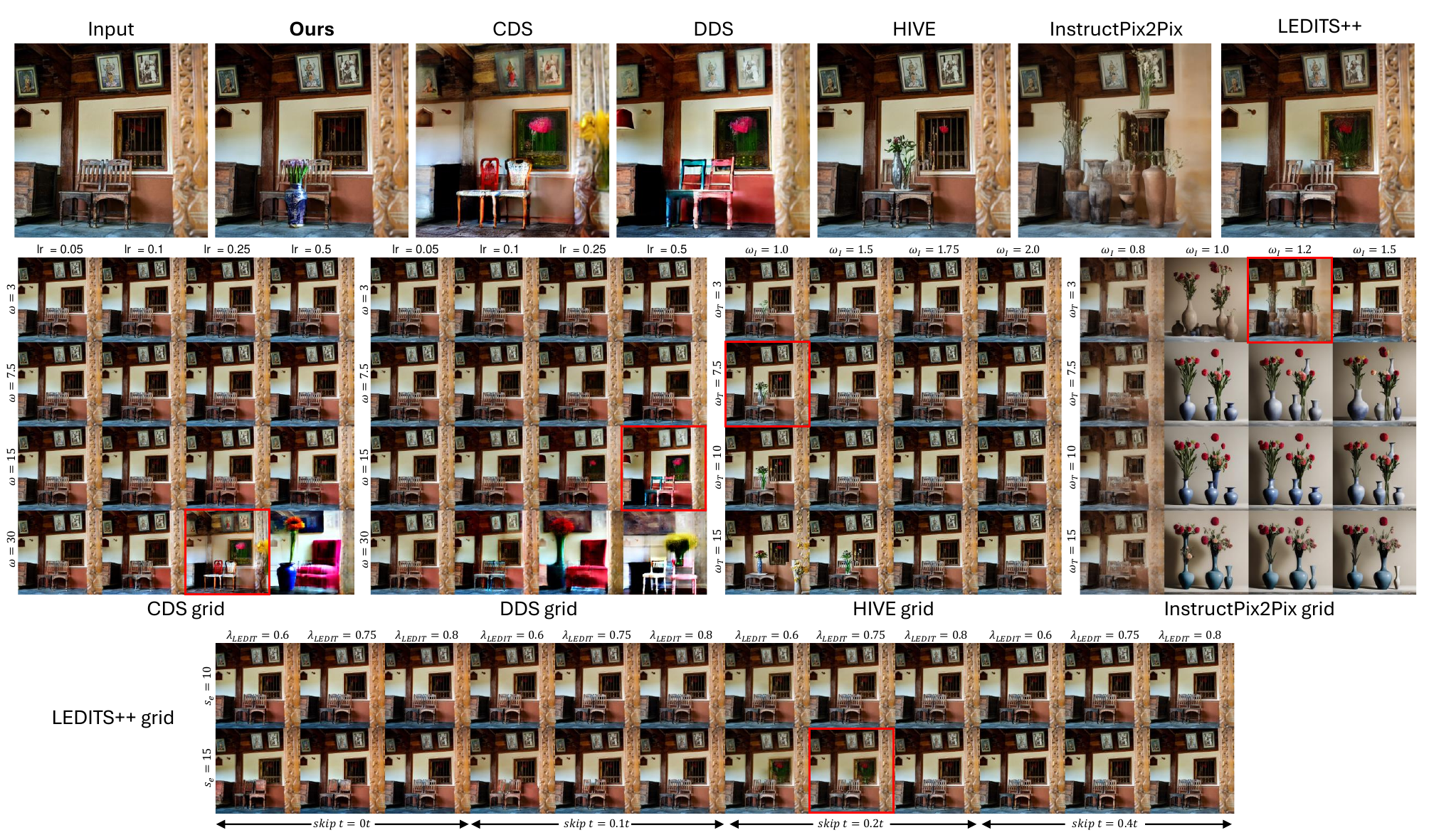}
        \caption{Prompt: two chairs \textcolor{red}{and a vase}}
        
    \end{subfigure}
    
    \begin{subfigure}[b]{0.99\textwidth}
        \centering
        \includegraphics[width=\textwidth]{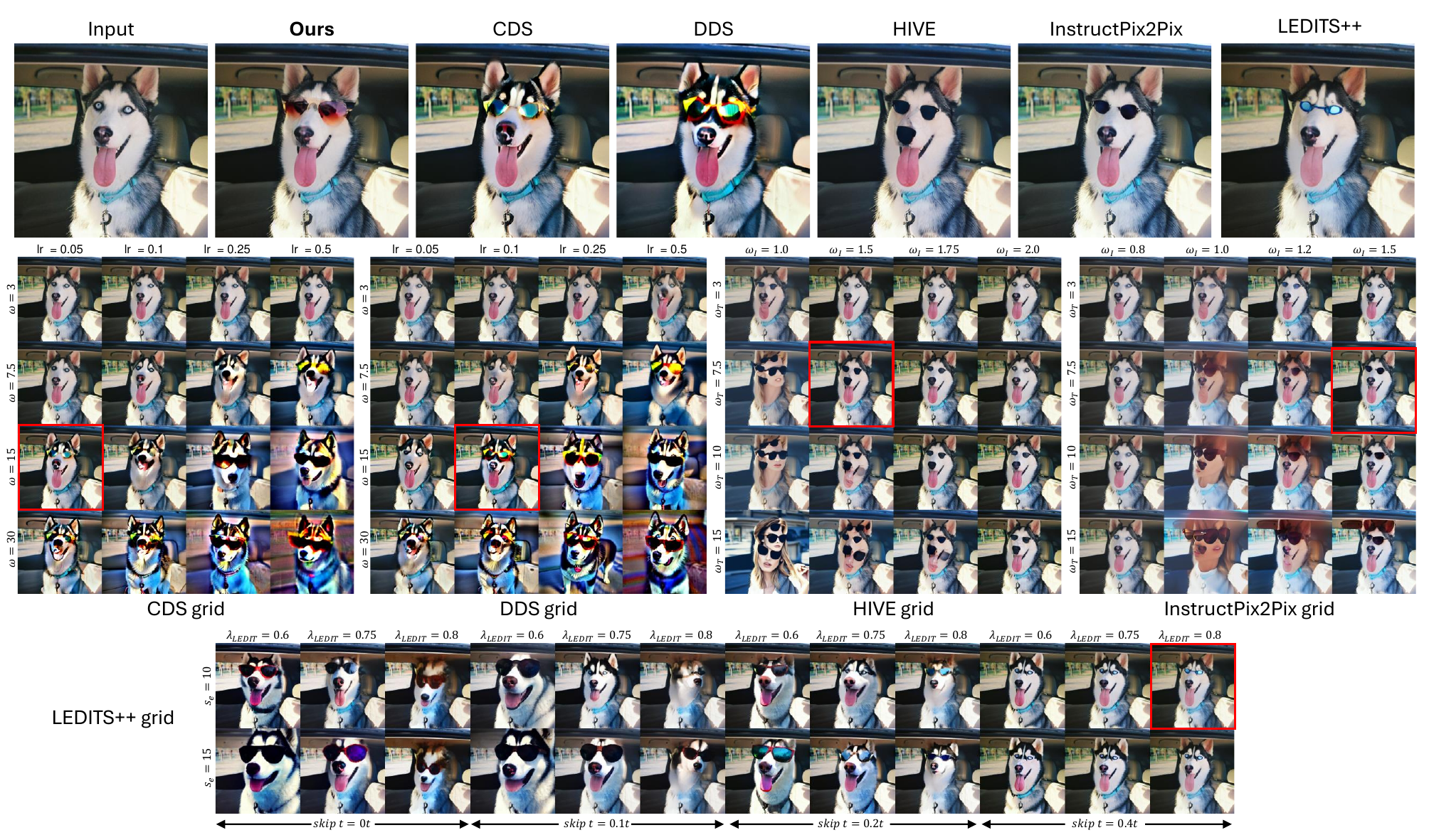}
        \caption{Prompt: a siberian husky \textcolor{red}{wearing stylish sunglasses}}
        
    \end{subfigure}
    
    \caption{Hypertuning grids on various in-the-wild editing tasks. We compare the best results (prioritizing object appearance) from state-of-the-art methods, optimized through hyperparameter tuning (highlighted by red boxes), with our results all generated using a single configuration. When several configurations perform equally, we choose the one that best preserves the background.}
    \vspace{-12pt}
    \label{fig:supp_grid5}
\end{figure*}

\begin{figure*}
    \centering
    \begin{subfigure}[b]{0.99\textwidth}
        \centering
        \includegraphics[width=\textwidth]{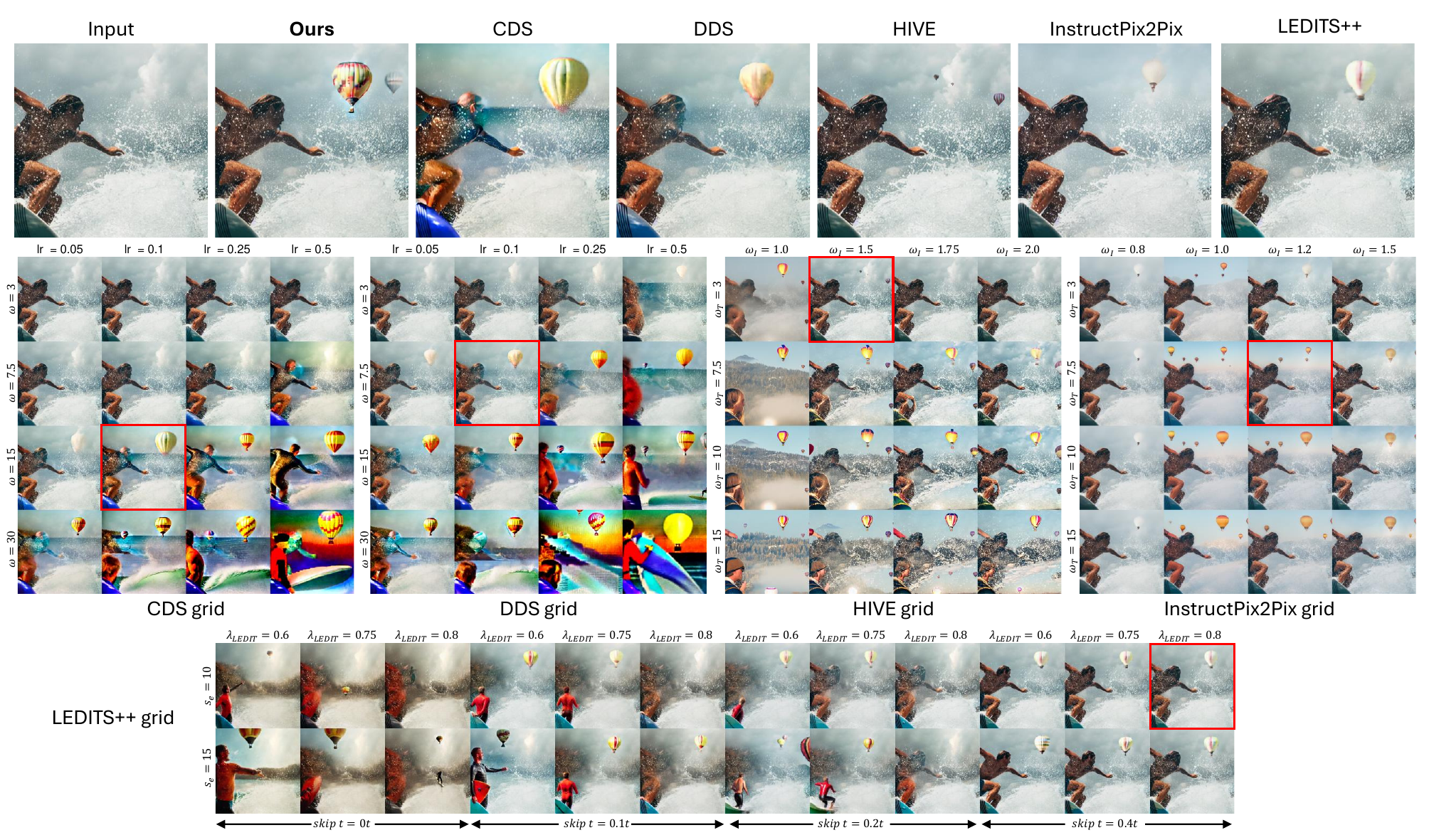}
        \caption{Prompt: a man surfing \textcolor{red}{with a hot air balloon at the background}}
        
    \end{subfigure}
    
    \begin{subfigure}[b]{0.99\textwidth}
        \centering
        \includegraphics[width=\textwidth]{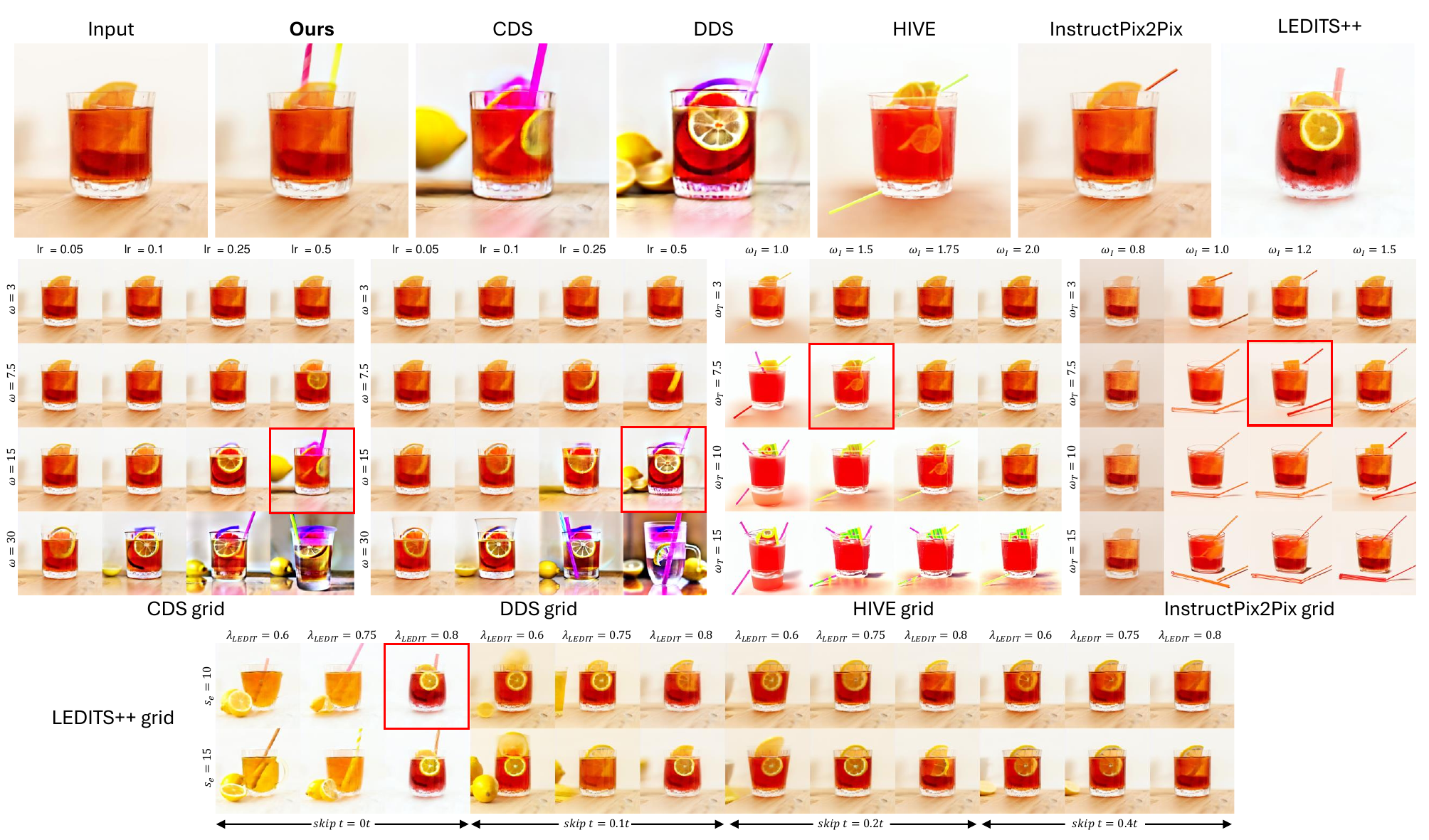}
        \caption{Prompt: a glass of lemon tea \textcolor{red}{with a straw}}
        
    \end{subfigure}
    
    \caption{Hypertuning grids on various in-the-wild editing tasks. We compare the best results (prioritizing object appearance) from state-of-the-art methods, optimized through hyperparameter tuning (highlighted by red boxes), with our results all generated using a single configuration. When several configurations perform equally, we choose the one that best preserves the background.}
    \vspace{-12pt}
    \label{fig:supp_grid6}
\end{figure*}

\begin{figure*}
    \centering
    \begin{subfigure}[b]{0.99\textwidth}
        \centering
        \includegraphics[width=\textwidth]{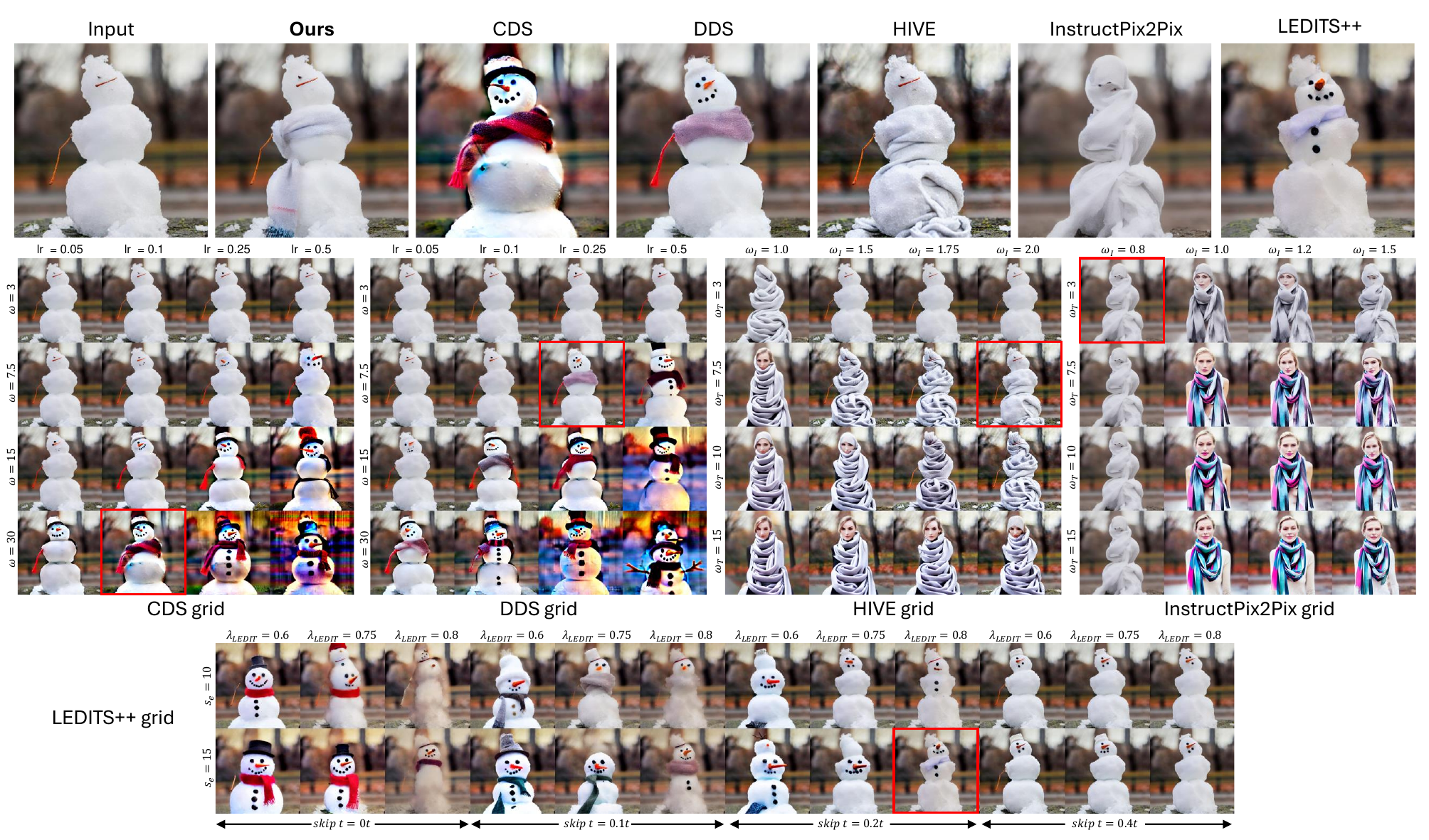}
        \caption{Prompt: a snowman \textcolor{red}{with a scarf}}
        
    \end{subfigure}
    
    \begin{subfigure}[b]{0.99\textwidth}
        \centering
        \includegraphics[width=\textwidth]{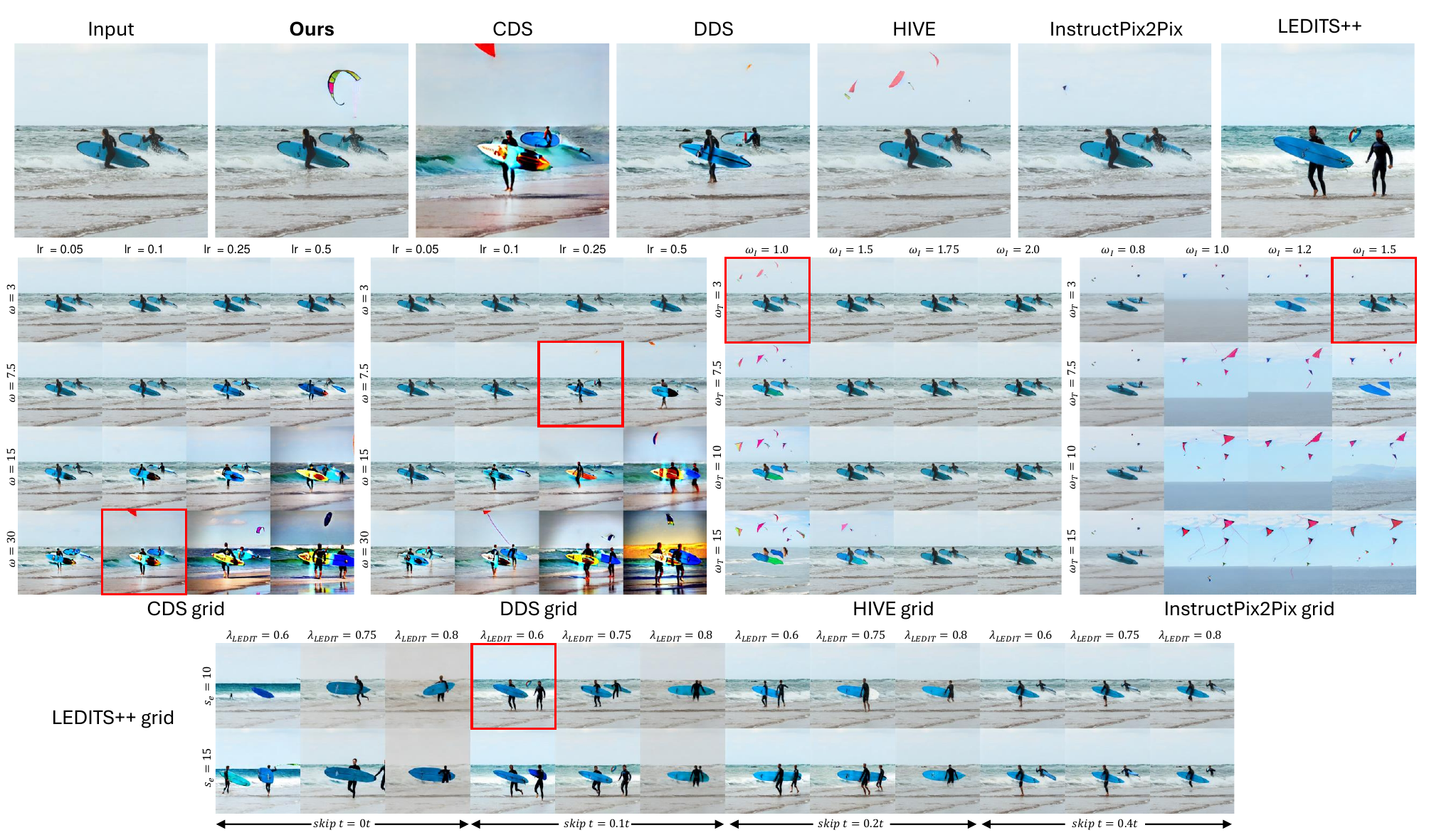}
        \caption{Prompt: two surfers carrying bright blue boards wade into the ocean \textcolor{red}{and a kite in the sky}}
        
    \end{subfigure}
    
    \caption{Hypertuning grids on various in-the-wild editing tasks. We compare the best results (prioritizing object appearance) from state-of-the-art methods, optimized through hyperparameter tuning (highlighted by red boxes), with our results all generated using a single configuration. When several configurations perform equally, we choose the one that best preserves the background.}
    \vspace{-12pt}
    \label{fig:supp_grid7}
\end{figure*}

\begin{figure*}
    \centering
    \begin{subfigure}[b]{0.99\textwidth}
        \centering
        \includegraphics[width=\textwidth]{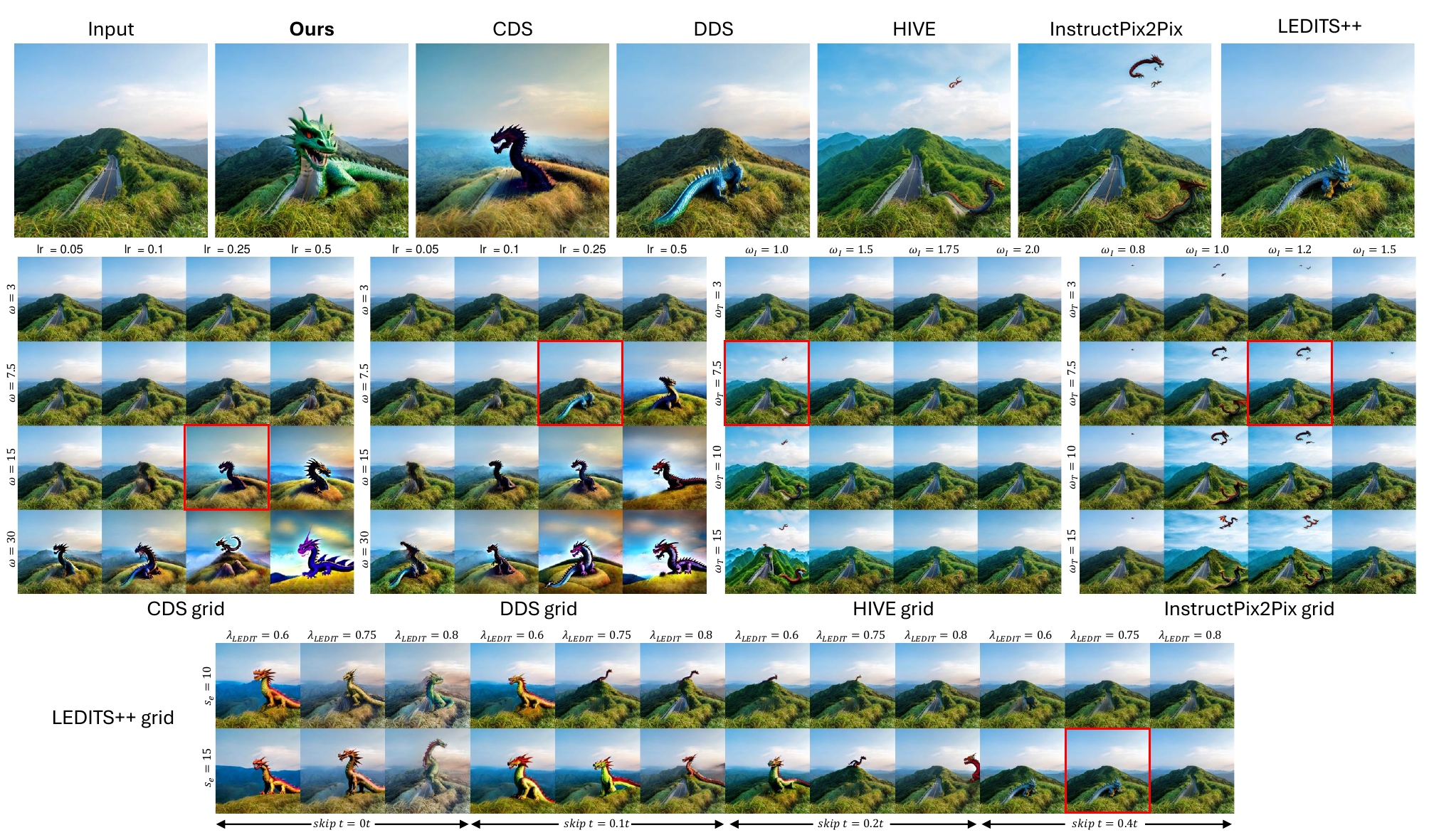}
        \caption{Prompt: \textcolor{red}{a dragon on} a hill}
        
    \end{subfigure}
    
    \begin{subfigure}[b]{0.99\textwidth}
        \centering
        \includegraphics[width=\textwidth]{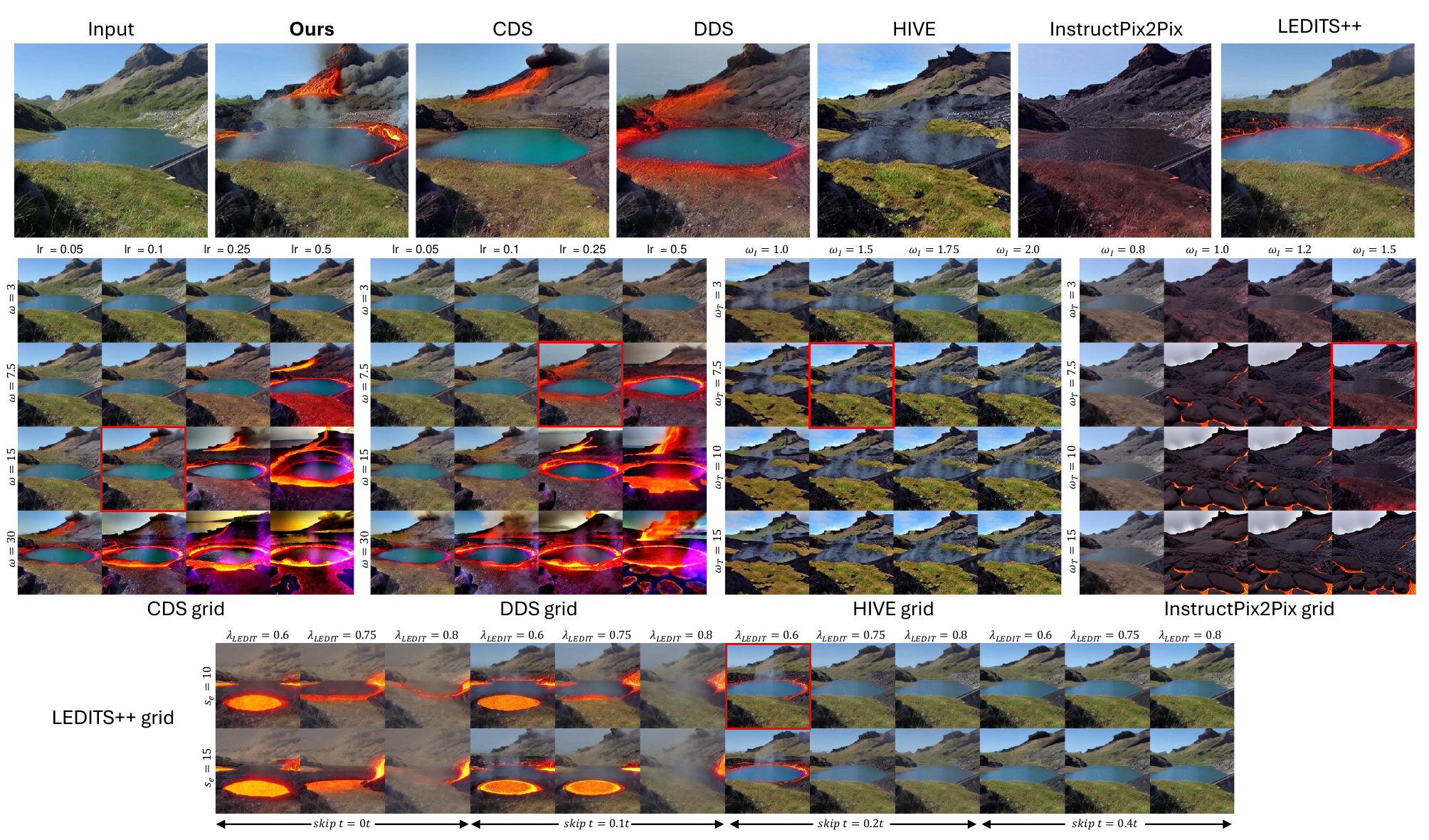}
        \caption{Prompt: a lake \textcolor{red}{of lava}}
        
    \end{subfigure}
    
    \caption{Hypertuning grids on various in-the-wild editing tasks. We compare the best results (prioritizing object appearance) from state-of-the-art methods, optimized through hyperparameter tuning (highlighted by red boxes), with our results all generated using a single configuration. When several configurations perform equally, we choose the one that best preserves the background.}
    \vspace{-12pt}
    \label{fig:supp_grid8}
\end{figure*}

\begin{figure*}
    \centering
    \begin{subfigure}[b]{0.99\textwidth}
        \centering
        \includegraphics[width=\textwidth]{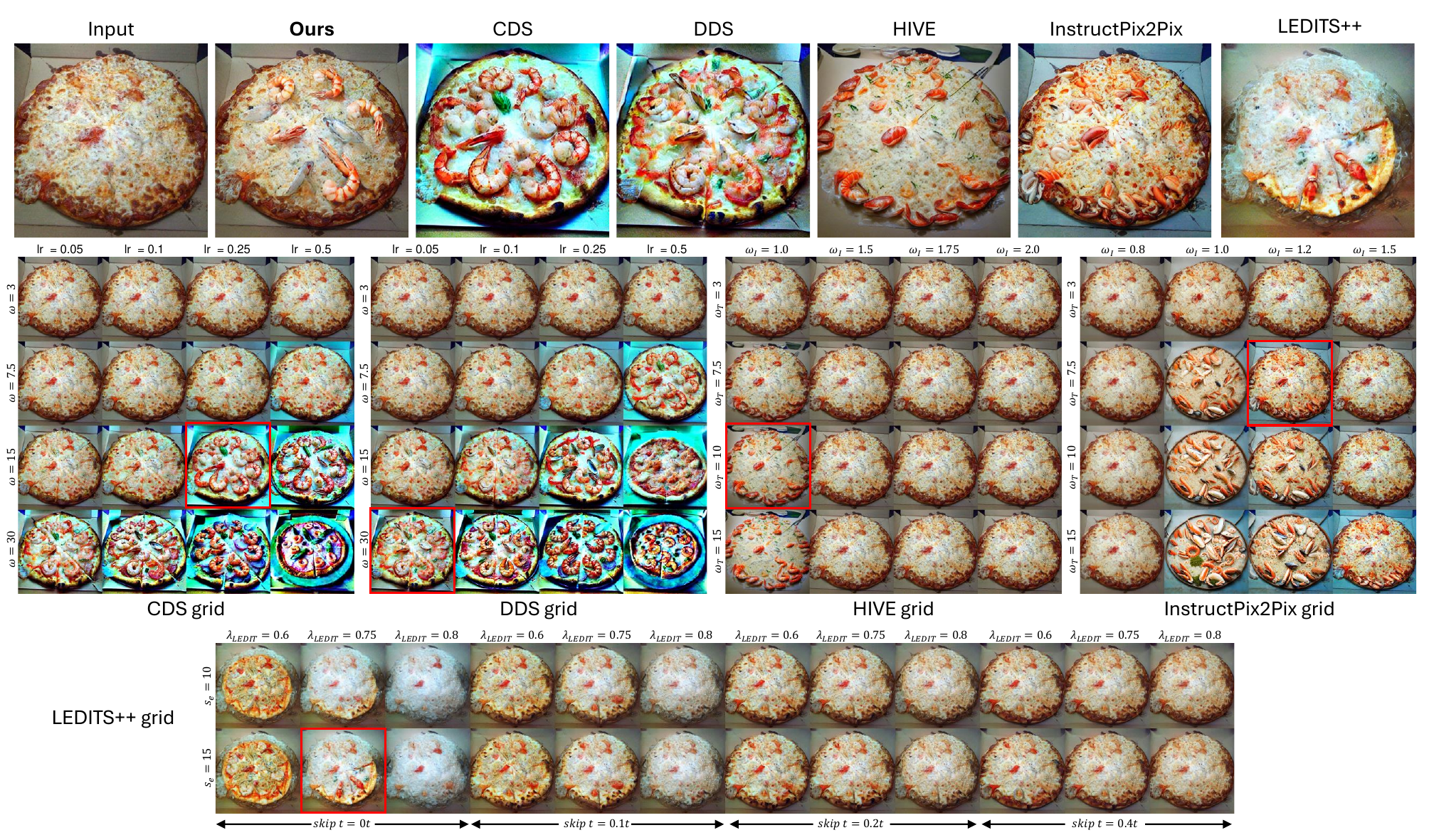}
        \caption{Prompt: \textcolor{red}{seafood} pizza}
        
    \end{subfigure}
    
    \begin{subfigure}[b]{0.99\textwidth}
        \centering
        \includegraphics[width=\textwidth]{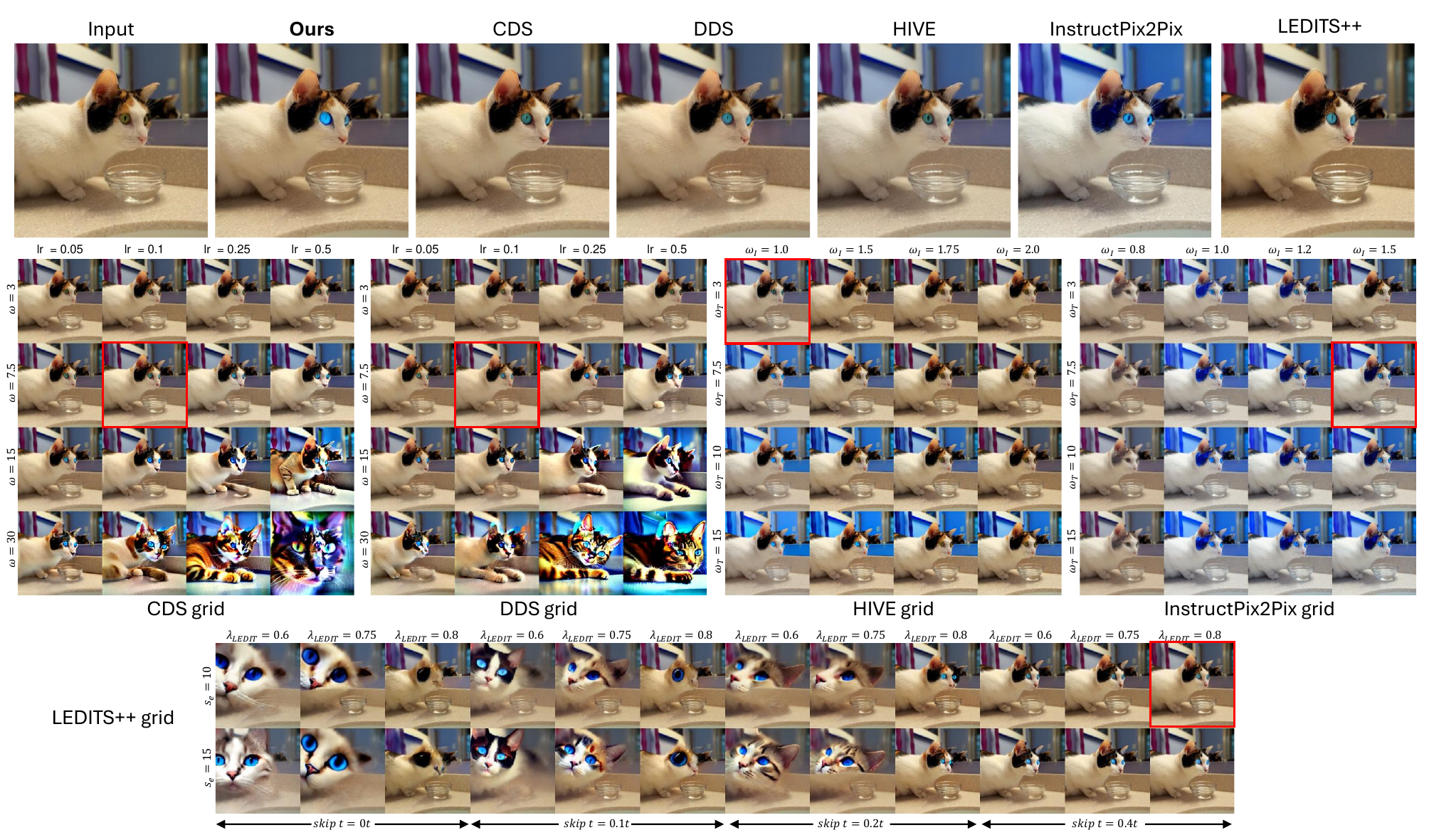}
        \caption{Prompt: a cat \textcolor{red}{with blue eyes}}
        
    \end{subfigure}
    
    \caption{Hypertuning grids on various in-the-wild editing tasks. We compare the best results (prioritizing object appearance) from state-of-the-art methods, optimized through hyperparameter tuning (highlighted by red boxes), with our results all generated using a single configuration. When several configurations perform equally, we choose the one that best preserves the background.}
    \vspace{-12pt}
    \label{fig:supp_grid9}
\end{figure*}

\begin{figure*}
    \centering
    \begin{subfigure}[b]{0.99\textwidth}
        \centering
        \includegraphics[width=\textwidth]{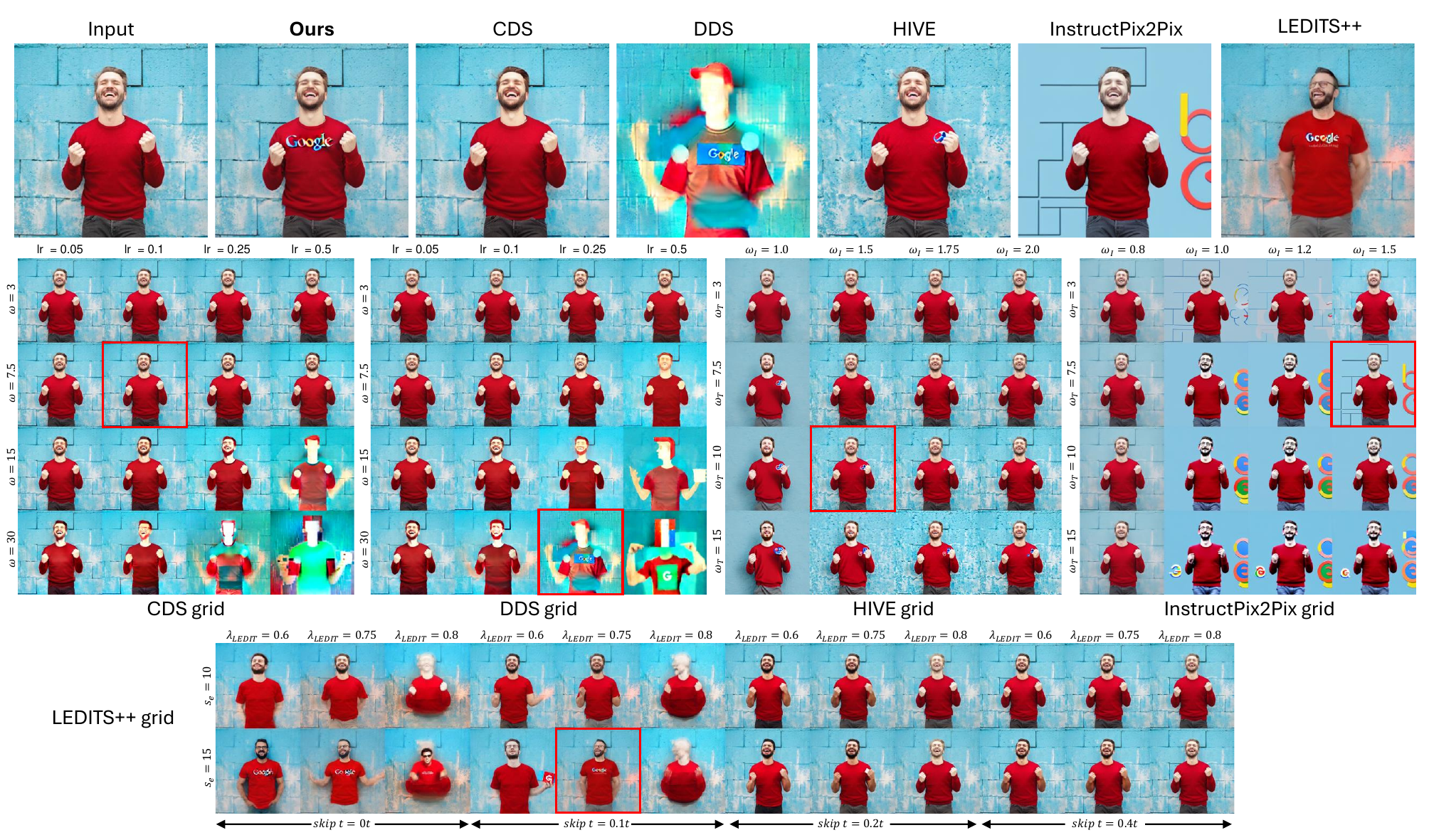}
        \caption{Prompt: a man wearing a red t-shirt \textcolor{red}{with a google logo}}
        
    \end{subfigure}
    
    \begin{subfigure}[b]{0.99\textwidth}
        \centering
        \includegraphics[width=\textwidth]{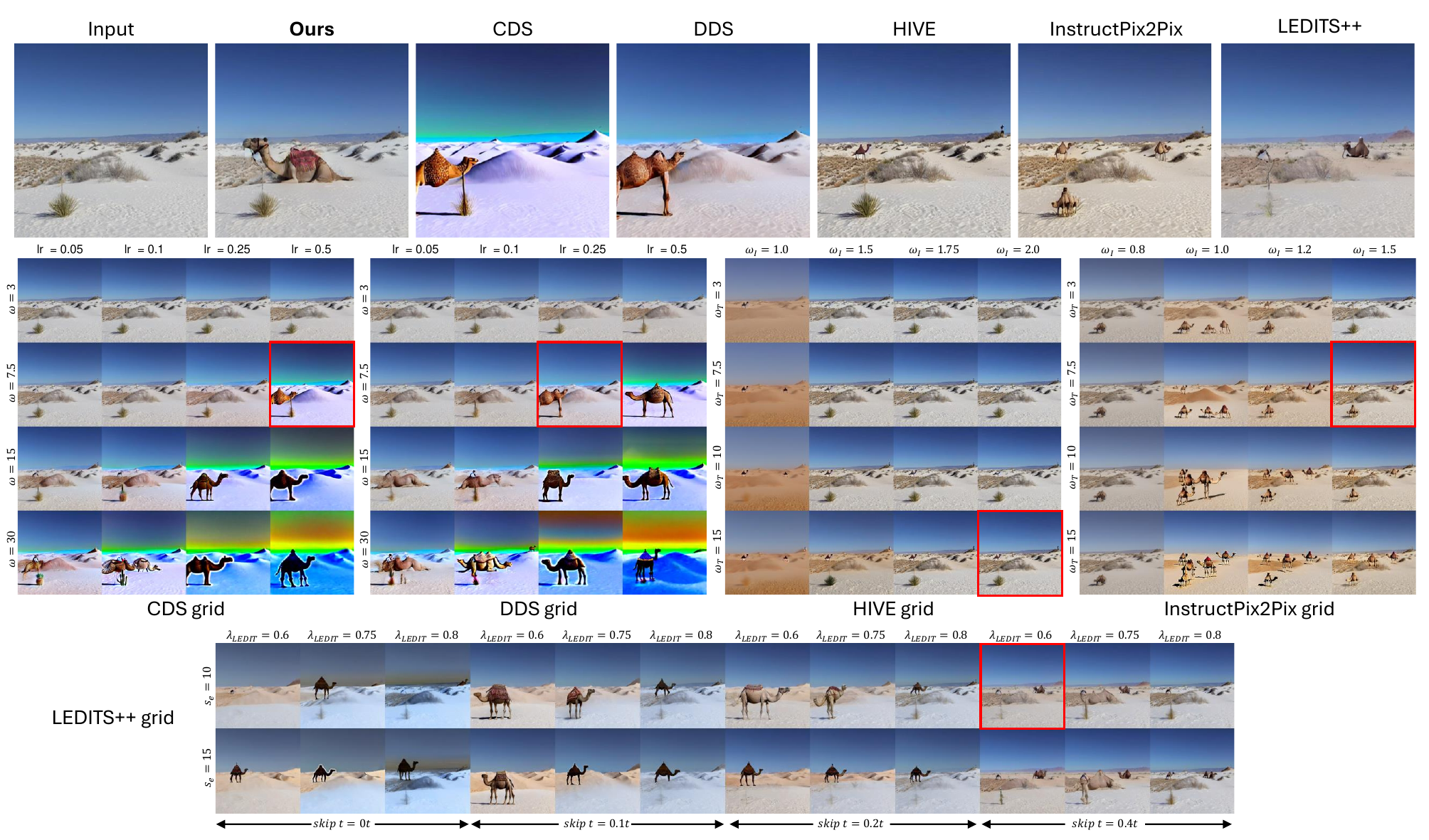}
        \caption{Prompt: a desert \textcolor{red}{with a camel}}
        
    \end{subfigure}
    
    \caption{Hypertuning grids on various in-the-wild editing tasks. We compare the best results (prioritizing object appearance) from state-of-the-art methods, optimized through hyperparameter tuning (highlighted by red boxes), with our results all generated using a single configuration. When several configurations perform equally, we choose the one that best preserves the background.}
    \vspace{-12pt}
    \label{fig:supp_grid10}
\end{figure*}

\begin{figure*}
    \centering
    \begin{subfigure}[b]{0.99\textwidth}
        \centering
        \includegraphics[width=\textwidth]{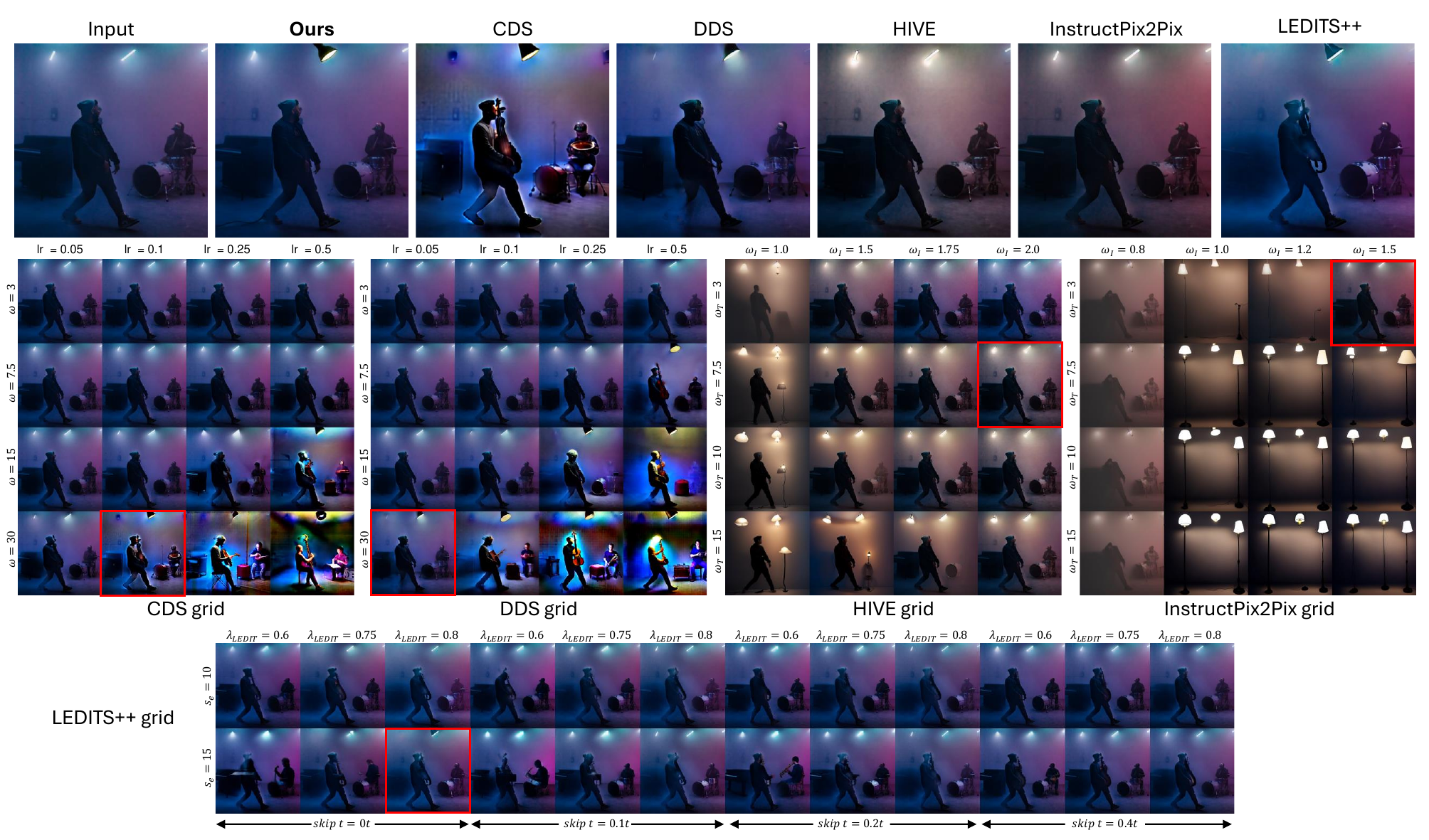}
        \caption{Prompt: musicians playing instruments inside a room \textcolor{red}{with a lamp}}
        
    \end{subfigure}
    
    \begin{subfigure}[b]{0.99\textwidth}
        \centering
        \includegraphics[width=\textwidth]{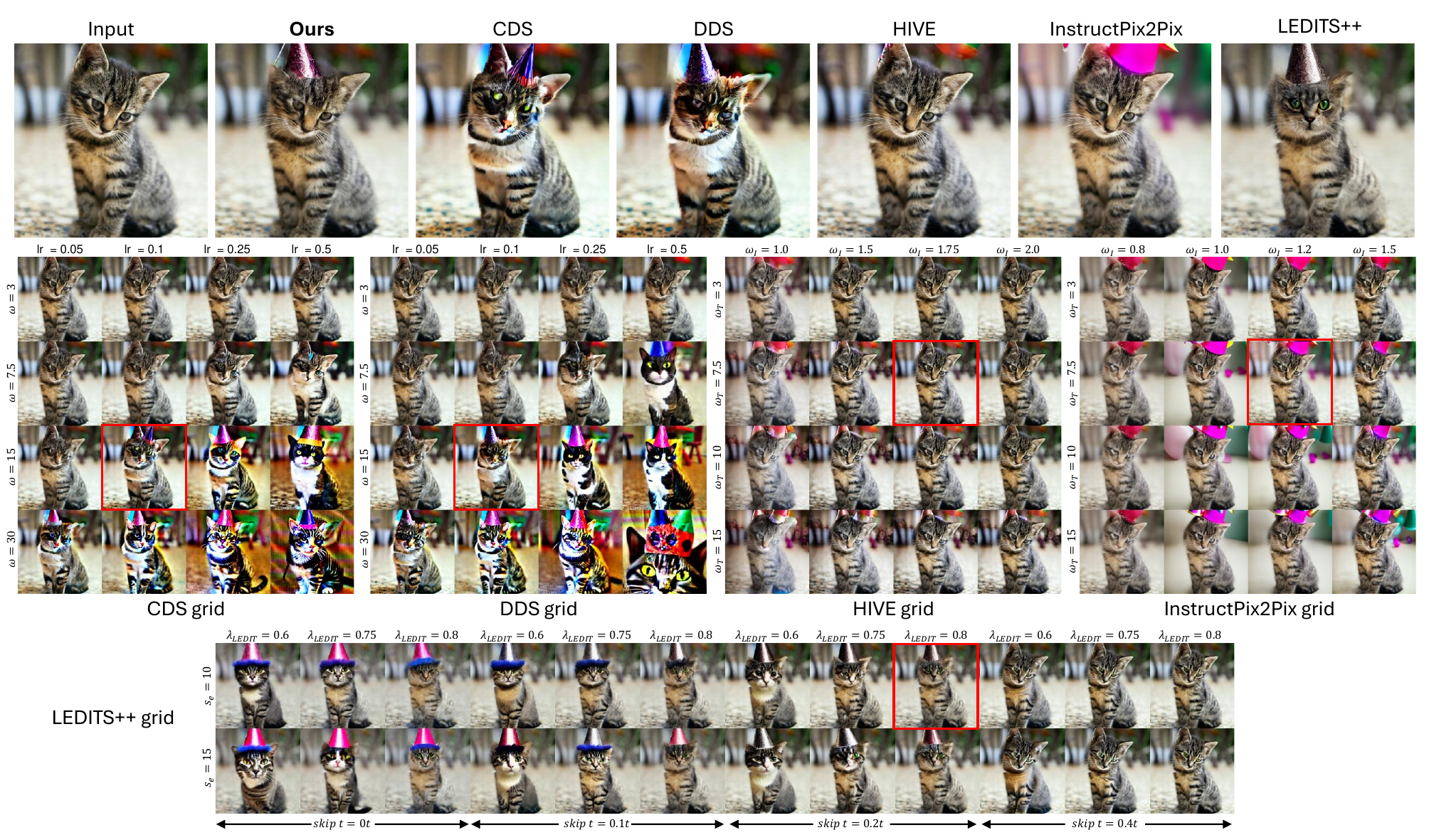}
        \caption{Prompt: a cat \textcolor{red}{wearing a party hat}}
        
    \end{subfigure}
    
    \caption{Hypertuning grids on various in-the-wild editing tasks. We compare the best results (prioritizing object appearance) from state-of-the-art methods, optimized through hyperparameter tuning (highlighted by red boxes), with our results all generated using a single configuration. When several configurations perform equally, we choose the one that best preserves the background.}
    \vspace{-12pt}
    \label{fig:supp_grid11}
\end{figure*}

\begin{figure*}
    \centering
    \begin{subfigure}[b]{0.99\textwidth}
        \centering
        \includegraphics[width=\textwidth]{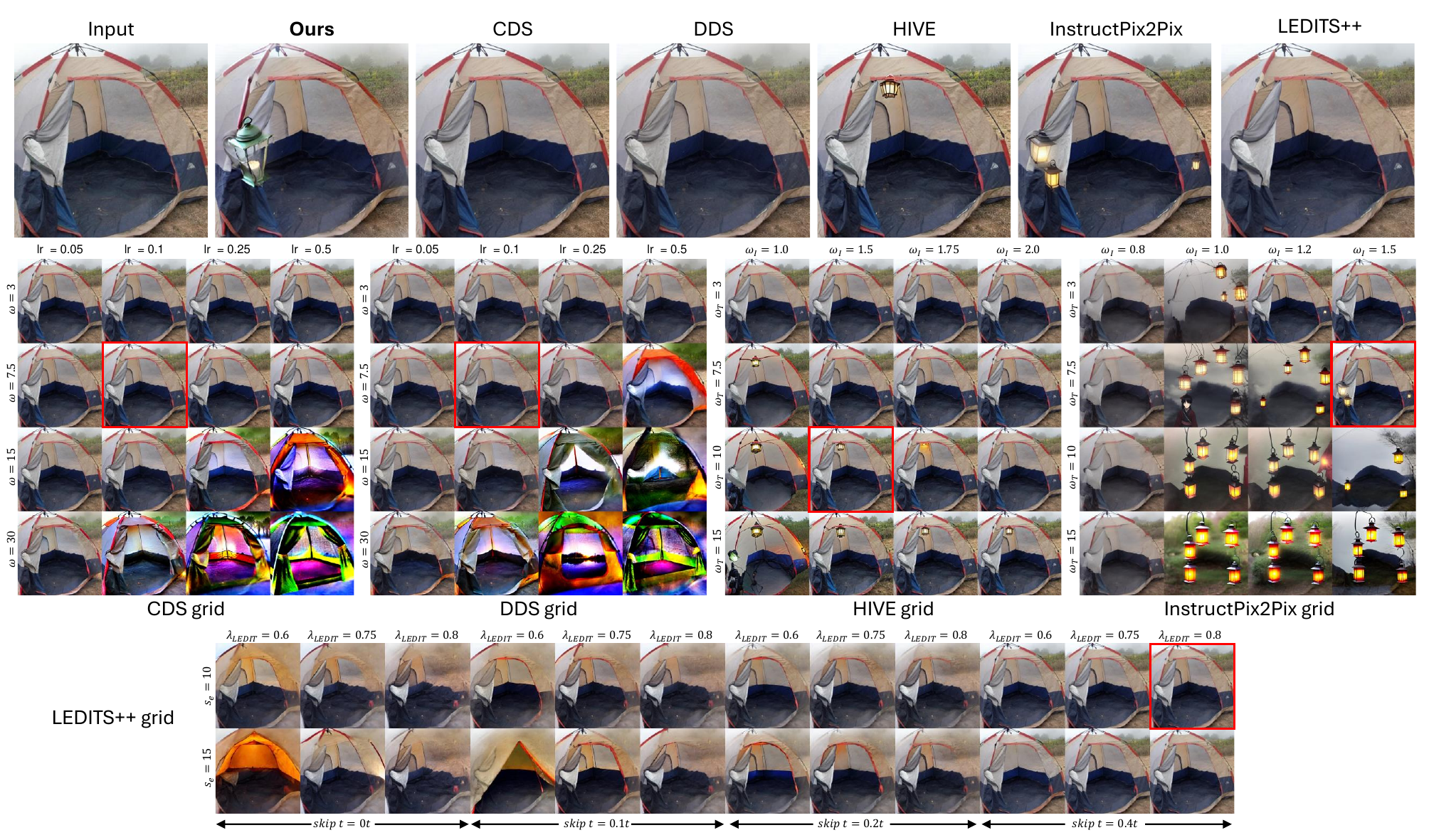}
        \caption{Prompt: a tent \textcolor{red}{with a lantern}}
        
    \end{subfigure}
    
    \begin{subfigure}[b]{0.99\textwidth}
        \centering
        \includegraphics[width=\textwidth]{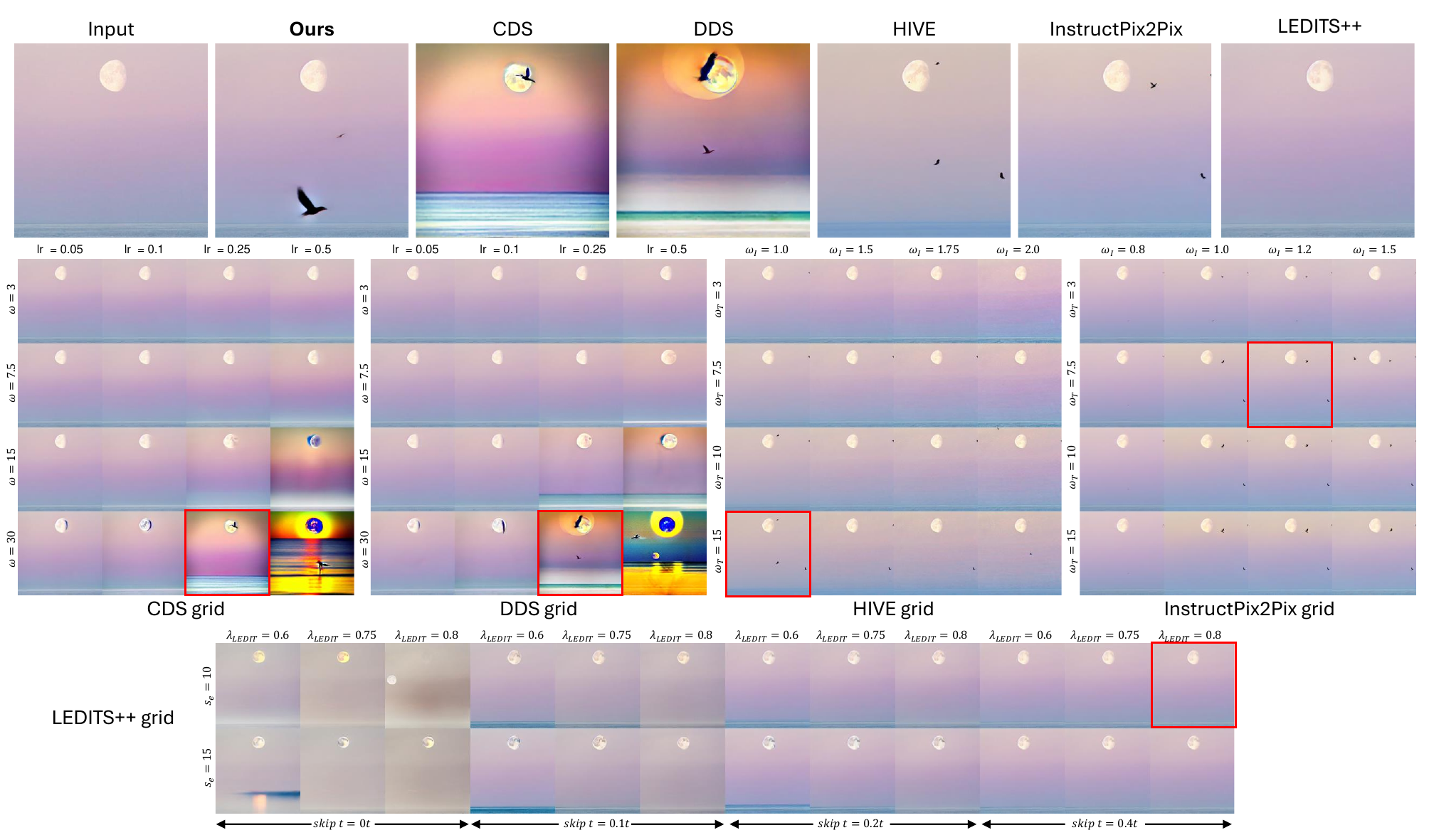}
        \caption{Prompt: an ocean with a clear blue sky and a full moon \textcolor{red}{and a bird}}
        
    \end{subfigure}
    
    \caption{Hypertuning grids on various in-the-wild editing tasks. We compare the best results (prioritizing object appearance) from state-of-the-art methods, optimized through hyperparameter tuning (highlighted by red boxes), with our results all generated using a single configuration. When several configurations perform equally, we choose the one that best preserves the background.}
    \vspace{-12pt}
    \label{fig:supp_grid12}
\end{figure*}

\begin{figure*}
    \centering
    \begin{subfigure}[b]{0.99\textwidth}
        \centering
        \includegraphics[width=\textwidth]{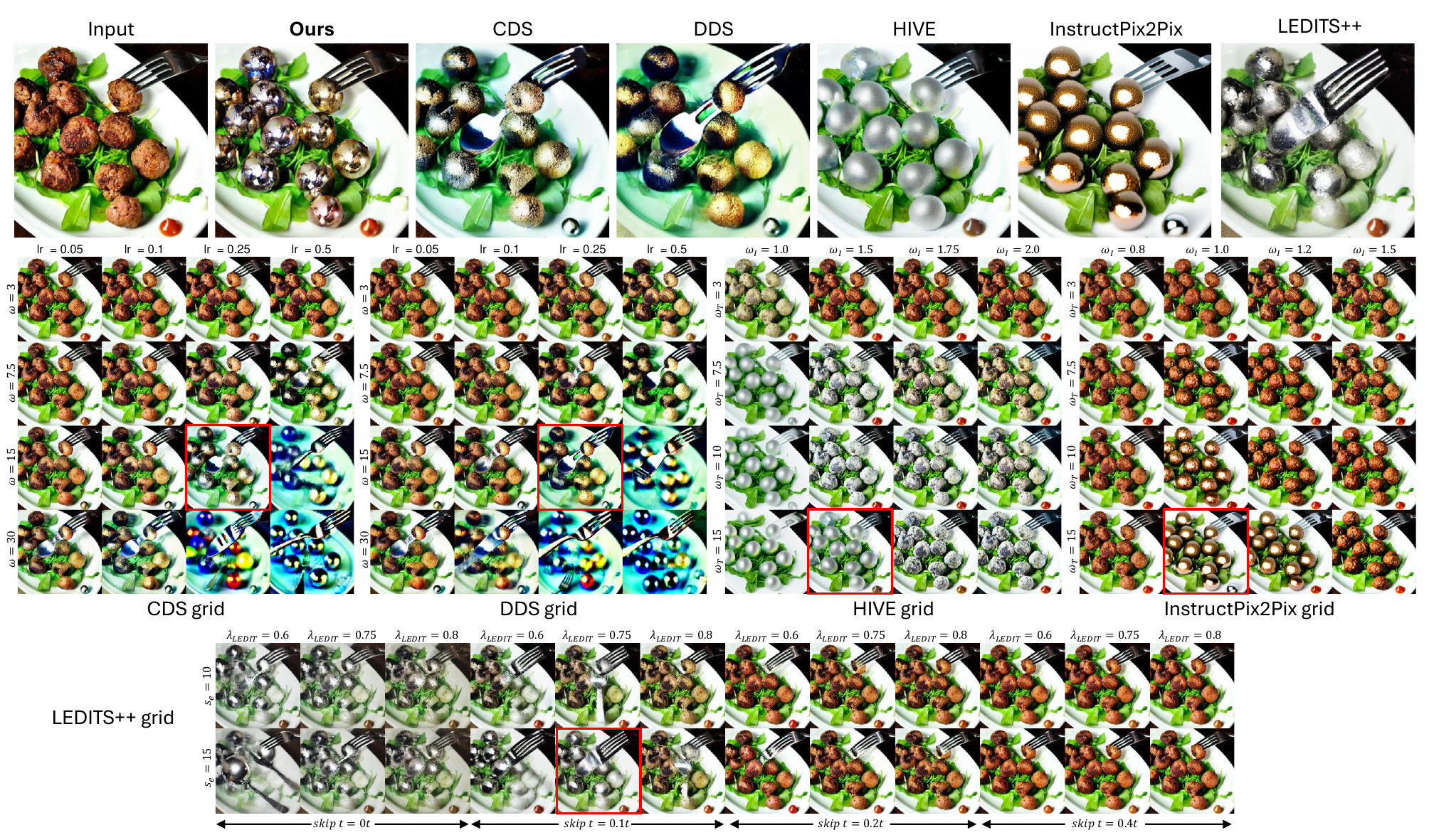}
        \caption{Prompt: (``meatballs'' $\rightarrow$ \textcolor{red}{``chrome balls''})}
        
    \end{subfigure}
    
    \begin{subfigure}[b]{0.99\textwidth}
        \centering
        \includegraphics[width=\textwidth]{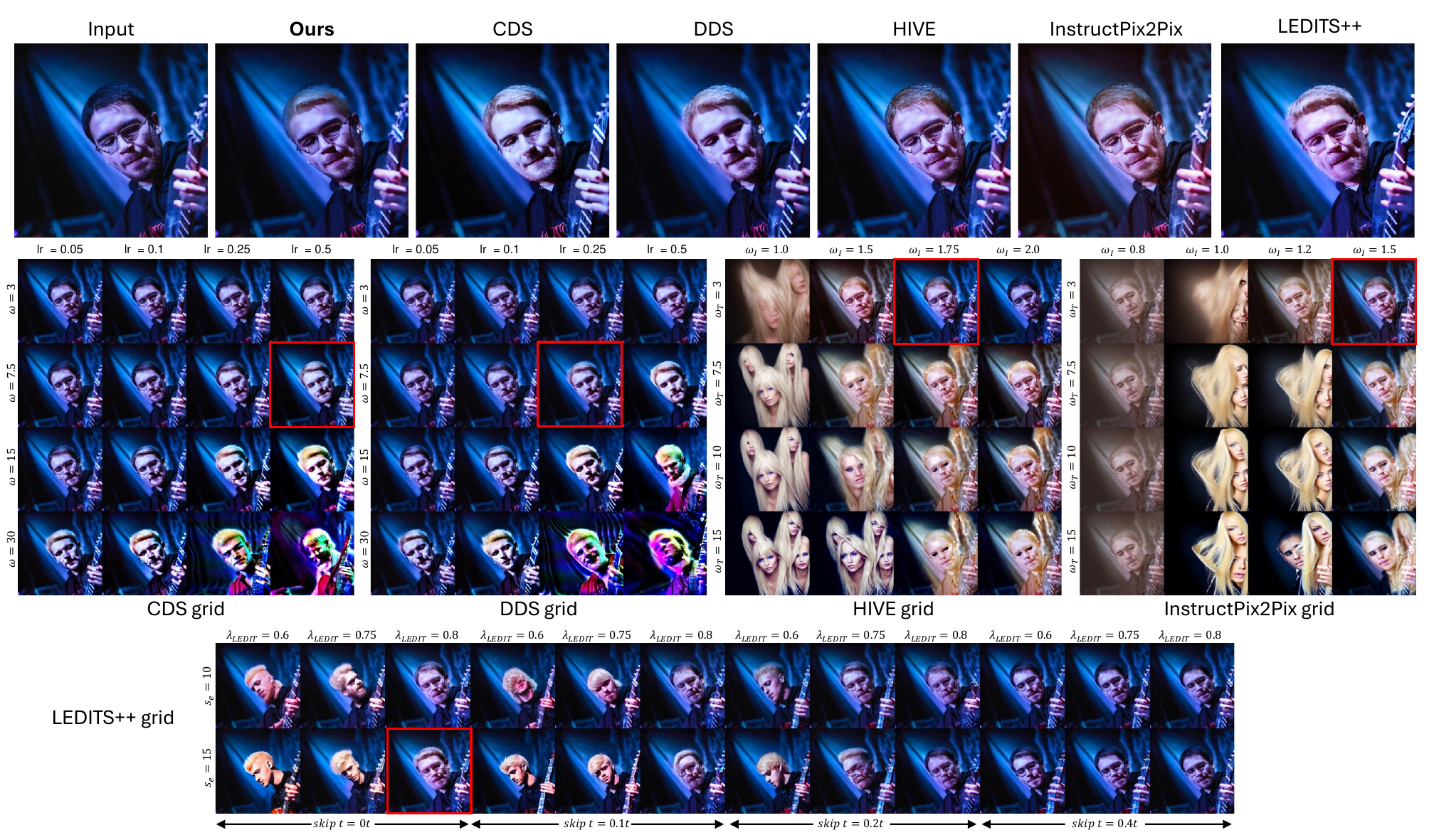}
        \caption{Prompt: a musician \textcolor{red}{with blonde hair}}
        
    \end{subfigure}
    
    \caption{Hypertuning grids on various in-the-wild editing tasks. We compare the best results (prioritizing object appearance) from state-of-the-art methods, optimized through hyperparameter tuning (highlighted by red boxes), with our results all generated using a single configuration. When several configurations perform equally, we choose the one that best preserves the background.}
    \vspace{-12pt}
    \label{fig:supp_grid13}
\end{figure*}

\begin{figure*}
    \centering
    \begin{subfigure}[b]{0.99\textwidth}
        \centering
        \includegraphics[width=\textwidth]{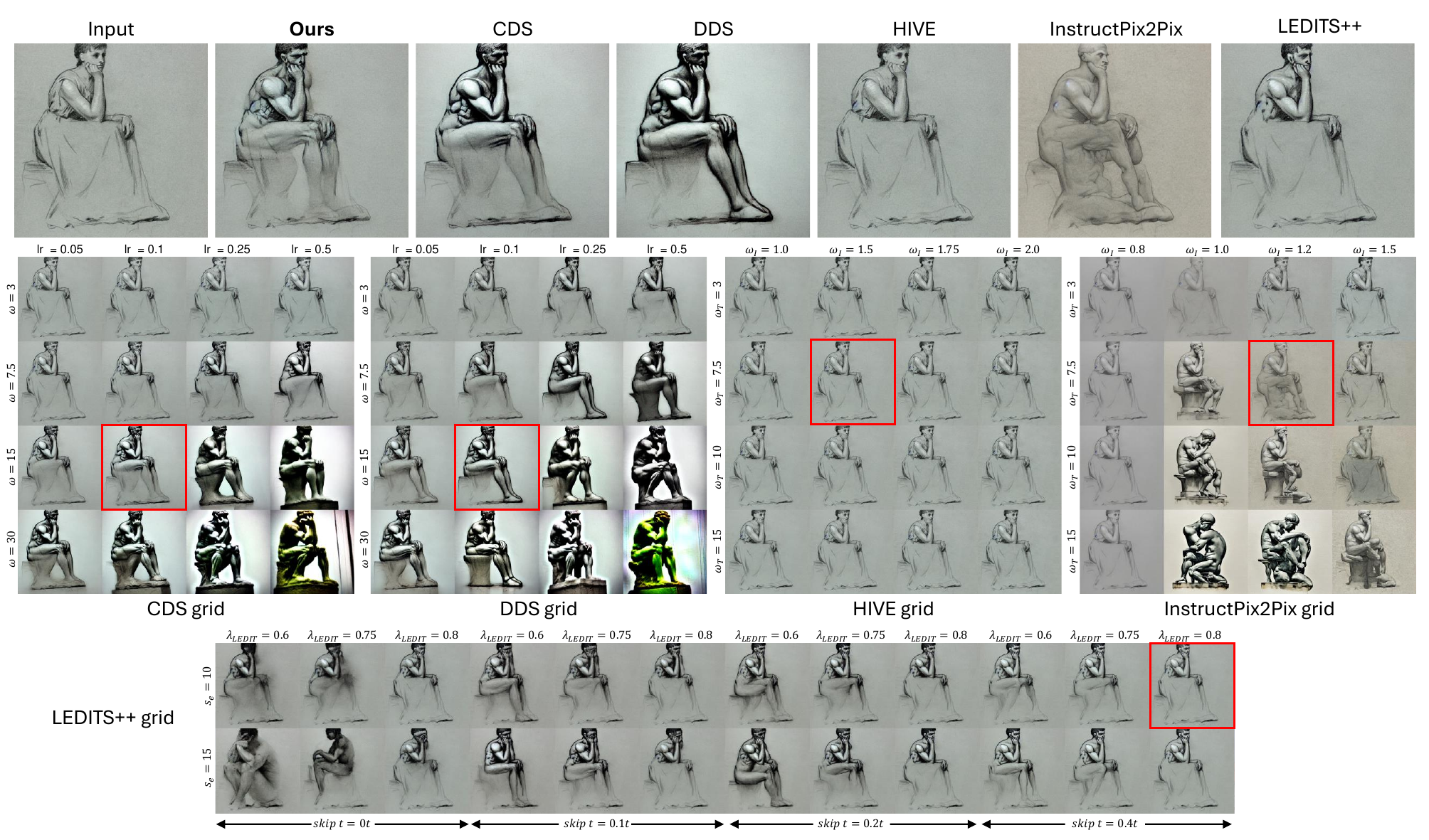}
        \caption{Prompt: (`` '' $\rightarrow$ \textcolor{red}{``the thinker''})}
        
    \end{subfigure}
    
    \caption{Hypertuning grids on various in-the-wild editing tasks. We compare the best results (prioritizing object appearance) from state-of-the-art methods, optimized through hyperparameter tuning (highlighted by red boxes), with our results all generated using a single configuration. When several configurations perform equally, we choose the one that best preserves the background.}
    \vspace{-12pt}
    \label{fig:supp_grid15}
\end{figure*}

In Figures \ref{fig:teaser}, \ref{fig:related_work_score_distillation} and \ref{fig:qual_results} in the main paper, along with Figures \ref{fig:supp_seed1} to \ref{fig:supp_seed3} in Appendix \ref{supp:qual_other}, we present qualitative comparison between our method and various SOTA approaches: CDS \cite{nam2024contrastive}, DDS \cite{hertz2023delta}, LEDITS++ \cite{brack2023ledits}, HIVE \cite{zhang2024hive}, and InstructPix2Pix \cite{brooks2023instructpix2pix}. In this section, we provide the hyperparameter tuning grids for all methods in Figures \ref{fig:supp_grid1} to \ref{fig:supp_grid15}. For each method, we tune the following hyperparameters:
\begin{enumerate}
    \item \textbf{InstructPix2Pix}:
    \begin{itemize}
        \item text guidance scale $\omega_T \in \{3, 7.5, 10, 15\}$
        \item image guidance scale $\omega_I \in \{0.8, 1.0, 1.2, 1.5\}$
    \end{itemize}
    \item \textbf{HIVE}:
    \begin{itemize}
        \item text guidance scale $\omega_T \in \{3, 7.5, 10, 15\}$
        \item image guidance scale $\omega_I \in \{1.0, 1.5, 1.75, 2.0\}$
    \end{itemize}
    \item \textbf{DDS} and \textbf{CDS}:
    \begin{itemize}
        \item learning rate $lr \in \{0.05, 0.10, 0.25, 0.50\}$
        \item guidance scale $\omega \in \{3, 7.5, 15, 30\}$
    \end{itemize}
    \item \textbf{LEDITS++}:
    \begin{itemize}
        \item skip time step $skip ~t \in \{0.0, 0.1, 0.2, 0.4\}$
        \item masking threshold $\lambda_{\text{LEDIT}} \in \{0.6, 0.75, 0.8\}$
        \item guidance scale $s_e \in \{10, 15\}$
    \end{itemize}
\end{enumerate}
Unlisted hyperparameters are set to their default values.




\section{More Comparison with DDS and CDS} 
\label{supp:compare_score_distillation}

\begin{figure*}
    \centering
    \includegraphics[width=1.0\textwidth, trim=0 {0.06\textheight} 0 0, clip]{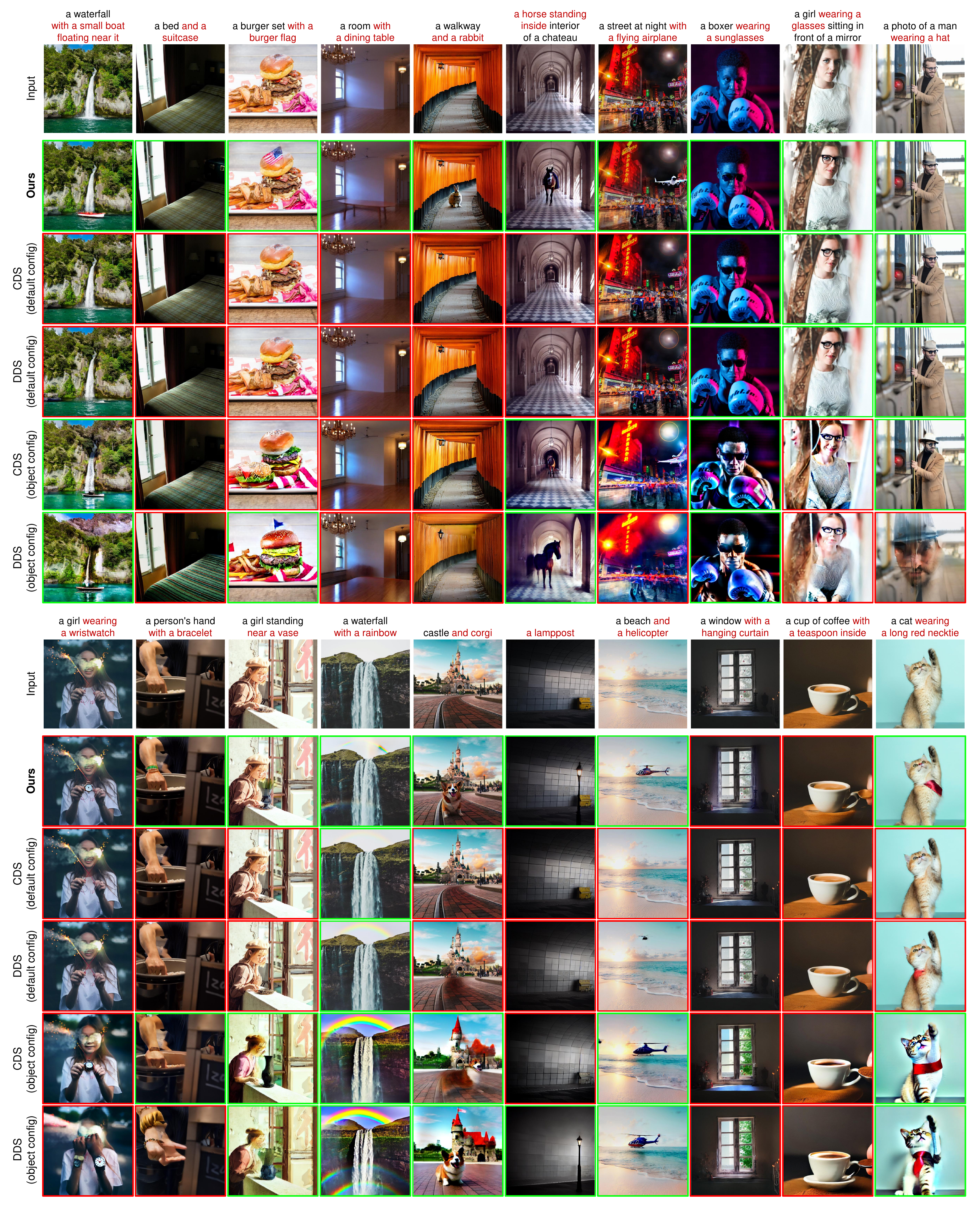}
    \caption{Qualitative results of SDS-based methods for object addition. For CDS \cite{nam2024contrastive} and DDS \cite{hertz2023delta}, we present results from both the default configuration and an alternative configuration (object config) that encourages object appearance but compromises background preservation. Successful cases are highlighted in green, while failed cases are highlighted in red. Our method has a higher success rate.}
    \vspace{-12pt}
    \label{fig:supp_object1}
\end{figure*}

In Figure \ref{fig:related_work_score_distillation} in Section \ref{sec:related_work}, we provide qualitative comparison for object insertion between our approach and existing SDS-based methods: CDS \cite{nam2024contrastive} and DDS \cite{hertz2023delta}. 
We present 20 additional object insertion results in Figures \ref{fig:supp_object1}.
For DDS and CDS, we include results from both their default configurations and a configuration optimized for better object insertion, manually selected based on hyperparameter tuning detailed in Appendix \ref{supp:tuning_grid}. This \textit{object configuration} employs a higher learning rate (0.25 instead of 0.1) and a higher classifier-free guidance value (15 instead of 7.5).

As shown in Figures \ref{fig:supp_object1}, 
our method and other SDS-based methods show competitive performance in common scenarios (e.g., adding sunglasses or a hat to a person). However, the default configurations of existing approaches fail to add objects in more challenging cases, such as inserting a horse into a chateau or putting a necktie on a cat. While the \textit{object configuration} alleviates this issue to some extent, it comes at the cost of poorer background preservation, particularly in earlier common scenarios. Additionally, this configuration still fails in certain instances, such as adding a rabbit to a walkway. In contrast, our method produces good results in most cases, albeit with some minor issues with minimal-effort regions (see Appendix \ref{supp:minimal_effort}).

\section{Extension to Other Score Distillation} \label{supp:extension_dds}

\begin{figure*}[ht]
    \centering
    \includegraphics[width=0.9\textwidth]{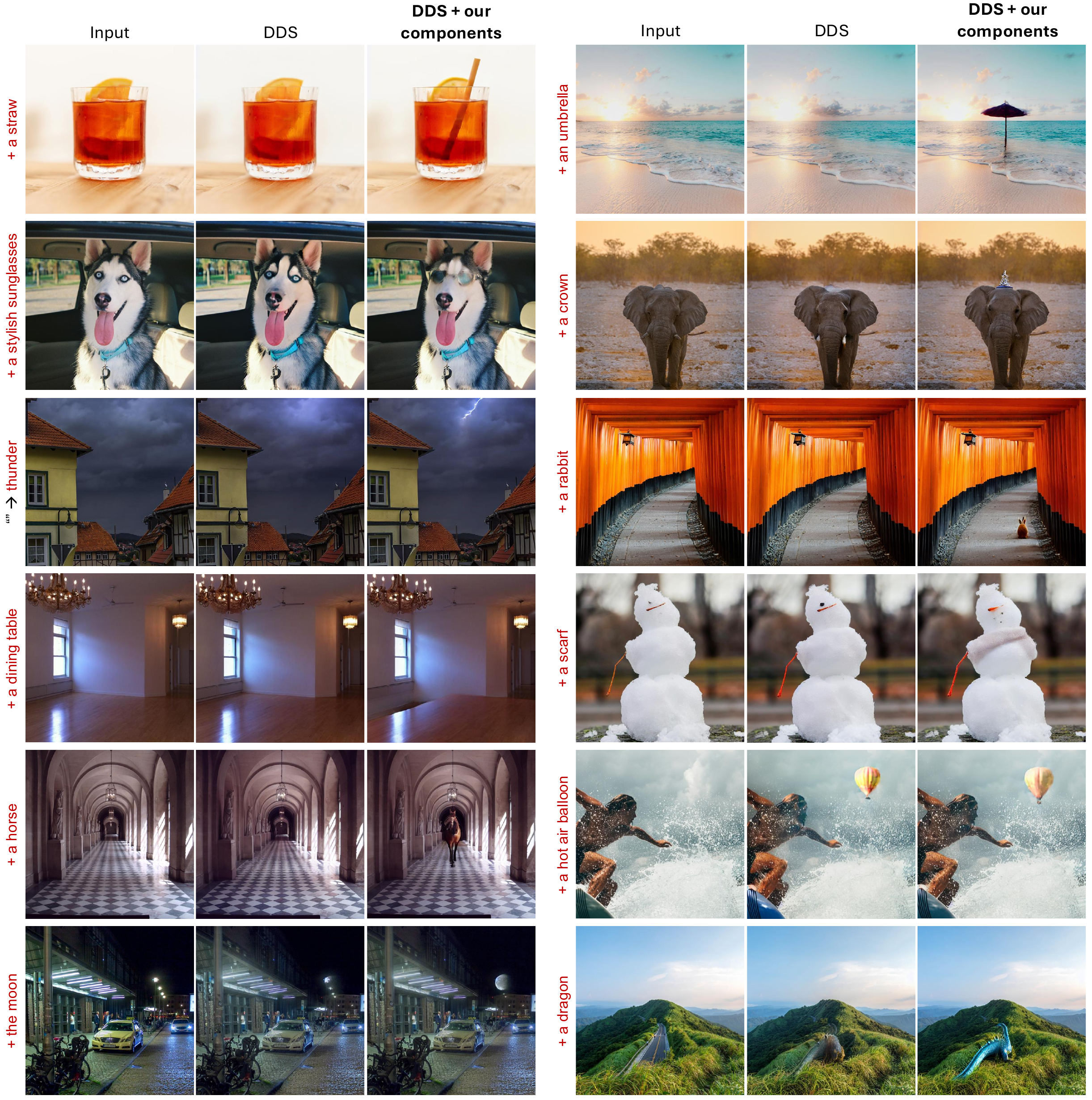}
    \captionsetup{skip=5pt}
    \caption{Our regularizer and gradient filtering/normalization help improve DDS's success rate and its background preservation in the default configuration.}
    \label{fig:aba_dds_improvement}
    \vspace{-0.8em}
\end{figure*}

In this work, we introduce attention-based spatial regularization, along with gradient filtering and normalization, to enhance prompt fidelity while preserving the background in SBP \cite{mcallister2024rethinking} for image editing. Nonetheless, our preliminary study suggests that these components can also effectively improve other distillation algorithms, such as DDS \cite{hertz2023delta}, as illustrated in Figure \ref{fig:aba_dds_improvement}. This can be done by simply modifying the noise prediction step (line 8-9 in Algorithm \ref{algo:lusd}) to reflect the DDS loss:
\begin{equation}
    \nabla_{\vect{z}} \mathcal{L}_{\text{DDS}} = \epsilon_{\phi}(\vect{z}_t, y^{\text{tgt}}, t) - \epsilon_{\phi}(\vect{z}^{\text{src}}_t, y^{\text{src}}, t),
\end{equation}
where $\vect{z}^{\text{src}}_t$ denotes a noisy latent code of the original image.

\begin{figure*}
    \centering
    \includegraphics[width=0.99\textwidth]{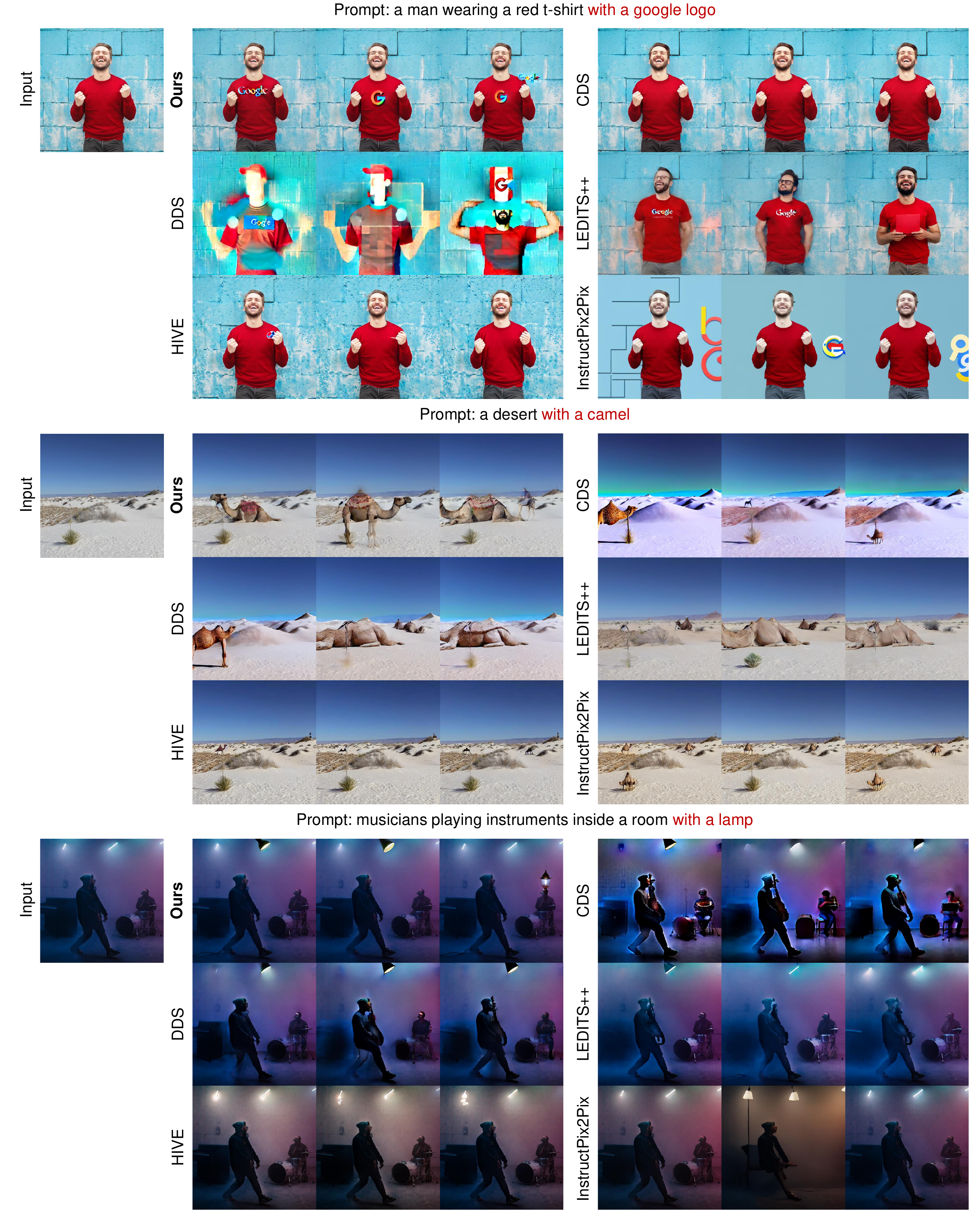}
    \captionsetup{skip=5pt}
    \caption{Qualitative results on various in-the-wild editing tasks. We compare the best results (prioritizing object appearance) from state-of-the-art methods, optimized through hyperparameter tuning (see Appendix \ref{supp:tuning_grid}), with our results all generated using a single configuration. For each input image, we show results from 3 different random seeds.}
    \vspace{-12pt}
    \label{fig:supp_seed1}
\end{figure*}

\begin{figure*}
    \centering
    \includegraphics[width=0.99\textwidth]{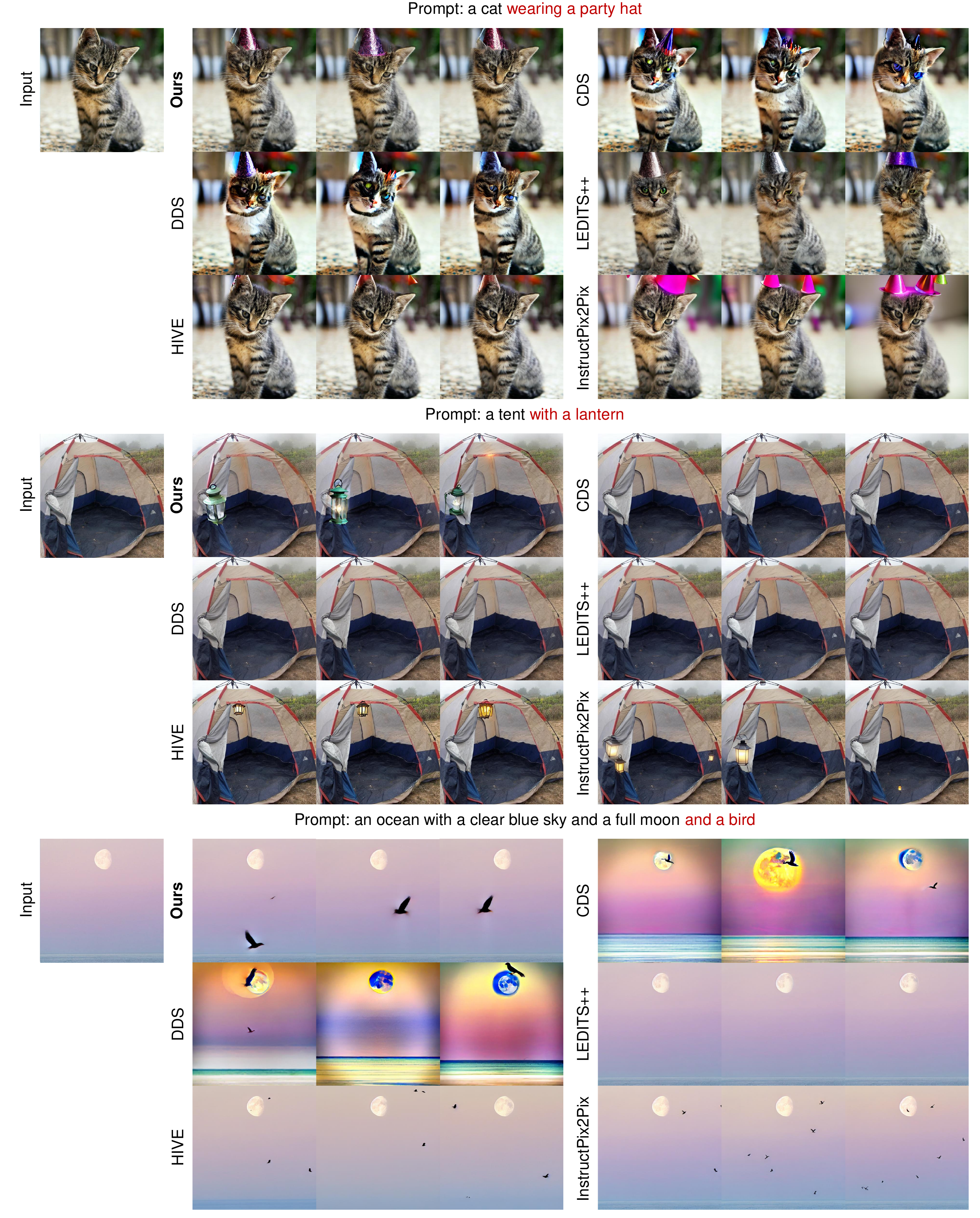}
    \captionsetup{skip=5pt}
    \caption{Qualitative results on various in-the-wild editing tasks. We compare the best results (prioritizing object appearance) from state-of-the-art methods, optimized through hyperparameter tuning (see Appendix \ref{supp:tuning_grid}), with our results all generated using a single configuration. For each input image, we show results from 3 different random seeds.}
    \vspace{-12pt}
    \label{fig:supp_seed2}
\end{figure*}

\begin{figure*}
    \centering
    \includegraphics[width=0.99\textwidth]{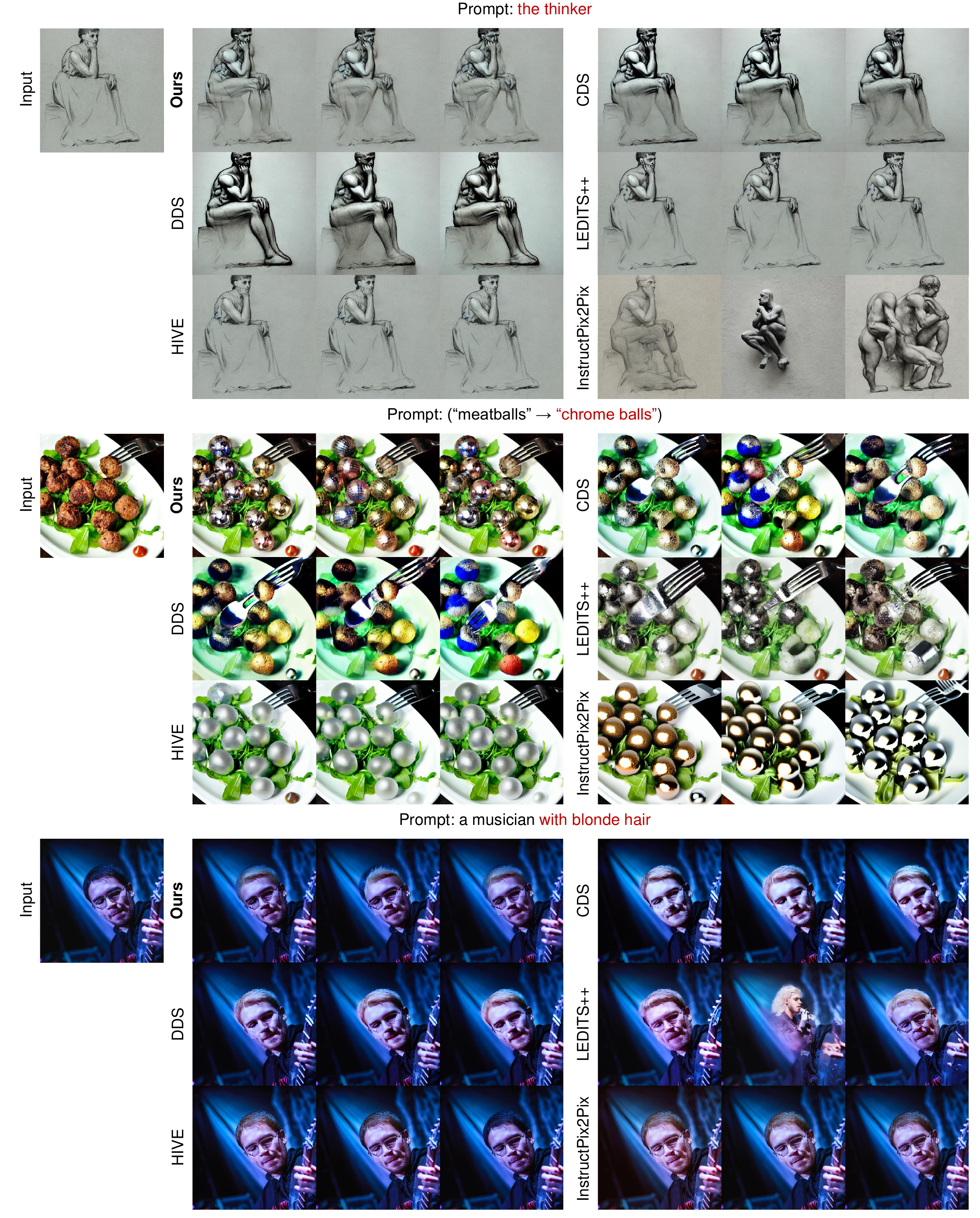}
    \captionsetup{skip=5pt}
    \caption{Qualitative results on various in-the-wild editing tasks. We compare the best results (prioritizing object appearance) from state-of-the-art methods, optimized through hyperparameter tuning (see Appendix \ref{supp:tuning_grid}), with our results all generated using a single configuration. For each input image, we show results from 3 different random seeds.}
    \vspace{-12pt}
    \label{fig:supp_seed3}
\end{figure*}

\end{document}